\documentclass[article]{aastex631}

\usepackage{hyperref}
\usepackage{color}
\usepackage{soul}
\usepackage{natbib}
\usepackage{hyperref}
\usepackage{amsmath,bm}
\usepackage{xspace}
\usepackage{graphicx}
\usepackage{enumitem}
\usepackage{amssymb}
\usepackage{xifthen}
\usepackage{hyperref}
\usepackage[normalem]{ulem}
\usepackage{comment}
\usepackage{amsmath}

\hypersetup{    
  colorlinks      = {true},
  linkcolor       = {blue},
  citecolor       = {blue},
  urlcolor        = {blue},
}

\graphicspath{{./}{figures/}}

\definecolor{orange}{rgb}{1.0,0.5,0.}

\DeclareMathOperator{\sech}{sech}

\def\MDM{\ifmmode{\>M_{\textnormal{\sc dm}}}\else{$$M_{\textnormal{\sc dm}}}\fi}

\def\XH{\ifmmode{\>X_{\textnormal{\sc h}}} \else{$X_{\textnormal{\sc h}}$}\fi}
\def\nH{\ifmmode{\>n_{\textnormal{\sc h}}} \else{$n_{\textnormal{\sc h}}$}\fi}

\def\maspyr{\ifmmode{\>\textnormal{mas~yr}^{-1}}\else{mas~yr$^{-1}$}\fi}

\def\mG{\ifmmode{\>\mu\mathrm{G}}\else{$\mu$G}\fi}
\def\erg{\ifmmode{\> {\rm erg}}\else{erg}\fi}
\def\keV{\ifmmode{\> {\rm keV}}\else{keV}\fi}

\def\deg{\ifmmode{\>^{\circ}}\else{$^{\circ}$}\fi}
\def\onedeg{\ifmmode{\>1^{\circ}}\else{$1^{\circ}$}\fi}

\def\xvir{\ifmmode{\>\!x_{vir}}\else{$x_{vir}$}\fi}
\def\Mvir{\ifmmode{\>\!M_{vir} }\else{$M_{vir} $}\fi}
\def\rvir{\ifmmode{\>\!r_{vir}}\else{$r_{vir}$}\fi}
\def\vvir{\ifmmode{\>\!v_{vir}}\else{$v_{vir}$}\fi}
\def\Vvir{\ifmmode{\>\!V_{vir} }\else{$V_{vir} $}\fi}

\def\tratio{\ifmmode{\>\tau}\else{$\tau$}\fi}

\def\rms{\ifmmode{\>r_{\textnormal{\sc ms}}}\else{$r_{\textnormal{\sc ms}}$}\fi}

\def\Mpc{\ifmmode{\>\!{\rm Mpc}} \else{Mpc}\fi}
\def\kpc{\ifmmode{\>\!{\rm kpc}} \else{kpc}\fi}
\def\pkpc{\ifmmode{\>\!{\rm kpc}^{-1}} \else{kpc$^{-1}$}\fi}
\def\pc{\ifmmode{\>\!{\rm pc}} \else{pc}\fi}

\def\Gyr{\ifmmode{\>\!{\rm Gyr}} \else{Gyr}\fi}
\def\Myr{\ifmmode{\>\!{\rm Myr}} \else{Myr}\fi}
\def\yr{\ifmmode{\>\!{\rm yr}} \else{yr}\fi}
\def\pyr{\ifmmode{\>\!{\rm yr}^{-1}}\else{yr$^{-1}$} \fi}
\def\s{\ifmmode{\>\!{\rm s}}\else{s}\fi}
\def\ps{\ifmmode{\>\!{\rm s}^{-1}}\else{s$^{-1}$}\fi}
\def\Hz{\ifmmode{\>\!{\rm Hz}}\else{Hz}\fi}

\def\kms{\ifmmode{\>\!{\rm km\,s}^{-1}}\else{km~s$^{-1}$}\fi}

\def\K{\ifmmode{\>\!{\rm K}}\else{K}\fi}

\def\sr{\ifmmode{\>\!{\rm sr}}\else{sr}\fi}
\def\psr{\ifmmode{\>\!{\rm sr}^{-1}}\else{sr$^{-1}$}\fi}
\def\arcs{\ifmmode{\>\!{\rm arcsec}}\else{arcsec}\fi}
\def\parcs{\ifmmode{\>\!{\rm arcsec}^{-1}}\else{arcsec${-1}$}\fi}
\def\parcss{\ifmmode{\>\!{\rm arcsec}^{-2}}\else{arcsec${-2}$}\fi}

\def\cm{\ifmmode{\>\!{\rm cm}}\else{cm}\fi}
\def\cc{\ifmmode{\>\!{\rm cm}^{3}}\else{cm$^{3}$}\fi}
\def\sqc{\ifmmode{\>\!{\rm cm}^{2}}\else{cm$^{2}$}\fi}
\def\pcc{\ifmmode{\>\!{\rm cm}^{-3}}\else{cm$^{-3}$}\fi}
\def\psc{\ifmmode{\>\!{\rm cm}^{-2}}\else{cm$^{-2}$}\fi}

\def\g{\ifmmode{\>\!{\rm g}}\else{g}\fi}
\def\Msun{\ifmmode{\>\!{\rm M}_{\odot}}\else{M$_{\odot}$}\fi}
\def\hMsun{\ifmmode{\> h^{-1}{\rm M}_{\odot}}\else{$h^{-1}$M$_{\odot}$}\fi}

\def\Zsun{\ifmmode{\>\!{\rm Z}_{\odot}}\else{Z$_{\odot}$}\fi}

\def\Lsun{\ifmmode{\>\!{\rm L}_{\odot}}\else{L$_{\odot}$}\fi}

\def\rayl{\ifmmode{\>\!{\rm R}}\else{R}\fi}
\def\mR{\ifmmode{\>\!{\rm mR}}\else{mR}\fi}

\def\lya{\ifmmode{\>\!{\rm Ly}\alpha}\else{Ly$\alpha$}\fi}

\def\Ha{\ifmmode{\>\!{\rm H}\alpha}\else{H$\alpha$}\fi}
\def\Hb{\ifmmode{\>\!{\rm H}\beta}\else{H$\beta$}\fi}

\def\HI{\ifmmode{\> \textnormal{\ion{H}{1}}} \else{\ion{H}{1}}\fi}
\def\HII{\ifmmode{\> \textnormal{\ion{H}{2}}} \else{\ion{H}{2}}\fi}
\def\CIV{\ifmmode{\> \textnormal{\ion{C}{4}}} \else{\ion{C}{4}}\fi}
\def\SiIV{\ifmmode{\> \textnormal{\ion{S}{4}}} \else{\ion{Si}{4}}\fi}

\def\NH{\ifmmode{\> {\rm N}_{\rm H}} \else{N$_{\rm H}$}\fi}
\def\Ng{\ifmmode{\> {\rm N}_{\rm gas}} \else{N$_{\rm gas}$}\fi}
\def\NHI{\ifmmode{\> {\rm N}_{\HI}} \else{N$_{\HI}$}\fi}
\def\MHI{\ifmmode{\> {\rm M}_{ \HI}} \else{M$_{\HI}$}\fi}

\def\mua{\ifmmode{\>\mu_{ \textnormal{\Ha}}}\else{$\mu_{ \textnormal{\Ha}}$}\fi}
\def\alphabha{\ifmmode{\>\alpha_{B}^{(\textnormal{\Ha})}}\else{$\alpha_{B}^{(\textnormal{\Ha})}$}\fi}

\newcommand{\ramses}{{\sc Ramses}}
\newcommand\agama{{\sc agama}}
\newcommand\STARBURST{{\sc starburst99}}

\newcommand{\unit}[1]{\ensuremath{\mathrm{\,#1}}\xspace}

\newcommand{\Msol}{\unit{M_\odot}}

\newcommand{\sfrff}{\mathrm{SFR}_\mathrm{ff}}
\newcommand{\sigs}{\sigma_s}
\newcommand{\scrit}{s_\mathrm{crit}}
\newcommand{\mach}{\mathcal{M}}
\newcommand{\alphavir}{\alpha_\mathrm{vir}}
\newcommand{\eps}{\epsilon}
\newcommand{\phit}{\phi_t}
\newcommand{\phix}{\phi_x}
\newcommand{\tff}{t_\mathrm{ff}}

\def\ex#1{\langle#1\rangle}

\defcitealias{pla20a}{Planck (2020)}

\shorttitle{Baryon domination in early disk galaxies}
\shortauthors{Bland-Hawthorn, Tepper-Garc\'ia, Agertz, Federrath}

\begin{document}

\title{Turbulent gas-rich disks at high redshift: bars \& bulges in a radial shear flow}


\author[0000-0001-7516-4016]{Joss~Bland-Hawthorn}
\affiliation{Sydney Institute for Astronomy, School of Physics, A28, The University of Sydney, NSW 2006, Australia}
\affiliation{Centre of Excellence for All-Sky Astrophysics in Three Dimensions (ASTRO 3D), Australia}

\author[0000-0002-1081-883X]{Thor Tepper-Garcia}
\affiliation{Sydney Institute for Astronomy, School of Physics, A28, The University of Sydney, NSW 2006, Australia}
\affiliation{Centre of Excellence for All-Sky Astrophysics in Three Dimensions (ASTRO 3D), Australia}

\author[0000-0002-4287-1088]{Oscar Agertz}
\affiliation{Lund Observatory, Division of Astrophysics, Department of Physics, Lund University, Box 43, SE-221 00 Lund, Sweden}

\author[0000-0002-0706-2306]{Christoph Federrath}
\affiliation{Research School of Astronomy and Astrophysics, Australian National University, Canberra, ACT 2611, Australia}
\affiliation{Centre of Excellence for All-Sky Astrophysics in Three Dimensions (ASTRO 3D), Australia}


\correspondingauthor{J. Bland-Hawthorn}
\email{jonathan.bland-hawthorn@sydney.edu.au}
 
\begin{abstract}
Recent observations of high-redshift galaxies ($z\lesssim 7$) reveal that a substantial fraction have turbulent, gas-rich disks with well-ordered rotation and elevated levels of star formation. In some instances, disks show evidence of spiral arms, with bar-like structures. These remarkable observations have encouraged us to explore a new class of dynamically self-consistent models using
our \agama/\ramses\ hydrodynamic N-body simulation framework that mimic a plausible progenitor of the Milky Way at high redshift. We explore disk gas fractions of $f_{\rm gas} = 0, 20, 40, 60, 80, 100\%$ and track the creation of stars and metals. The high gas surface densities encourage vigorous star formation, which in turn couples with the gas to drive turbulence. We explore three distinct histories: (i) there is no ongoing accretion and the gas is used up by the star formation; (ii) the star-forming gas is replenished by cooling in the hot halo gas; (iii) in a companion paper, we revisit these models in the presence of a strong perturbing force.
At low $f_{\rm disk}$ ($\lesssim 0.3$), where $f_{\rm disk}$ is the baryon mass fraction of the disk relative to dark matter within 2.2 $R_{\rm disk}$, a bar does not form in a stellar disk; this remains true even when gas dominates the inner disk potential. For a dominant baryon disk ($f_{\rm disk} \gtrsim 0.5$) at all gas fractions, the turbulent gas forms a strong {\it radial shear flow} that leads to an intermittent star-forming bar within about 500~Myr; turbulent gas speeds up the formation of bars compared to gas-free models. For $f_{\rm gas} \lesssim 60\%$, all bars survive, but for higher gas fractions, the bar devolves into a central bulge after 1~Gyr. The star-forming bars are reminiscent of recent discoveries in high-redshift ALMA observations of gaseous disks.
\end{abstract}


\section{Introduction} \label{s:intro}

Stellar bars have long intrigued astronomers, not least because they are a common feature of galaxies in the local universe \citep{erwin18}, but what do they teach us? A cold stellar disk with ordered rotation is a low-entropy system that wants to transport angular momentum outwards. Bar instabilities are routinely observed in N-body disk simulations \citep{hoh71,athan86}, and are a genuine feature of synthetic disks supported by other approaches to dynamical analysis \citep{sel93}.
{\it All} galactic disks are unstable to the onset of bars ultimately $-$
just why they arise was first illustrated by \cite{too81}. Disk perturbations are amplified into outward-propagating (leading) waves by the shearing action of the disk; if the inner galactic potential is not too deep, the trailing waves can tunnel through the centre and emerge as leading waves, thereby ``swing amplifying" the instability in a positive feedback loop, analogous to the action of a resonant laser cavity. Thus, bars probe the inner galactic potential (including the influence of any supermassive black hole), the dynamical stability and maturity of the stellar disk, the outer halo through the bar's angular velocity and, most notably, the degree of baryon domination over dark matter within the domain of the stellar disc.

In light of recent developments, our view of disk formation in the early universe is undergoing a radical overhaul. With the launch of the James Webb Space Telescope (JWST), a new window into the distant universe has revealed that (mostly thin) disk galaxies\footnote{We caution that the JWST disk fraction is inferred from multiband photometry. At $z<1$, \citet{neichel08} found a high degree of correlation between photometric ``disks'' and ordered rotation in kinematic studies, but the relevance of the S\'{e}rsic fitting function out to such high redshifts has yet to be established.} dominate the population out to at least $z\approx 6$ \citep{ferreira22,kartaltepe23,robert23}, although further analysis is needed \citep{huert23}. 
We can use bars (and bulges) to learn about the early history of galaxies, particularly when the phenomenon can be traced to the highest redshifts. The recent discovery of high-redshift stellar bars \citep{guo22,leconte23,costat23} is one of the early successes of the JWST. If baryons dominate the local gravitational potential \citep[e.g.][]{Price2021}, stellar bars can form in $1-2$ Gyr \citep{bla23}.

The early disks have provided a number of unexpected results. Most disks have well-defined spiral arms up to the redshift cut-off where this is possible ($z\lesssim 3$), with only a small proportion showing signs of strong tidal interaction \citep{kuhn24}.
While the statistics are incomplete at present, a substantial fraction of the early disks are very rich in gas, exhibit high levels of star formation and show signs of turbulence in the dense molecular gas.
In the redshift range $z=1.1 - 2.3$, \cite{guo22} showed that turbulent star-forming disks can have well-developed stellar bars that look much like their mature low-redshift counterparts. Remarkably, spiral arms are also observed in turbulent (``hot") disk galaxies at $z\approx 2-2.5$ \citep{daw03a,Law12,yuan17,mar22}; in one instance, the galaxy disk appears to have three spiral arms \citep{Law12}, a rare phenomenon in the local universe.

Our new work is in response to ALMA, VLA and integral-field spectroscopic observations of gas-rich, star-forming disks in the redshift range $z\sim 1-3$ \citep{chap04,gen06h,forster06}, supported by subsequent observations \citep{shapiro08,swinbank12,hodge19,nee21c,riz22w}.
Intriguingly, there are clear
cases where the gas entirely dominates the baryon disk fraction \citep[e.g.][]{hodge19,riz22w}, and even here bar-like features and/or velocity fields are observed \citep{tsukui24,smail23,huang23,neel23,amvro24}. In these instances, {\it stellar} bars are either weak or non-existent.

So what are these ``gaseous bars''?
Very little is known about the evolution of non-axisymmetric perturbations in gas-rich disks \citep[q.v.][]{shlos89,chr95}. For earlier publications on bar formation that include gas, only a narrow range in gas fraction is treated and few consider prescriptions for star formation. A notable example that does consider bar formation in star-forming disks is \citet{seo19}, but here $f_{\rm gas}\lesssim 10\%$. Gas-dominated disks are a difficult problem that requires the gas to be internally supported if it is to avoid collapsing on a dynamical timescale ($\lesssim 100$ Myr).

But these must be considered because there is a tendency for the disk mass fraction in the form of gas to increase with redshift \citep{carilli13,tac20}. Star formation rates, gas turbulence and clumpiness also increase, such that the mean circular rotation speed divided by the internal gas velocity dispersion $V_c/\sigma$ declines with increasing redshift \citep{wis15x,zhou17}. Importantly, gas dispersions measured with cool molecular lines tend to be narrower than dispersions measured with emission lines arising from warmer ionized gas. Both measures of dispersion, however, show an increasing trend with redshift \citep{ejdet22}.
\citet[][]{wis15x} argued that these galaxies were in quasi-equilibrium, being fed by smooth accretion of cool gas from the environment, with galaxy mergers playing a lesser role. The fraction of these disks that are dynamically settled is unclear \citep{forster06}, with some authors suggesting that half \citep{bell12,rod17} or all \citep{simons19} of these systems could be experiencing strong mergers, within the limits of the integral-field spectroscopic observations. 

\vspace{0.2cm}
We have developed
a new high-resolution (parsec scale) simulation framework that is ideally suited to discovering how turbulent disks evolve dynamically. In high-redshift galaxies, the observed high gas fraction and enhanced gas surface densities lead to elevated star-formation rates and the output energy and momentum couples with the gas and drives the turbulence \citep{agertz09}. A turbulent (compressible) gas disk with ordered rotation, perhaps surprisingly, shares dynamical similarities with (collisionless) stellar disks. It may even be susceptible to bar instabilities, although there is very little published work on this problem \citep{cazes00}.

Our first paper is focussed on objects with halo masses of order 10$^{11}$ M$_\odot$, a dynamical mass that is characteristic of the recent JWST discoveries at $z\sim 1-3$ \citep{guo22,leconte23,costat23}. This is the expected mass of
a Milky Way progenitor at that epoch. (We consider more massive systems in a later paper because these are more representative of the ALMA disks observed to date.)
Our first step is to consider a galactic ecosystem in some form of dynamical equilibrium, with and without smooth accretion from the ambient hot corona. As we show, these ecosystems manifest intermittent and long-lived behaviour, including disk-halo interaction, bars and spiral arms, bulge formation, and so forth. We discover ``bar-like'' phenomena for the first time at all gas fractions, even in fully gas-dominated turbulent disks, including a remarkable ``radial shear flow'' with its own unique signatures. Once again, as for the stellar disks, these bar-like signatures only arise in disks that dominate the local gravitational potential \citep{bla23}.

The structure of the paper is as follows. In Sec.~\ref{s:simul}, we introduce and motivate the model parameters for a Milky Way progenitor galaxy. To carry out this work, we update the \agama\ self-consistent modelling module \citep{vas19a} to allow for the inclusion of gas components (disk, halo; details to be presented in a companion paper; Tepper-Garc\'ia et al., in prep). All of the computations are carried out with the \ramses\ N-body/hydrodynamics code \citep[][]{tey02a} at low ($N_{\rm lo}\sim 10^7$ elements) and high ($N_{\rm hi}\sim 10^8$ elements) resolution, including star formation and metal production. We also conduct repeated simulations at the same initial conditions (different random seeds) to examine stochastic effects given the nature of turbulent media. In Sec.~\ref{sec:sfrdefs}, we outline the main processes that lead to star formation in turbulent media.
In Sec.~\ref{s:results}, we present the main findings from the new simulations and provide links to the simulation movies. In Sec.~\ref{s:discuss}, we discuss the simulations and their implications for high-redshift disks, before drawing our conclusions in Sec.~\ref{s:main}.

\section{Galaxy models and simulations} \label{s:simul}

\subsection{Framework and initial conditions} \label{s:frame}

In defining a Milky Way progenitor, we adopt a model with three key components $-$ a live dark matter halo, a massive stellar/gaseous disk and, for a subset of models,
a hot coronal gas filling the live dark matter halo, which serves to supply the disk with a smooth flow of accreting gas after it cools. The hot coronal gas also compresses the disk-halo interaction maintained by feedback processes. Our intent is to minimize the number of free parameters in the model while retaining its usefulness. Small central bulges tend to form during the evolution of the disk, or from gas accreting from the halo, but we do not start with central bulges when simulating such early disks.

In the gaseous disk, the star formation rate is set by the gas surface density and this activity feeds back sufficient energy and momentum to maintain the turbulent pressure support (see below). We explore two distinct accretion histories: (i) there is no ongoing accretion and the gas is used up by the star formation; (ii) the star-forming gas is replenished by cooling in the hot halo gas. In a later paper, we explore the response to a strong impulse triggered by an interacting, massive companion. 


The initial conditions (particle positions and velocities) for each of the components making up our Galaxy model are created with the Action-based GAlaxy Modelling Architecture software package \citep[\agama; ][]{vas19a}. The reader is referred to \citet[][their section 3]{tep21v} for a detailed description of \agama{}'s self-consistent module for the collisionless components in our model (dark matter [DM] halo, stellar disk). We have complemented the \agama\ framework to include the gas phase \citep[cf.][]{tep22x}; our methodology is described in detail in a companion paper (Tepper-Garcia et al., in preparation). Here only a brief description is provided for completeness.

Our approach for setting up a gas disk follows \citet[][]{wan10a} who give a prescription for isothermal gas disks in equilibrium. The gas disk is initially perfectly isothermal and axisymmetric, with a surface density profile described by a radially declining function proportional to $\exp (-R/R_{\rm disk})$, where $R_{\rm disk}$ is the scalelength of the disk. Its vertical structure is dictated by the total gravitational potential of the system, but roughly follows a $\sech^2(z/z_0)$ profile, characteristic of a gas distribution in hydrostatic equilibrium, with a scaleheight $z_0$ that varies with cylindrical radius (a `flaring' disk). The azimuthal velocity profile of the gas disk ensures rotational support against radial instabilities \citep[see][their equation 13]{wan10a}. The initial configuration of the gas disk does not greatly affect its long-term evolution, but it is important however to define the model through its initial gas fraction $f_{\rm gas}$.

In setting up the hot gaseous halo, we focus our attention on pressure-supported, spinning gas configurations, embedded within a DM halo of mass $M_{\rm halo}$. For simplicity, the hot halo is assumed to follow initially the same profile as its host DM halo \citep[e.g.][]{mo98a,asc03a,tey13a}. The hot halo mass is chosen so as to close the cosmic baryon budget in the box. The cosmic baryon fraction $f_b = 0.16$ \citep{pla20a} implies a hot halo mass of approximately $5.5\times10^9$~\Msun. This is appropriate for a Milky Way progenitor at $z\approx 3$.

Since the mass distribution of the hot halo is fixed by construction, all that remains to be calculated is its velocity structure and the internal energy (temperature). The velocity structure, i.e. rotation speed, is calculated assuming that the specific angular momentum of the gas, $j_h$, follows that of the dark matter \citep[c.f.][]{bul01b,kau06a}. We adopt \mbox{$v_\phi(r)~=~j_h(r)~/~r~\propto~M_{\rm halo}(<r)~/~r$}. The specific internal energy of the gas, corrected for its net rotation speed $v_\phi(r)$, is given by \mbox{$e~=~(\gamma -1)^{-1}~(c_s^2 - v_\phi^2 )$}, where $c_s$ is the hydrostatic (i.e. $v_\phi \equiv 0$) sound speed. This approach results in gas temperatures of about a few $10^6$~K close to the centre and $\sim10^4$~K in the outer regions, and rotation speeds in the range 10-50~\kms.

It is worth emphasising that, as with the gas disk, the initial configuration of the hot halo is largely irrelevant to the subsequent evolution of the galaxy, with the exception of its mass, which will impact the amount of accretion onto the disk. Finally, we note that the mass of the hot halo does not count towards the gas fraction $f_{gas}$ that defines each model; the latter is entirely defined by the initial cool gas assigned to the disk.


\subsection{Progenitor galaxy parameters} \label{s:fund}

We now describe the choice of parameters for our progenitor analogue at high redshift. Within the context of the CDM hierarchy, it is possible to estimate the likely mass and size evolution of a Milky Way progenitor with cosmic time. \citet[][their figure 1]{bla16a} use a Press-Schechter code to estimate the $1\sigma$ range in 
total halo mass and size with redshift across all merger trees that lead to the Milky Way's estimated mass ($\log M_{\rm vir}/{\rm M}_\odot = 12.1$) and size ($R_{\rm vir}\approx 280$ kpc) at $z=0$. Over the redshift range from $z\approx 3$ to the present day, the halo mass and size increase by an order of magnitude. In support of this, \citet[][their figure 1]{kartaltepe23}
show the distribution of baryon stellar masses for disk systems over the redshift interval $z\approx 1-6$. These exhibit a large scatter about a mean baryon mass of $\log M_\star/{\rm M}_\odot = 9.5$ as compared with $\log M_\star/{\rm M}_\odot \approx 10.6$ for the Milky Way at $z=0$. 
In summary, for our Milky Way progenitor, we adopt halo parameters of $R_{\rm vir}\approx 40$ kpc and $\log M_{\rm vir}/{\rm M}_\odot \approx 11$. The exact values are given in Tab.~\ref{t:mod}. The halo concentration is calculated for its given mass and redshift with the help of the COMMAH package \citep{cor15h}.

The next important parameter is the disk mass fraction $f_{\rm disk}$ that determines how dominant the disk baryons are with respect to the underlying dark matter halo. This is defined as
\begin{equation}
\label{e:fd}
    f_{\rm disk}= \left(\frac{V_{\rm c, disk}(R_s)}{V_{\rm c, tot}(R_s)}\right)_{R_s=2.2 R_{\rm disk}}^2 \, .
\end{equation}
Here
$V_c(R)$ is the circular velocity at a radius $R$, $R_{\rm disk}$ is the exponential disk scalelength, and $R_s=2.2 R_{\rm disk}$ is the traditional scalelength adopted in studies of disks. The Milky Way today has a dominant central disk with $f_{\rm disk}=0.5-0.65$ depending on the chosen bar model
\citep[][their Fig. 17]{bla16a}. Here we adopt $f_{\rm disk}=0.5$. There is evidence for a substantial fraction of disks dominating their local gravitational potential in the redshift interval $z\approx 1-2.5$ \citep{Price2021}. Baryon domination has an impact on the subsequent disk evolution as it becomes more susceptible to internal and external perturbations \citep{fuj18a,bla23}.

The third model parameter is the gas mass fraction within the same radial scale:
\begin{equation}
\label{e:fdd}
    f_{\rm gas}=
    \left(\frac{M_{\rm disk, gas}}{M_{\rm disk}}\right)_{R_s=2.2 R_{\rm disk}} \,
\end{equation}
where $M_{\rm disk}$ is the total disk mass and $M_{\rm disk,gas}$ is the gas contribution. This definition is not universal and care must be taken. The gas fraction declines as more of the mass is locked up in stars.
The main parameters are listed in Tab.~\ref{t:mod}. The synthetic galaxy is composed of a host DM halo with a fixed mass, a pre-existing (`old') stellar disk, and a gaseous disk. The summed masses of the stellar and gaseous component are set to $\log M_{\rm disk}/{\rm M}_\odot \approx 10$, yielding $f_{\rm disk} \approx 0.5$, but their individual masses (either gas or stars) are adjusted so as to attain different $f_{\rm gas}$ values in the range 0\% to 100\%. The evidence for a broad range in gas fractions is presented in \citet{carilli13}: the overall trend is a rising gas fraction with increasing redshift, but the scatter is large. 

The baryon mass is preserved in both the accreting and non-accreting halo models. In the former case, the total baryon mass of the disk increases with time. Note that the initial scalelength of the disk (both gas and stars) is roughly maintained
at $R_{\rm disk}\approx 1.8$~kpc across models. At $z\sim 0$, the gaseous and stellar disk scalelengths can differ by a factor of two \citep[e.g. Milky Way;][]{bla16a}
due to the cumulative effect of gas accretion at later times.
The above parameters are broadly consistent with arguably the best
Milky Way progenitor analogue to date, the object CEERS-2112 at a photometric redshift of $z\approx 3.0$ \citep{costat23}.

\begin{table*}
\begin{tabular}{rrccclcc}
\hline
$M_\star$ & $M_{\rm gas}$ & $R_{\rm disk}$ &  $f_{\rm disk}$   & $f_{\rm gas}$ & Label & High & Low \\
($10^{8}$~\Msun)  & ($10^{8}$~\Msun) &  (ckpc) &  &  &  &  & \\
\hline
9.80 &  0.0  &  0.7  &  {\bf 0.3}  & 0.0 &  fd30\_fg00\_nac & 0 & 1 \\
7.80 &  2.0  &  0.7  &  {\bf 0.3}  & 0.2 &  fd30\_fg20\_nac & 0 & 1 \\
5.90 &  3.90  &  0.7  &  {\bf 0.3}  & 0.4 &  fd30\_fg40\_nac & 0 & 1 \\
3.90 &  5.90  &  0.7  &  {\bf 0.3}  & 0.6 &  fd30\_fg60\_nac & 0 & 1 \\
\hline
112 &  0.0  &  1.8  &  0.5    & 0.0 &  fd50\_fg00\_nac & 0 & 1 \\
88.8 &  22.2  &  1.8  &  0.5  & 0.2 &  fd50\_fg20\_nac & 2 & 1 \\
66.6 &  44.4  &  1.8  &  0.5  & 0.4 &  fd50\_fg40\_nac & 2 & 3 \\
44.4 &  66.6  &  1.8  &  0.5  & 0.6 &  fd50\_fg60\_nac & 2 & 1 \\
22.2 &  88.8  &  1.8  &  0.5  & 0.8 &  fd50\_fg80\_nac & 0 & 1 \\
0.0 &  112  &  1.8  &  0.5  & 1.0 &  fd50\_fg100\_nac  & 0 & 1 \\
\hline
88.8 &  22.2  &  1.8  &  0.5  & ((0.2)) &  fd50\_fg20\_ac & 0 & 1 \\
66.6 &  44.4  &  1.8  &  0.5  & ((0.4)) &  fd50\_fg40\_ac & 0 & 1 \\
44.4 &  66.6  &  1.8  &  0.5  & ((0.6)) &  fd50\_fg60\_ac & 0 & 1 \\
22.2 &  88.8  &  1.8  &  0.5  & ((0.8)) &  fd50\_fg80\_ac & 0 & 1 \\
0.0 &  112  &  1.8  &  0.5  & ((1.0)) &  fd50\_fg100\_ac & 0 & 1 \\
\hline
555 &  0.0  &  3.0  &  {\bf 0.7}  & 0.0 &  fd70\_fg00\_nac & 0 & 1 \\
444 &  111  &  3.0  &  {\bf 0.7}  & 0.2 &  fd70\_fg20\_nac & 0 & 1 \\
333 &  222  &  3.0  &  {\bf 0.7}  & 0.4 &  fd70\_fg40\_nac & 0 & 1 \\
222 &  333  &  3.0  &  {\bf 0.7}  & 0.6 &  fd70\_fg60\_nac & 0 & 1 \\
\hline\\
\end{tabular}
\caption{Overview of galaxy models. The DM host halo properties ($M_{\rm halo} = 10^{11}$~\Msun; $R_{\rm vir} = 37$ ckpc; $r_{\rm s} = 9.2$ ckpc) are identical across models, and they are consistent with a MW-progenitor analogue at $z \approx 3$. Table columns are as follows: (1) Disk stellar mass; (2) Disk gas mass; (3) Disk scalelength; (4) Disk-to-total mass ratio (Eq.~\ref{e:fd}); (5) Gas to total disk mass fraction at $t=0$ (Eq.~\ref{e:fdd}) $-$ the double brackets indicate accreting gas from a hot corona with a mass $M \approx 5.5\times10^9$~\Msun\ at $t = 0$; note that the hot halo mass is not accounted for in $f_{gas}$. (6) Model designation; (7) No.~of high-resolution models; (8) No.~of low-resolution models with different random seeds.}
\label{t:mod}
\end{table*}
\begin{figure}[!htb]
    \centering
\includegraphics[width=0.8\textwidth]{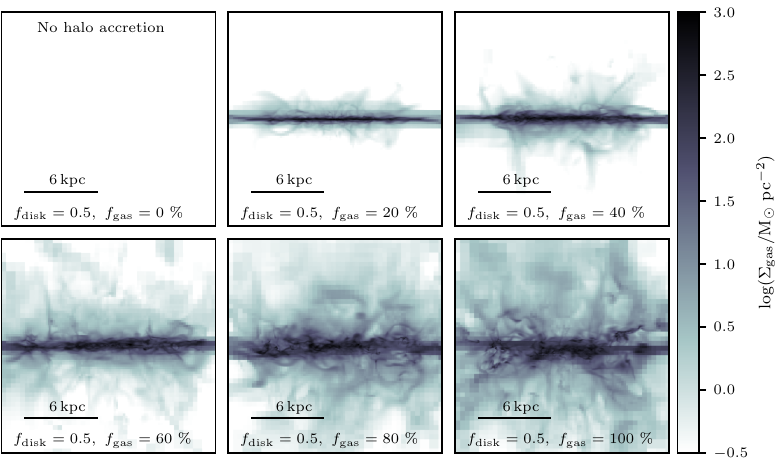}
\includegraphics[width=0.8\textwidth]{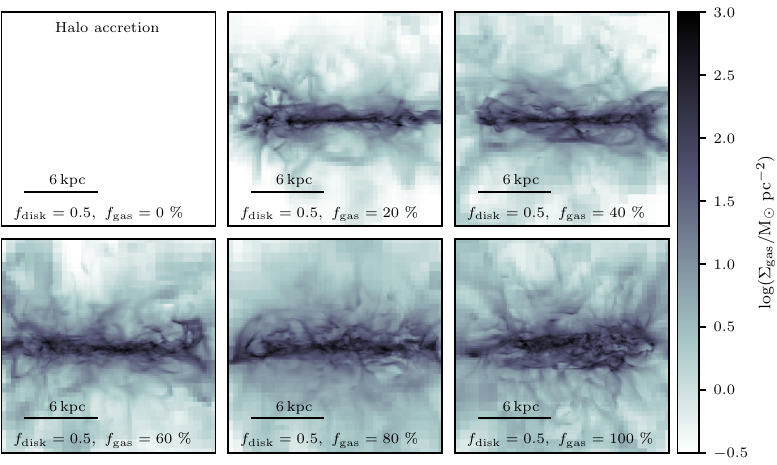}
\caption{Edge-on gas surface density maps taken from our simulations at $t_{\rm o}=1\,{\rm Gyr}$ for a disk mass fraction comparable to the Milky Way today. The panels are $18\,{\rm ckpc}$ (comoving kpc) across. The top six panels are for six different initial gas fractions with {\it no} halo accretion; the bottom six are the matching simulations that include halo accretion.
}
   \label{f:edgeon}
\end{figure} 
\begin{figure}[!htb]
    \centering
\includegraphics[width=0.8\textwidth]{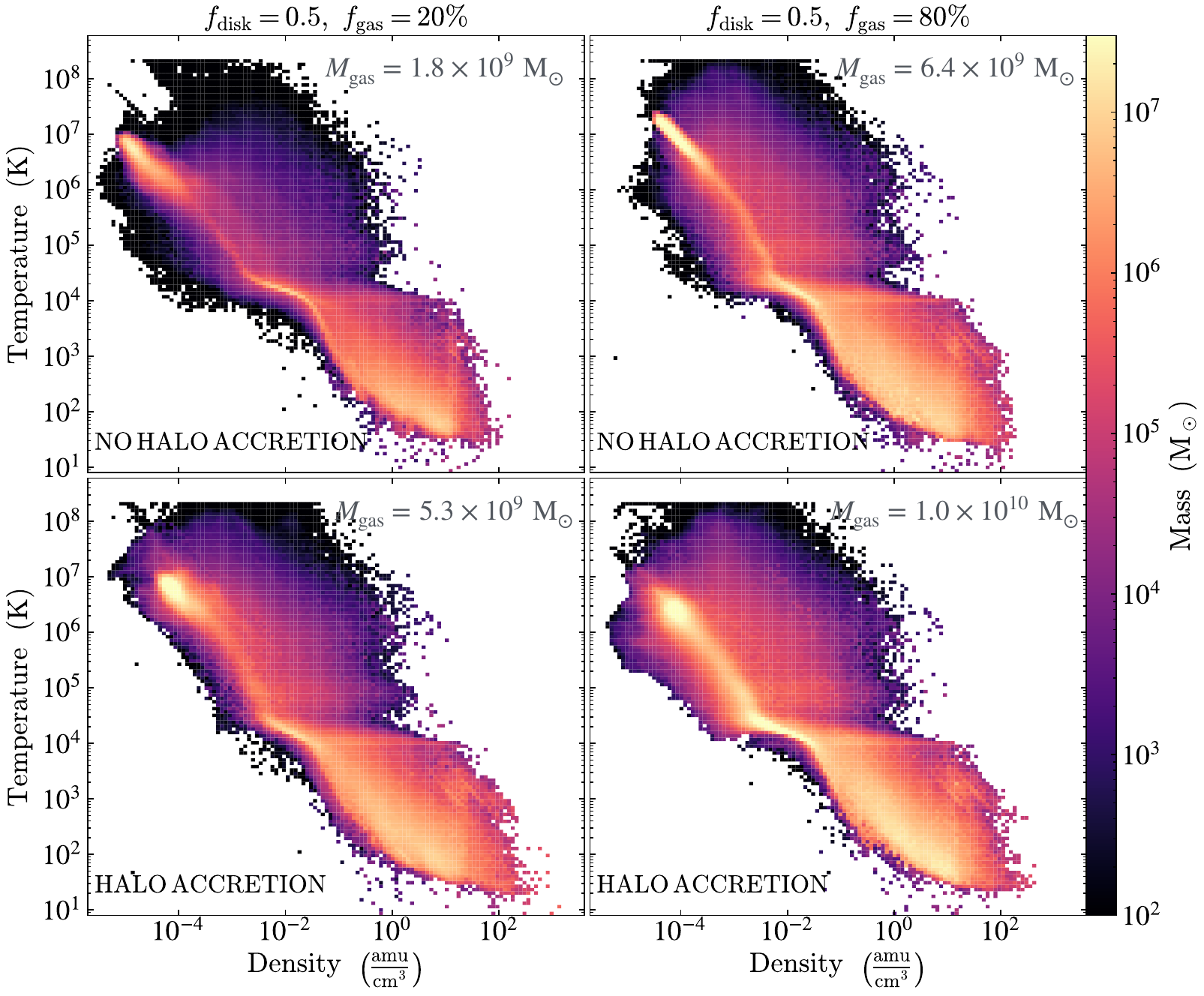}
\caption{Gas phase structure after 1 Gyr in the $f_{\rm gas}=20\%$ (left panels) and $80\%$ (right panels) simulations with and without halo accretion (top and bottom rows respectively). The indicated gas mass is the total gas mass in the simulation volume at this instance.
}
   \label{f:phases}
\end{figure} 
\subsection{Galaxy formation physics}

The galaxies are evolved using the hydrodynamics+$N$-body code \ramses\ \cite[][]{tey02a}. \ramses\ adopts an oct-tree hierarchical grid that provides adaptive resolution and solves the fluid equations using a second-order unsplit Godunov method with the HLLC approximate Riemann solver \citep[][]{Toro1999}. We apply a MinMod slope limiter to reconstruct the piecewise linear solution for the Godunov solver. To close the relation between gas pressure and internal energy, we use an ideal gas equation of state with an adiabatic index $\gamma=5/3$. Mesh refinement is based on a quasi-Lagrangian approach, where a cell is split into 8 sub-cells if its mass exceeds $94,\!000~{\rm M}_\odot$ and $11,\!750~{\rm M}_\odot$ in the low- and high-resolution models, respectively. In addition, a cell is allowed to refine if it contains more than 8~dark matter particles. We allow for 12 (14) levels of grid refinement in the low (high) resolution suite, leading to a finest grid resolution of 24 (6) parsecs.

The adopted star/galaxy formation physics is presented in \citet{age13} and \citet{agertz2021}. Briefly, star formation is treated as a Poisson process, sampled using $10^3~{\rm M}_\odot$ star particles, occurring on a cell-by-cell basis according to the star formation law,
\begin{equation}
	\dot{\rho}_{\star}= \epsilon_{\rm ff}\frac{ \rho_{\rm g}}{t_{\rm ff}} \quad {\mbox{for}} \quad  \rho_{\rm g}>\rho_{\rm SF}.
	\label{eq:schmidtH2}	
\end{equation}
Here $\dot{\rho}_{\star}$ is the star formation rate density, $\rho_{\rm g}$ the gas density, $t_{\rm ff}=\sqrt{3\pi/32G\rho_{\rm g}}$ is the free-fall time, and $\epsilon_{\rm ff}$ is the star formation efficiency per free-fall time of gas in the cell. The star formation threshold is set to $\rho_{\rm SF}=10(100)~{\rm cm}^{-3}$ in the low(high)-resolution models. \citet{Grisdale2017,Grisdale2018,Grisdale2019} demonstrate how high star formation efficiencies ($\epsilon_{\rm ff}\sim 10\%$) on scales of parsecs, coupled to the feedback models in this work, provide a close match to the observed density and velocity structure of the ISM in local spirals, as well as giant molecular cloud properties in good agreement with Milky Way observations. Motivated by these findings, we adopt $\epsilon_{\rm ff}=10\%$.

Each formed star particle is treated as a single-age stellar population with a \citet{chabrier03} initial mass function. We account for injection of energy, momentum, mass, and heavy elements over time from core-collapse SN and SNIa, stellar winds, and radiation pressure into the surrounding gas. 
Each mechanism depends on stellar age, mass and gas/stellar metallicity \citep[with main sequence lifetimes taken from][]{Raiteri1996}, calibrated on the stellar evolution code \STARBURST\ \citep{Leitherer1999}. 
The effect of supernova explosions is captured following the approach by \citet[]{KimOstriker2015}. Briefly, when the supernova cooling radius\footnote{The cooling radius in gas with density $n$ and metallicity $Z$ scales as $\approx 30 (n/1\cc)^{-0.43} (Z/Z_\odot + 0.01)^{-0.18} \pc$ for a SN explosion with energy $E_{\rm SN}=10^{51}$ erg \citep[e.g.][]{Cioffi1988, Thornton1998}.} is resolved by more than 6~cells, supernova explosions are initialized in the `energy conserving' phase by injecting $10^{51} \erg$ per SN into the nearest grid cell. 
When the cooling radius is resolved by less than 6 cells, the explosion is initialized in its `momentum conserving' phase, with the momentum built up during the Sedov-Taylor 
phase\footnote{The adopted relation for the momentum is $4\times 10^5 (E_{\rm SN}/10^{51}\erg)^{16/17} (n/1~{\rm cm}^{-3})^{-2/17} (Z/Z_\odot)^{-0.2} \Msol\; \kms$ \citep[e.g.][]{Blondin1998, KimOstriker2015, Hopkins2018}, where $E_{\rm SN}$ is the total energy injected by SNe in a cell with gas density $n$ and metallicity $Z$ compared to Solar ($Z_\odot=0.02$).} injected into cells surrounding the star particle.

The iron (Fe) and oxygen (O) abundances are tracked separately, with yields taken from \citet{woosley_heger07}. When computing the gas cooling rate, which is a function of the total metallicity, we construct a total metal mass following $M_{Z}=2.09M_{\rm O}+1.06M_{\rm Fe}$ \citep[see][]{kim_agora2014} according to the mixture of alpha and iron group elements for the Sun \citep{Asplund2009}. Metallicity dependent cooling is accounted for  using the cooling functions by \citet{sutherlanddopita93} for gas temperatures in the range $10^{4-8.5}$~K, with rates from \citet{rosenbregman95} used for cooling down to $\sim 10\,$K.
Heating from a cosmic UV background is modelled following \citet{haardtmadau96}, under the assumption that gas self-shields at high enough densities \citep[see][]{AubertTeyssier2010}. 

In order to provide a visual sense of what the different levels of disk-halo interaction look like, we present two collages in Fig.~\ref{f:edgeon}. The top collage shows the impact of star formation and feedback (for a range of gas fractions) when the disk is viewed edge-on after 1 Gyr. The gas is initially prescribed with no subsequent accretion. The bottom collage shows the matching models that have subsequent halo accretion of cool gas from the hot corona surrounding each galaxy. 

In the top row of Fig.~\ref{f:phases} we present the resulting gas phase structure after 1 Gyr from the $f_{\rm gas}=20\%$ and $80\%$ simulations without halo accretion, with the corresponding halo accretion models shown in the bottom row. In each panel, the total gas mass contained in the simulation volume is indicated. All models show a highly multiphase gas structure, and we note that even in the absence of an initial hot corona, a hot phase ($T\sim 10^6~{\rm K}$) develops from galactic outflows, albeit at a lower density than in the halo accretion simulations ($n\sim 10^{-5}~{\rm cm^{-3}}$ compared to $\sim 10^{-4}~{\rm cm^{-3}}$ in the $f_{\rm gas}=20\%$ models). 

In the $f_{\rm gas}=20\%$ suite, the halo accretion model features more gas mass in all gas phases, with a relative  contribution to the total mass of $\sim 40\%$ and $\sim 50\%$ from the cold ($T<5\times 10^3~{\rm K}$) and hot ($T>3\times 10^4~{\rm K}$) phases, respectively. Without halo accretion, only $\sim 25\%$ of the total mass resides in the hot phase, and $\sim 65\%$ is in the cold phase. The two $f_{\rm gas}=80\%$ simulations display a more similar behaviour in terms of the relative contribution from the hot phase ($\sim 30\%$ in both cases), indicating that the gas-rich disk can eject large amounts of gas into the halo via hot galactic winds. However, the warm phase ($5\times 10^3~{\rm K}<T<3\times 10^4~{\rm K}$) mass fraction is twice as high in the halo accretion model (and over 3 times in actual gas mass), which is visible in the more extended, warm disk-halo interface in Fig.~\ref{f:edgeon}.

\begin{figure}[!htb]
    \centering
\includegraphics[width=0.75\textwidth]{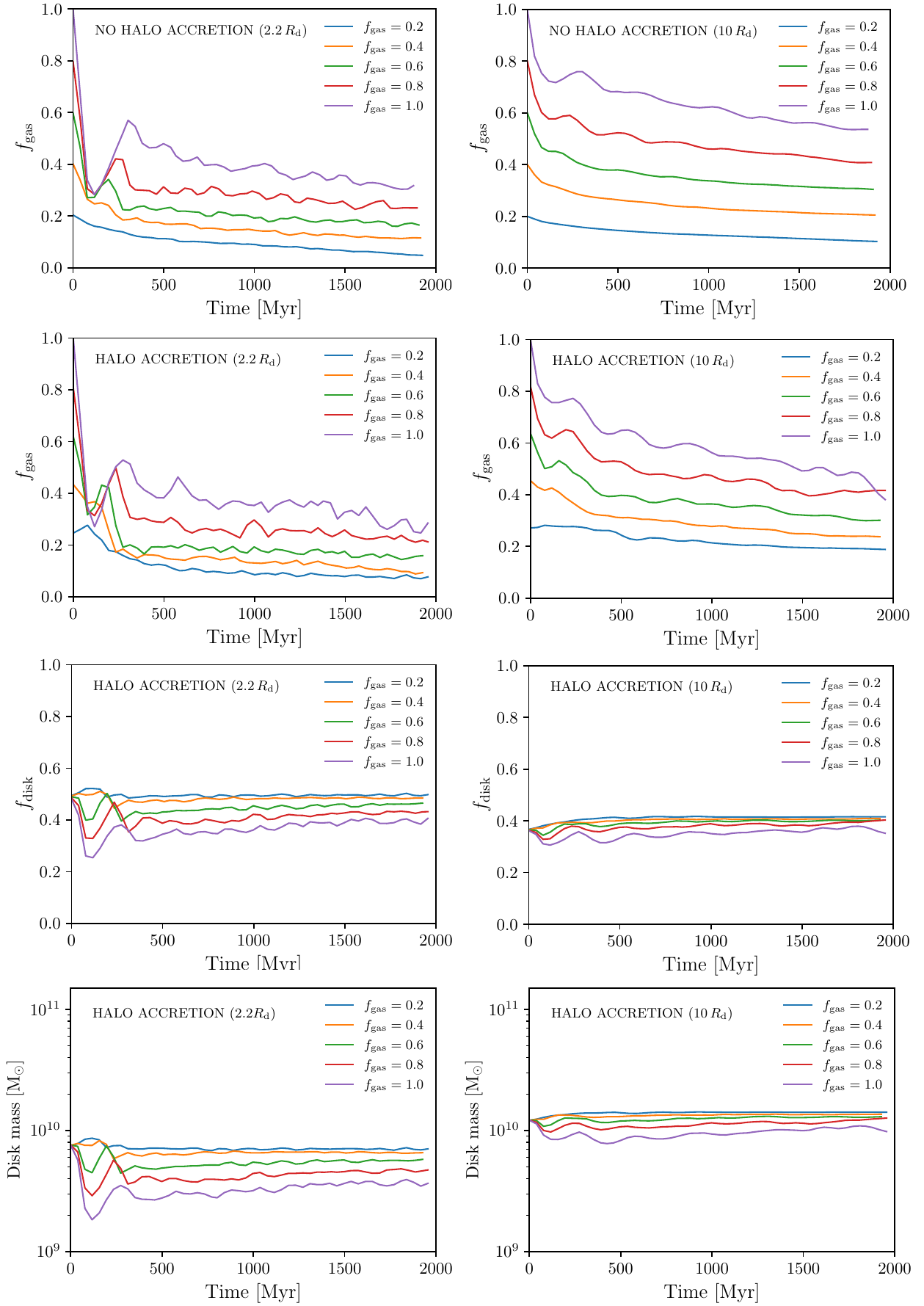}
\caption{The change in gas fraction, disk mass fraction and disk mass with cosmic time for our heavy disk ($f_{\rm disk}=0.5$) simulations. The top row is for ``no halo accretion" models; all other rows are for ``halo accretion" models, for which there is constant accretion onto the disk plane.
Two disk zones are shown: (Left) $R < 2.2 R_{\rm disk}$ ; (Right) $R < 10 R_{\rm disk}$. In each plot, we show the time evolution for 5 different initial gas fractions, as indicated. These are systematically lower in the halo-accretion models, because the enhanced accretion drives more vigorous star formation, which consumes the gas at a faster rate and drives mass-loaded, circulating winds $-$ see Fig.~\ref{f:sfr_dens}. In the bottom two rows, the winds are increasingly more effective at removing disk baryons at higher gas fractions, particularly across the inner disk.}
   \label{f:fgas}
\end{figure}


\section{Star formation in turbulent media: basic equations and concepts}
\label{sec:sfrdefs}

Many authors have considered star formation in turbulent media \citep{Padoan1995,KlessenHeitschMacLow2000,ElmegreenEtAl2003,KrumholzMcKee2005,PadoanNordlund2011,HennebelleChabrier2011,FederrathKlessen2012,Federrath2018,BurkhartMocz2019}, motivated by the fact that clouds in nearby galaxies and the Milky Way have non-thermal line widths \citep[e.g.,][]{Larson1981,SolomonEtAl1987,OssenkopfMacLow2002,HeyerBrunt2004,RosolowskyBlitz2005}, widely believed to be due to supersonic turbulence \citep{FederrathEtAl2021}. Both from observations and simulations, the turbulent velocity dispersion $\sigma_v$ averaged over a volume of diameter $\ell$ scales as $\sigma_v \propto \ell^p$ with $p\approx 0.5$ \citep{Larson1981,HeyerBrunt2004,RomanDuvalEtAl2011,Federrath2013,FederrathEtAl2021}.

\subsection{The gas density distribution}

Supersonically turbulent, isothermal gas has a lognormal density distribution \citep{Vazquez1994,PassotVazquez1998,PadoanNordlund2002,KritsukEtAl2007} such that
\begin{equation}
p(s) = \frac{1}{\sqrt{2\pi \sigs^2}}\exp\left[-\frac{(s-s_0)^2}{2\sigs^2}\right],
\end{equation}
for which $s=\ln(\rho/\rho_0)$ is the dimensionless log-density contrast
(i.e., the natural logarithm of the density divided by the mean density $\rho_0$), and the average log-density parameter is $s_0=-(1/2)\sigs^2$ \citep{LiKlessenMacLow2003,FederrathKlessenSchmidt2008,FederrathDuvalKlessenSchmidtMacLow2010}. The dispersion in density of $p(s)$ over the region is \citep{PadoanNordlund2011,MolinaEtAl2012}
\begin{equation}
\sigs=\sqrt{\ln\left(1+b^2\mach^2\frac{\beta}{\beta+1}\right)}\,,
\label{eq:sigs}
\end{equation}
where $\mach=\sigma_v/c_\mathrm{s}$ is the three-dimensional (3D) turbulent Mach number of the medium, $c_\mathrm{s}$ is the sound speed within the density fluctuation of diameter $\ell$, and $\beta$ is the plasma beta parameter (ratio of thermal to magnetic pressure; note that $\beta\to\infty$ in cases without magnetic fields). The parameter $b$ in Eq.~(\ref{eq:sigs}) is the turbulence driving parameter, which is controlled by the mixture of solenoidal vs.~compressive modes in the driving mechanism of the turbulence \citep{FederrathKlessenSchmidt2008}. Purely solenoidal (divergence-free) driving has $b\sim1/3$, while purely compressive (curl-free) driving is characterised by $b\sim1$ \citep{FederrathDuvalKlessenSchmidtMacLow2010,DhawalikarEtAl2022,GerrardEtAl2023}. Slight modifications of Eq.~(\ref{eq:sigs}) can be made to account for non-isothermal gas conditions \citep{NolanFederrathSutherland2015,FederrathBanerjee2015}. We do not consider magnetic fields in this early analysis; these supply local pressure support and therefore tend to slow down the evolution of collapsing filaments \citep[e.g.,][]{LiEtAl2004,PadoanNordlund2011,Federrath2015}. We consider the consequences of magnetohydrodynamics (MHD) and turbulence in a later paper.

\subsubsection{The star formation rate}
\label{s:sfr}

\citet{KrumholzMcKee2005} consider the fraction of the turbulent gas mass in collapsing density fluctuations. In these regions, gravity starts to become a dominant factor, as opposed to turbulence. This defines a critical density for star formation \citep[see][for a comprehensive comparison of different models for the cI ritical density]{FederrathKlessen2012} by comparing the Jeans length with the turbulent sonic scale \citep{FederrathEtAl2021}, which marks the transition from supersonic turbulence on cloud scales, to subsonic turbulence inside the dense star-forming cores and accretion discs. The total star formation rate (SFR) in a gas cloud with mass $M_{\rm cl}$ is thus given by \citep{KrumholzMcKee2005,FederrathKlessen2012} 
\begin{equation}
\label{eq:sfrff}
\dot{M}_\star = \sfrff \frac{M_{\rm cl}}{\tff},
\end{equation}
where $\tff$ is the freefall (collapse) time within the cloud, and $\sfrff$ is the star formation rate per freefall time, which takes account of the complex fractal hierarchy in a turbulent medium. It is a dimensionless SFR; for example, if $\sfrff=0.1$, then in one freefall time, 10\% of the gas mass in the cloud would have formed stars. From fits to simulations, \citet{KrumholzMcKee2005} arrived at
\begin{equation} \label{eq:km05}
\sfrff \approx 0.014 \left(\frac{\alphavir}{1.3}\right)^{-0.68}\left(\frac{\mach}{100}\right)^{-0.32},
\end{equation}
where $\alphavir$ is the virial parameter (ratio of twice turbulent to gravitational energy). Note that $\sfrff$ drops with increasing $\mach$ in this model, contrary to the expectation that increasing $\mach$ results in a higher dense-gas fraction, and thus higher SFR. Indeed, more recent calculations show that the model provided by Eq.~(\ref{eq:km05}) does not fit simulations in which the Mach number ($\mach$) is varied \citep{FederrathKlessen2012}. This is due to the fact that the \citet{KrumholzMcKee2005} model does not account for the density dependence of the freefall time. 

\citet{HennebelleChabrier2011} provide a multi-freefall framework of the SFR, in which the density dependence of the freefall time is taken into account by evaluating it inside the integral that defines $\sfrff$ \citep{FederrathKlessen2012}. This multi-freefall model of star formation provides excellent fits to numerical simulations with a wide range of different parameters, namely the virial parameter ($\alphavir$), the sonic Mach number ($\mach$), the plasma beta ($\beta$, or alternatively the Alfv\'en Mach number), and the turbulence driving parameter  \citep[$b$;][]{FederrathKlessen2012}. The strongest dependence of the SFR is on the turbulence driving mode, resulting in differences in $\sfrff$ by as much as an order of magnitude \citep{FederrathKlessen2012,Federrath2018}, and in the virial parameter $\alphavir$, which can completely shut off star formation at sufficiently high $\alphavir$. However, all 4~dimensionless parameters can play a critical role, depending on the exact position in the 4D parameter space. This space is illustrated in a set of figures presented in \citet[][see their figure 1]{FederrathKlessen2012}.

The full theoretical expression for the multi-freefall turbulence-regulated SFR$_\mathrm{ff}$ model is given by \citep{FederrathKlessen2012}
\begin{equation} \label{eq:fk12sfrff}
\sfrff = \frac{\eps}{2\phit} \exp\left(\frac{3}{8}\sigs^2\right) \left[1+\mathrm{erf}\left(\frac{\sigs^2-\scrit}{\sqrt{2\sigs^2}}\right)\right],
\end{equation}
with the critical log-density
\begin{equation} \label{eq:fk12scrit}
\scrit = \ln\left[(\pi^2/5)\phix^2\,\alphavir\,\mach^{2}\left(1+\beta^{-1}\right)^{-1}\right].
\end{equation}
The log-density dispersion $\sigs$ is given by Eq.~(\ref{eq:sigs}) and the numerical parameters $\eps$, $\phit$, and $\phix$ are of order of unity, and were determined in \citet{FederrathKlessen2012} by fitting a set of $\sim30$~MHD simulations with varying $\alphavir$, $\mach$, $b$, and $\beta$. The star-to-core ratio \mbox{$\eps\sim0.3$--$0.5$} \citep{FederrathEtAl2014}, and the best-fit $\phit\sim2$ and $\phix\sim0.2$ \citep[for details, see table~3 in][]{FederrathKlessen2012}. We return to these ideas in the next section when we examine star formation processes across each of our simulated disks.

Note that for a fixed value of $\sfrff=0.1$, Eq. \ref{eq:sfrff} reduces to the star formation law adopted on a cell-by-cell basis (Eq.~\ref{eq:schmidtH2}) in this work. In future work we will explore the galactic scale impact of self-consistently adopting turbulence based star formation models in our \agama/\ramses\ simulations. Preliminary result of such an approach is presented in section \ref{s:hinder}.
\begin{figure}[!htb]
\includegraphics[width=\textwidth]{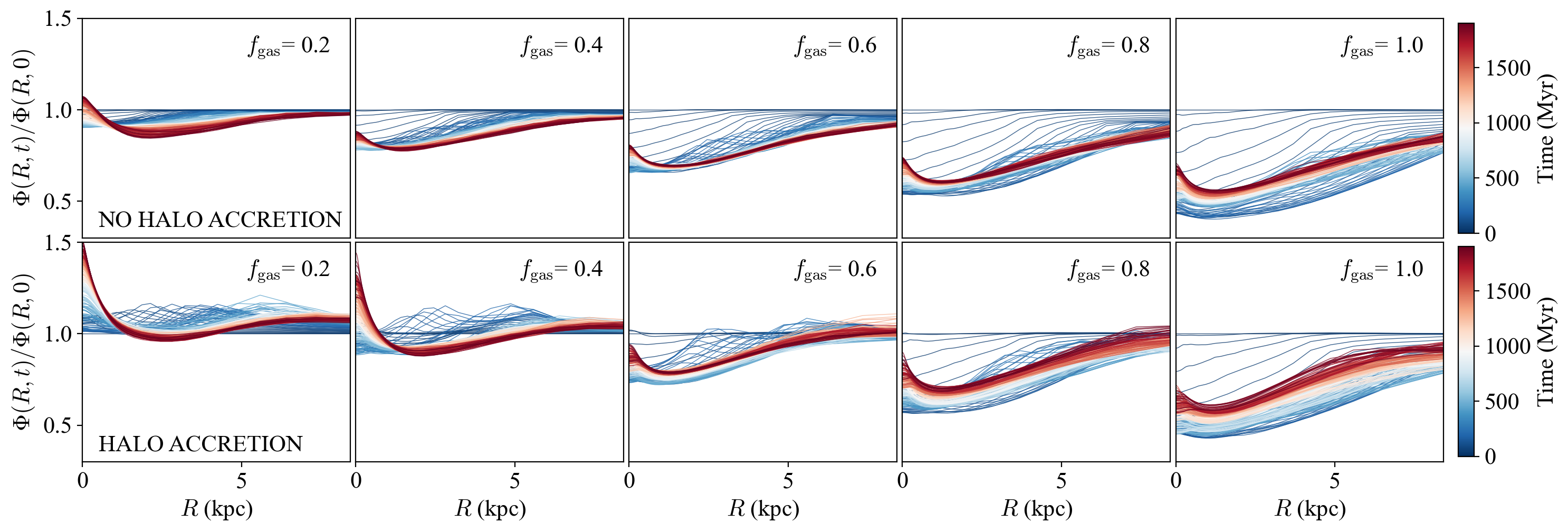}
    \caption{The evolution of the azimuthally averaged, radial profile of the
    total disk potential $\Phi(R,t)$ with cosmic time, normalised to the starting potential $\Phi(R,0)$ encoded in colour for (top) no halo accretion, (bottom) halo accretion. Blue tracks are early in cosmic time and red tracks are later times, as indicated. For $f_{\rm gas} < 0.5$, the disk potential is fairly constant but, above this limit, the loss of gas mass in circulating winds leads to a substantial weakening of the disk potential. Note that the vertical axis is a ratio of gravitational potentials, such that a curve moving downwards reflects a weaker potential.
    }
   \label{f:pots0}
\end{figure}
\begin{figure}[!htb]
    \centering
\includegraphics[width=0.49\textwidth]
{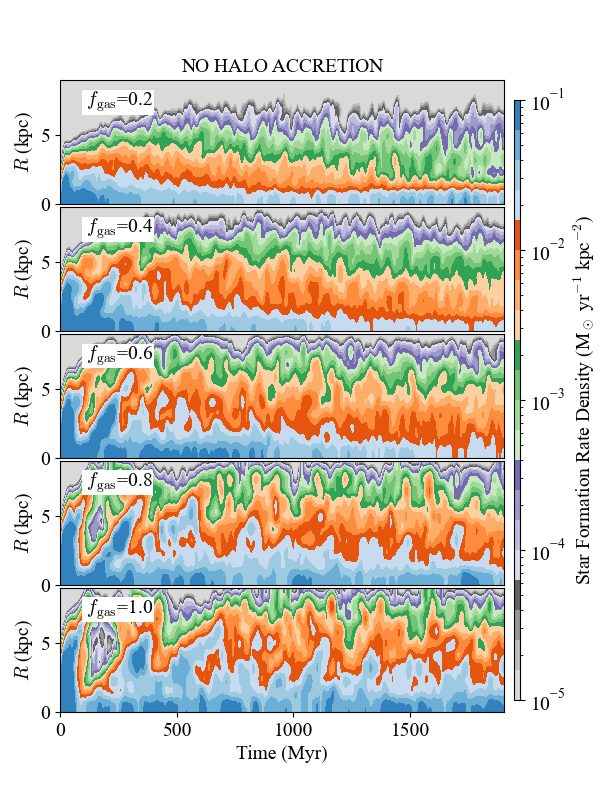}
\includegraphics[width=0.49\textwidth]
{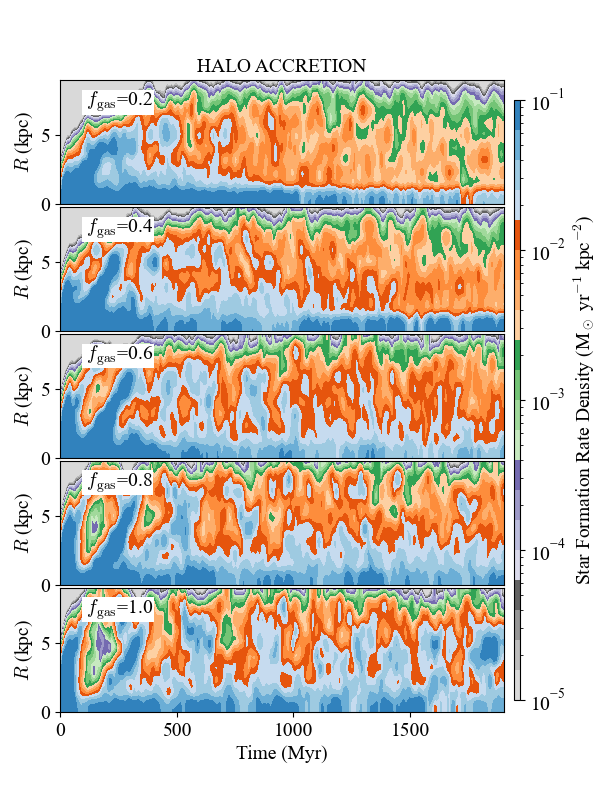}
\caption{
Periodograms drawn from our heavy disk ($f_{\rm disk}=0.5$) simulations that capture how the star formation rate (SFR) surface density (in units of M$_\odot$ yr$^{-1}$ kpc$^{-2}$) evolves as a function of radius and cosmic time: (Left) No halo accretion; (Right) Halo accretion. The five rows correspond to different gas fractions, in order from the top, $f_{\rm gas}=20,40,60,80,100$\%. All models commence with a short-lived, disk-wide starburst phase that moves outwards for the first 300 Myr. The higher $f_{\rm gas}$ models sustain a higher SFR surface density on average for both models. The halo-accretion models sustain higher star formation rates for all times compared to the matched ``no accretion" model.
   }
   \label{f:sfr_dens}
\end{figure}

\begin{figure}[!htb]
    \centering
    \includegraphics[width=1.05\textwidth]{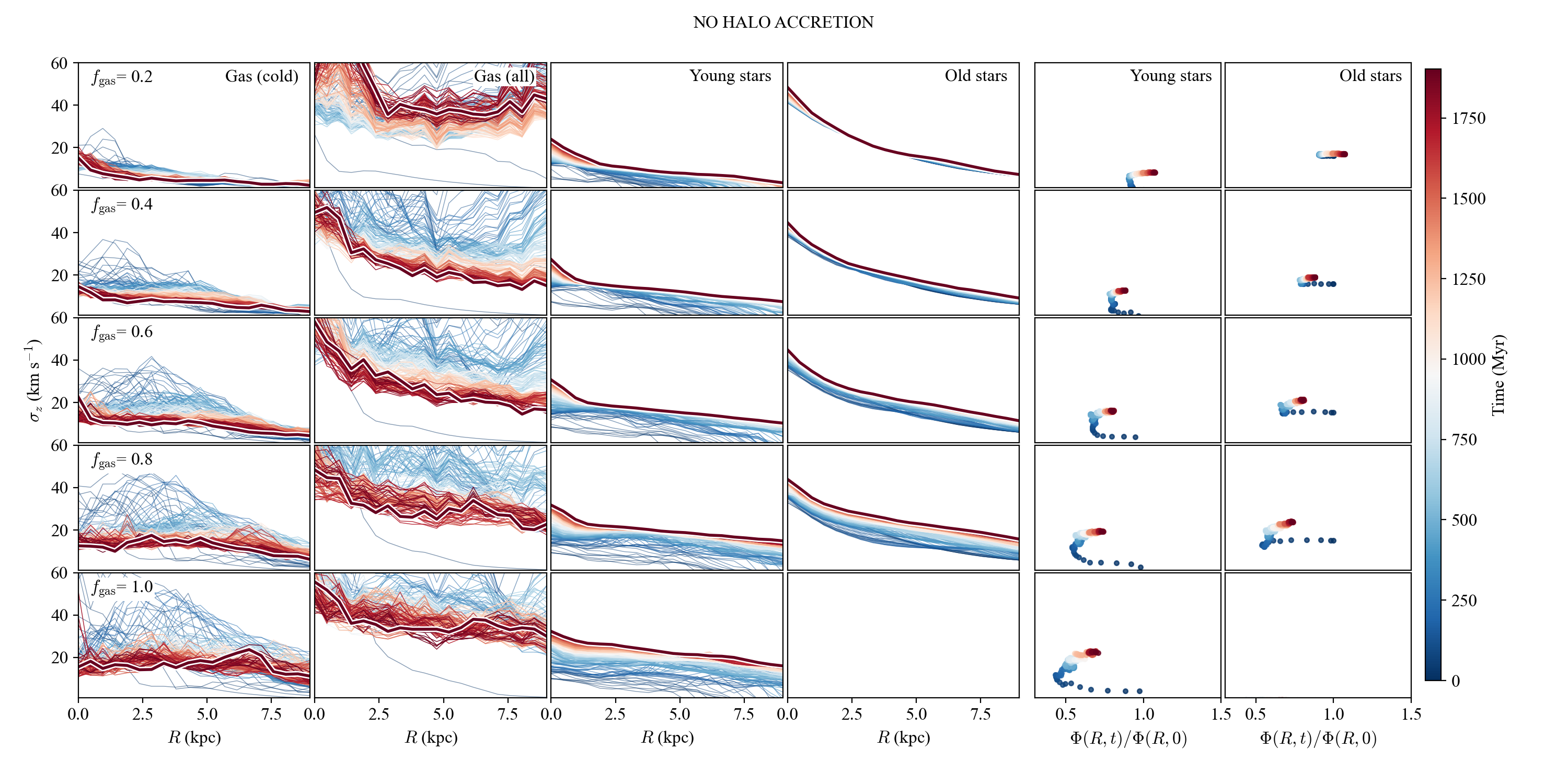}
    \includegraphics[width=1.05\textwidth]{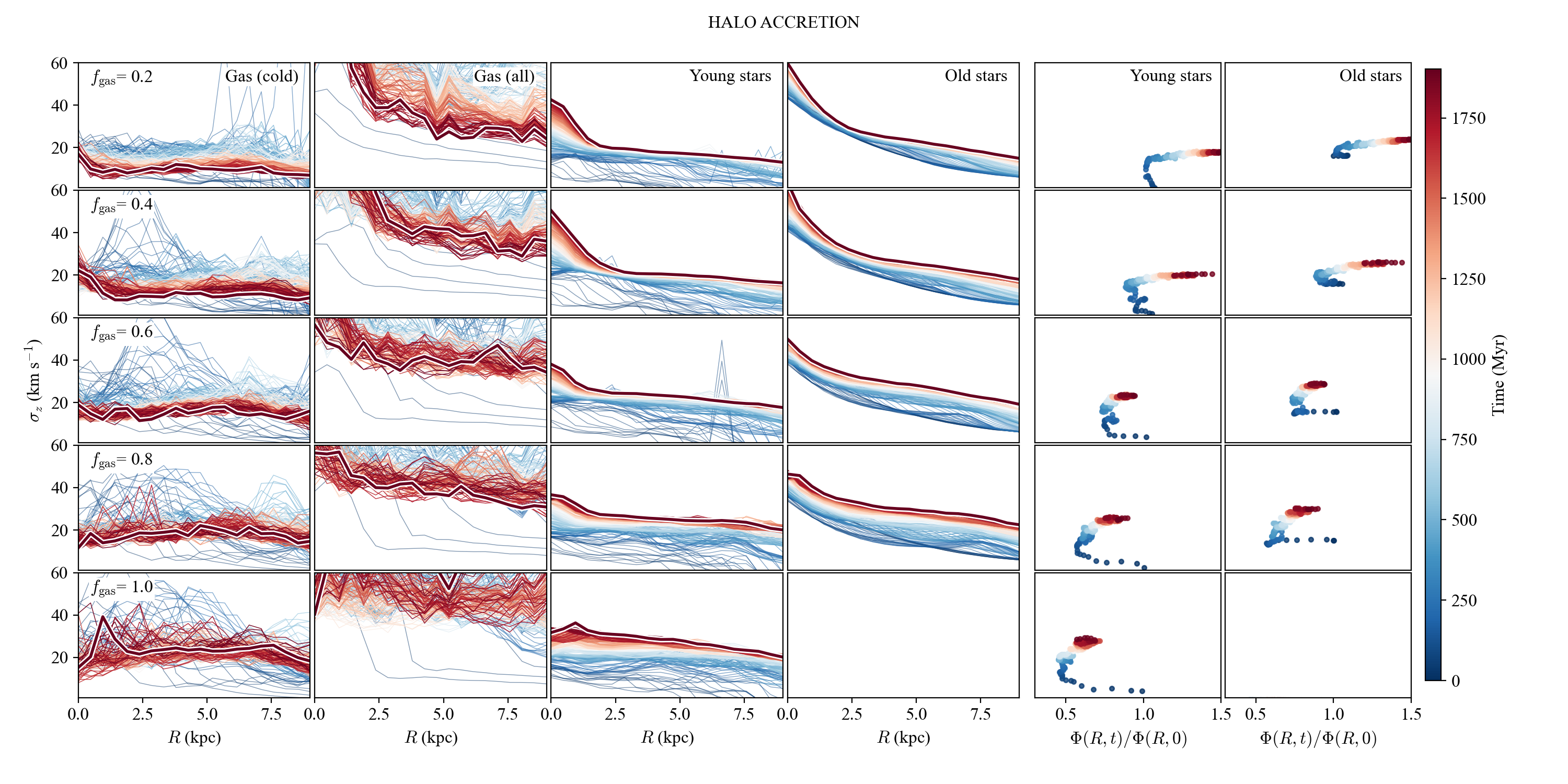}
   \caption{The evolution of the azimuthally averaged, vertical kinematic dispersion $\sigma_z$ with radius for all disk components: (top) no halo accretion, (bottom) halo accretion. Blue tracks are early in cosmic time and red tracks are later times. Within each collage, we show the evolution for all gas fractions: columns (1) cold gas, (2) all gas, (3) young stars, (4) pre-existing stars. The gas dispersions settle down from initially high values; interestingly, the stellar dispersions show the opposite trend. The superimposed thick line shows the profile at the last time step.
   In columns (5) and (6), the stellar $\sigma_z$ has a strong dependence on the depth of the disk's potential (see Fig.~\ref{f:pots0}) as a function of time (see text). The disk potential is weakening as points move to the left due to baryon mass loss.}
   \label{f:disp_evol}
\end{figure}

\bigskip
\section{Results: formation of gaseous and stellar bars} \label{s:results}

\subsection{Milky Way progenitor simulations and general properties}

In Table~\ref{t:mod}, the simulations are grouped into different categories $-$ low resolution, repeated simulations, and high resolution with different initial conditions. All movies are available at our website \href{http://www.physics.usyd.edu.au/turbo\_disks/}{http://www.physics.usyd.edu.au/turbo\_disks/}. There are 6 types of simulations distinguished by the different initial gas fractions ($f_{\rm gas}=0, 20, 40, 60, 80, 100\%$); these are run at low resolution for a total of 2 Gyr because we are focussed on the high-redshift universe ($z\gtrsim 3$). 

\smallskip
The filename convention is as follows:
\begin{quote}
We use {\tt fdXX\_fgYY\_nac} for ``no halo accretion'' models, and {\tt fdXX\_fgYY\_ac} for ``halo accretion'' models. Here, {\tt XX} is the disk mass fraction percentage and {\tt YY} is the gas fraction percentage. If the simulation has multiple versions generated by different random seeds, the file is referred to as {\tt fdXX\_fgYY\_nac.ZZ}, where {\tt ZZ} is the version number.
\end{quote}

In Fig.~\ref{f:fgas}, we present the evolution of $f_{\rm gas}$ for all 5 gas-rich disk models. The top row presents the models without halo accretion for radii within 2.2$R_{\rm disk}$ and within 10$R_{\rm disk}$. The second row shows the matching results for the halo accretion model. In both models, the gas is cooling from the outset and is characterised by an initial starburst forming stars, before declining rapidly to a lower gas fraction. The bottom two rows show that the disk loses progressively more of its mass after the initial burst as a function of $f_{\rm gas}$, but then regains some of the mass at later times. The mass loss clearly affects the inner disk (LHS figures) much more than the outer disk (RHS figures). This mass loss is seen more clearly with the evolution of the disk's gravitational potential (Fig.~\ref{f:pots0}). The radial traces are shown at different times, with blue corresponding to early times, and red curves depicting late times. For $f_{\rm gas} < 0.5$, the disk potential is fairly constant, but above this limit, the loss of gas mass in circulating winds leads to a substantial weakening of the disk potential. 

We also investigate if the initial setup influences the long-term behaviour of the simulations. We explore three different initial equations of state: (i) cooling ($t_{\rm start}=0$ Myr); (ii) short-term adiabaticity before cooling ($t_{\rm start}=50$ Myr); and (iii) long-term adiabaticity before cooling ($t_{\rm start}=400$ Myr). Here, $t_{\rm start}$ is the time at which gas cooling, and therefore heating, star formation and stellar feedback, are switched on. The first case leads to an initial starburst when the gas column density is at its peak; for the other two cases, the initial starburst event is more subdued. The long-term behaviour does not appear to be affected by the initial equation of state.

As we have seen, the gas fractions are systematically lower in the halo-accretion models, compared to the ``no accretion'' case, which is surprising at first glance. But an examination of the models reveals that the halo accretion drives more vigorous star formation, which consumes the gas at a faster rate, and some of the gas is carried away from the disk. We see this more clearly in the periodograms presented in Fig.~\ref{f:sfr_dens} by comparing the left and right figures at the same value of $f_{\rm gas}$. Periodograms simultaneously display the spatial and temporal behaviour of a quantity $q$. The spatial information is azimuthally averaged, such that a periodogram effectively represents a map $q(R, t)$, e.g. $\Sigma_{\rm SFR}(R,t)$. It is constructed by calculating a radial profile of $q$ at a given time, and concatenating each of these profiles for all available time steps.

All of the models in Fig.~\ref{f:sfr_dens} show high levels of burstiness and intermittent behaviour, becoming increasingly so at higher gas fraction. The models with the highest gas fractions exhibit other forms of intermittent behaviour, including spiral arms and bar-like structures that come and go. For this reason, we run the $f_{\rm gas}=40\%$ model three times with different random seeds to give us a better insight into the effects of stochasticity. (These models are {\tt fd50\_fg40\_noacc.01, fd50\_fg40\_noacc.02} and {\tt fd50\_fg40\_noacc.03}.) If the simulations are synchronized, we see how each of the models develop multiple flocculent spiral arms at different times, before they all settle into the bar-dominated phase within about 500 Myr. These are noisy systems.

\subsection{Gaseous and stellar velocity dispersions}
\label{s:sigmaz}

In Fig.~\ref{f:disp_evol}, we present the vertical kinematic dispersions $\sigma_z$ for all 5 gas-rich disk models measured through the simulated disks (averaged in azimuth) as a function of radius and time. Once again, the radial traces are shown at different times, with blue corresponding to early times, and red curves depicting late times. The top and bottom collages correspond to the ``no halo accretion'' and ``halo accretion'' models respectively. The left column shows the predicted results for cold gas ($T_{\rm gas} < 10^3$ K) and, in all cases, the vertical kinematics settle down to less than $\sigma_{z,\rm gas} \approx 20$ km s$^{-1}$ from a high of twice that value within a gigayear. For the halo accretion models, the settling dispersion is $20-40$ km s$^{-1}$ at the highest gas fractions.
For a baseline comparison, adopting specific mean molecular weights for the ISM comprising 74\% H, 24\% He and 2\% metals, the kinematic dispersions due to thermal motions are $0.6-0.9$ km s$^{-1}$ (\HI,
H$_2$ at $T_{\rm gas}=10^2$ K), $2-3$ km s$^{-1}$ (\HI,
H$_2$ at $T_{\rm gas}=10^3$ K), and $12-13$ km s$^{-1}$ (\HII\ at $T_{\rm gas}=10^4$ K).

Interestingly, in line with the observations, the simulated warm gas dispersions are systematically higher, and for all time. 
Gas dispersions measured with cool molecular lines tend to be narrower than dispersions measured with emission lines arising from warm gas, regardless of the galaxy's redshift \citep[e.g.][]{ejdet22}.
For all gas phases, our simulated dispersions are substantially broader than the thermal values given above.

For the ``no halo accretion'' models, the dispersions are a factor of two higher. The ``halo accretion'' models are systematically higher still, mostly driven by the stronger disk-halo circulation seen in all the higher $f_{\rm gas}$ models. Unlike the cold gas, this material is not in dynamical equilibrium. There are strong vertical wind flows in these models out to tens of kiloparsecs - we refer the reader to the movies on our website. Furthermore, the cool gas mass is typically larger than the warm gas mass, and therefore is more representative of the disk dynamics.

In Fig.~\ref{f:disp_evol}, the unexpected result is what is seen to occur with the vertical stellar dispersions, both for pre-existing stars and stars born within the gas in columns (3) and (4). In particular, the young stars are rapidly and kinematically heated from an initially very cold disk to a much warmer stellar disk. This does not reflect the high initial gas dispersions seen in the cold gas; these have exactly the opposite trend with time. Furthermore, the old stars are also heated at a time when the cold gas dispersions are settling down. Columns (5) and (6) illlustrate how the vertical dispersion $\sigma_z$ varies as the disk potential initially declines and then climbs again. We provide a novel interpretation of what is going on here in a later section (Sec.~\ref{s:diskmassloss}).

\begin{figure}[!htb]
    \centering
    \includegraphics[width=0.7\textwidth]{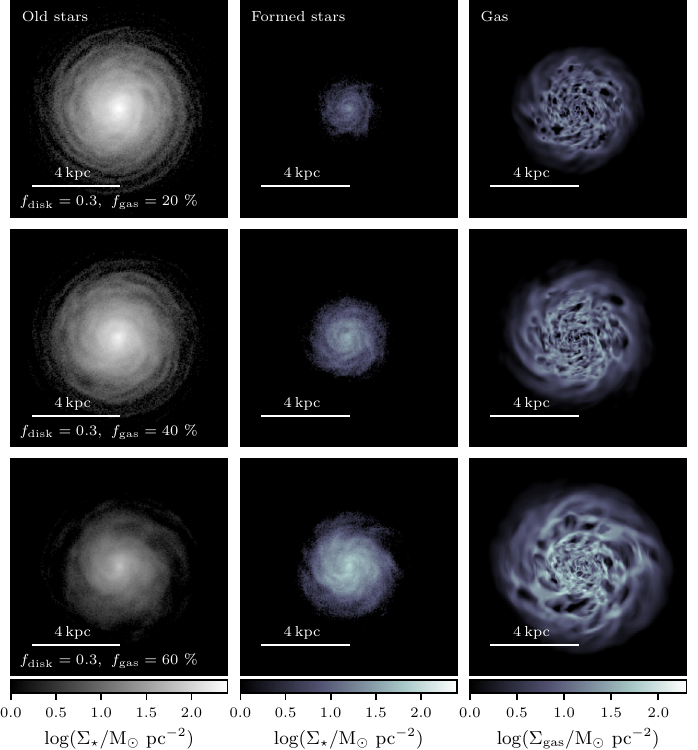}
    \caption{Final snapshots taken from three of our simulations in the low $f_{\rm disk}$ (=0.3) limit. The gas fractions from top to bottom are $f_{\rm gas} =$ 20\%, 40\% and 60\%; the pre-existing (old) stellar population has a smaller contribution, as we move to higher $f_{\rm gas}$, to maintain a constant disk mass.
    Even after 2~Gyr, a gas-rich disk does {\it not} form a bar  in contrast to what happens in heavy disks. 
    The box scale is 10$\times$10 ckpc.
    The {\it total} (stars$+$gas) disk baryon fraction $f_{\rm disk}$ is the primary agent that determines the onset timescale of the bar; $f_{\rm gas}$ becomes important only in the high $f_{\rm disk}$ limit (see text).
    }
   \label{f:lowfdisk}
\end{figure}

\begin{figure}[!htb]
    \centering
\includegraphics[width=0.49\textwidth]
{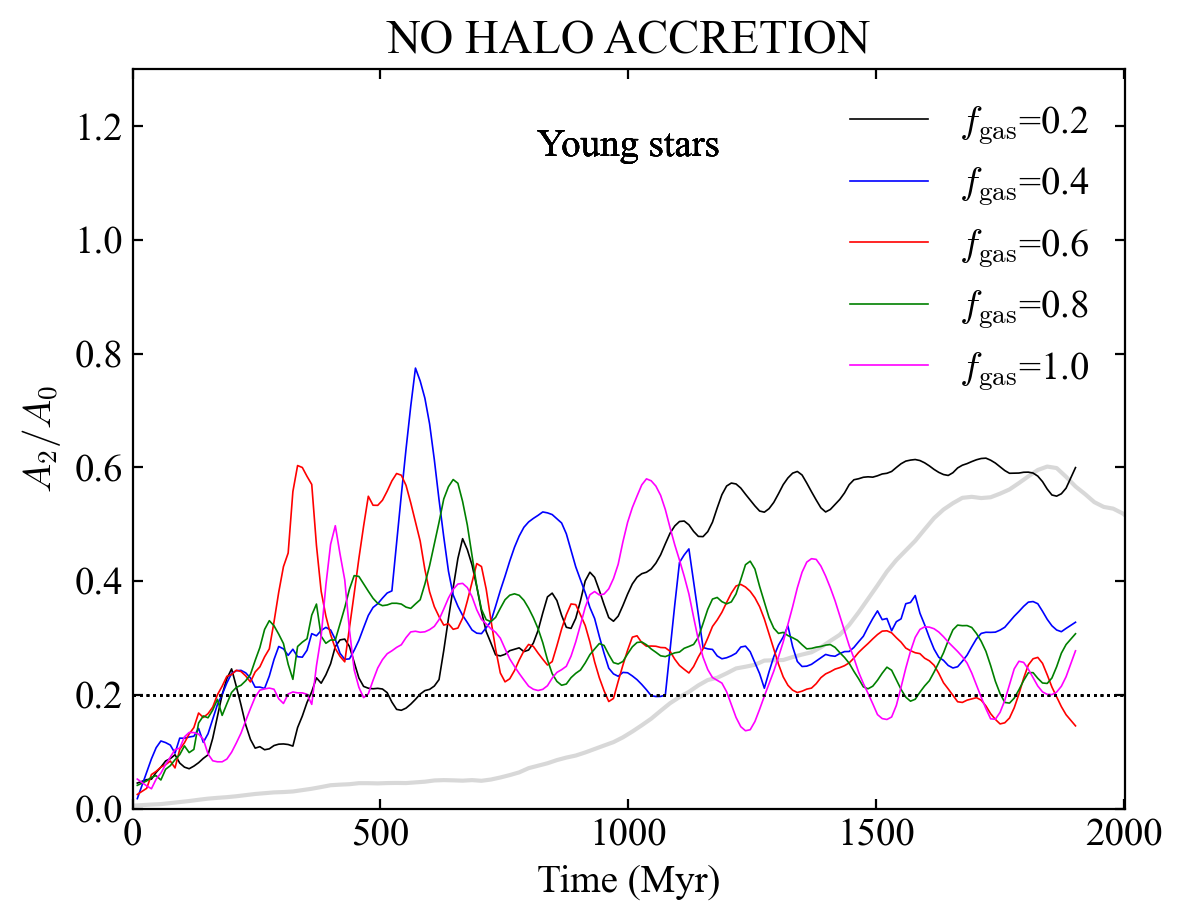}
\includegraphics[width=0.49\textwidth]
{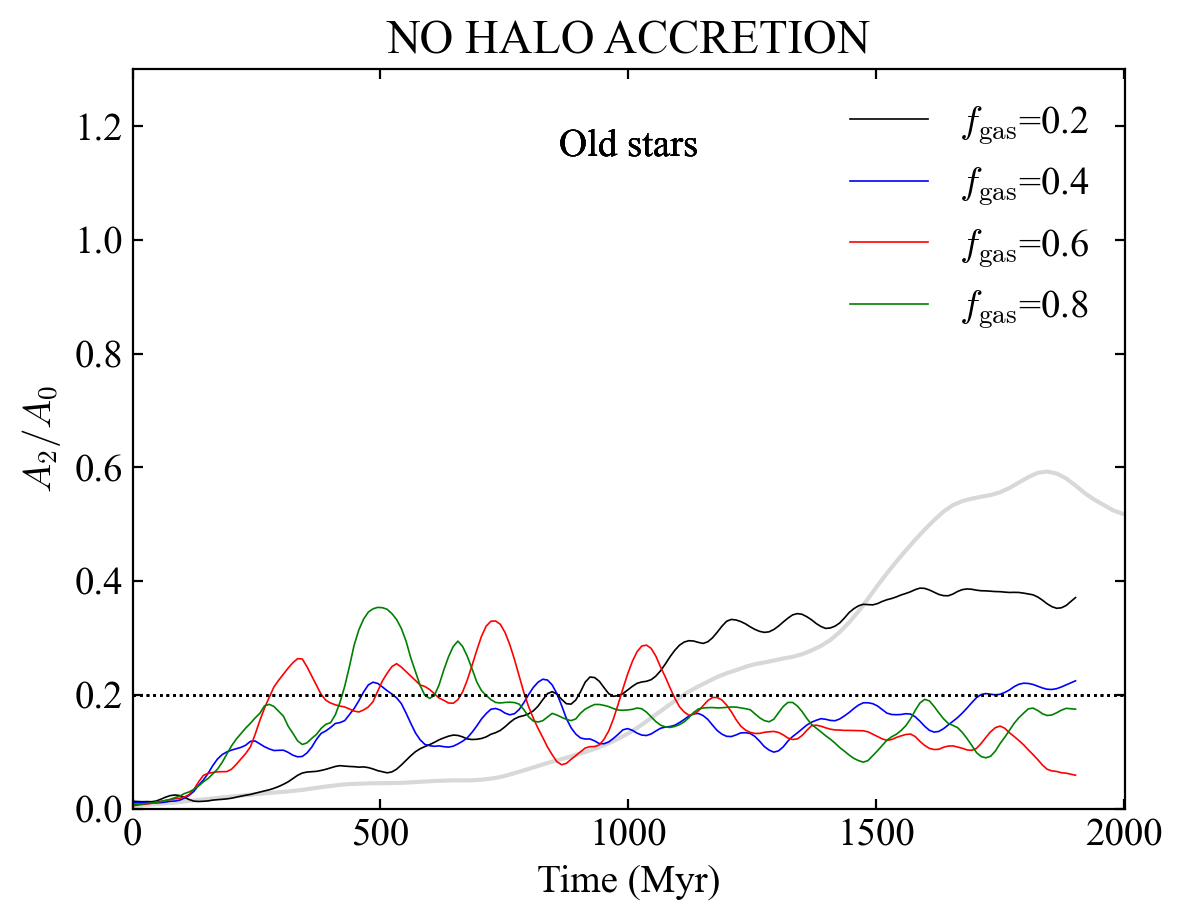}
\includegraphics[width=0.49\textwidth]
{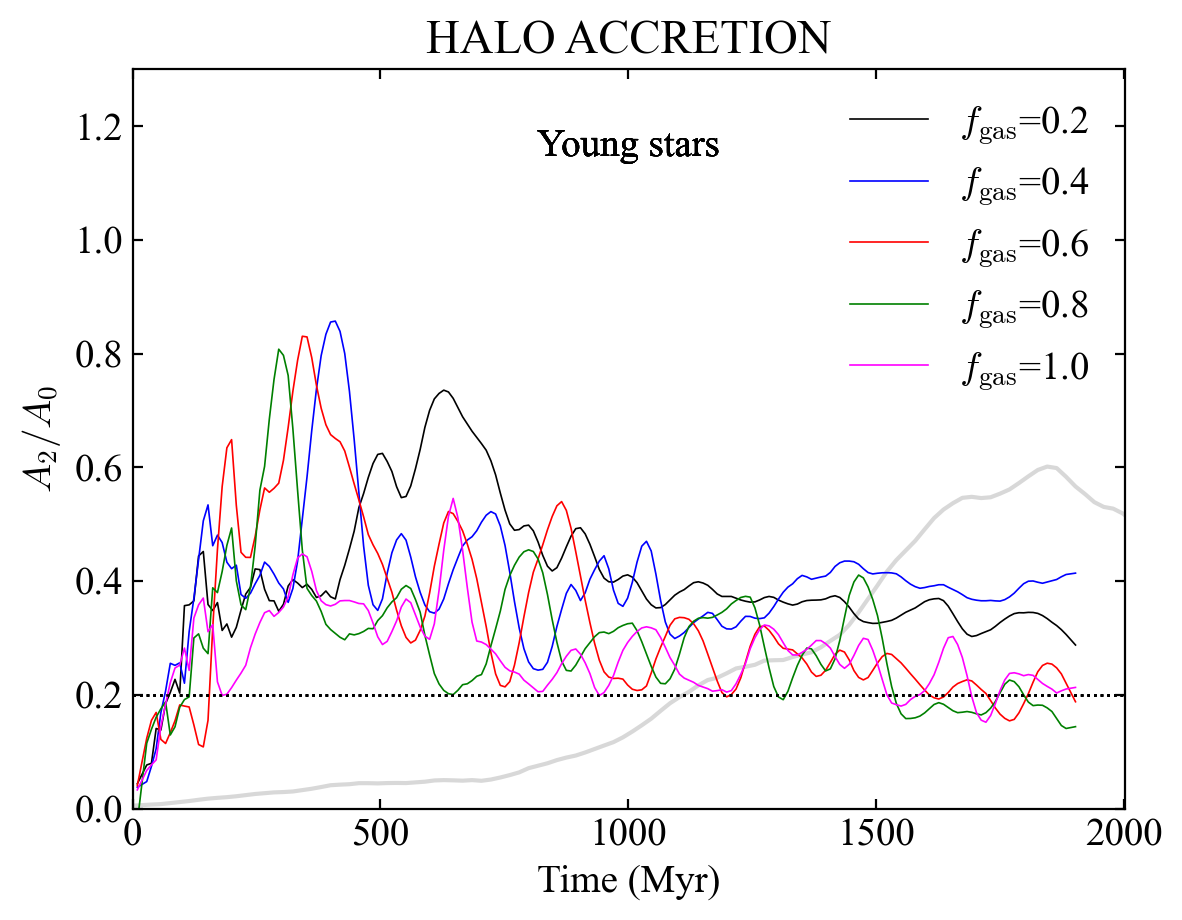}
\includegraphics[width=0.49\textwidth]
{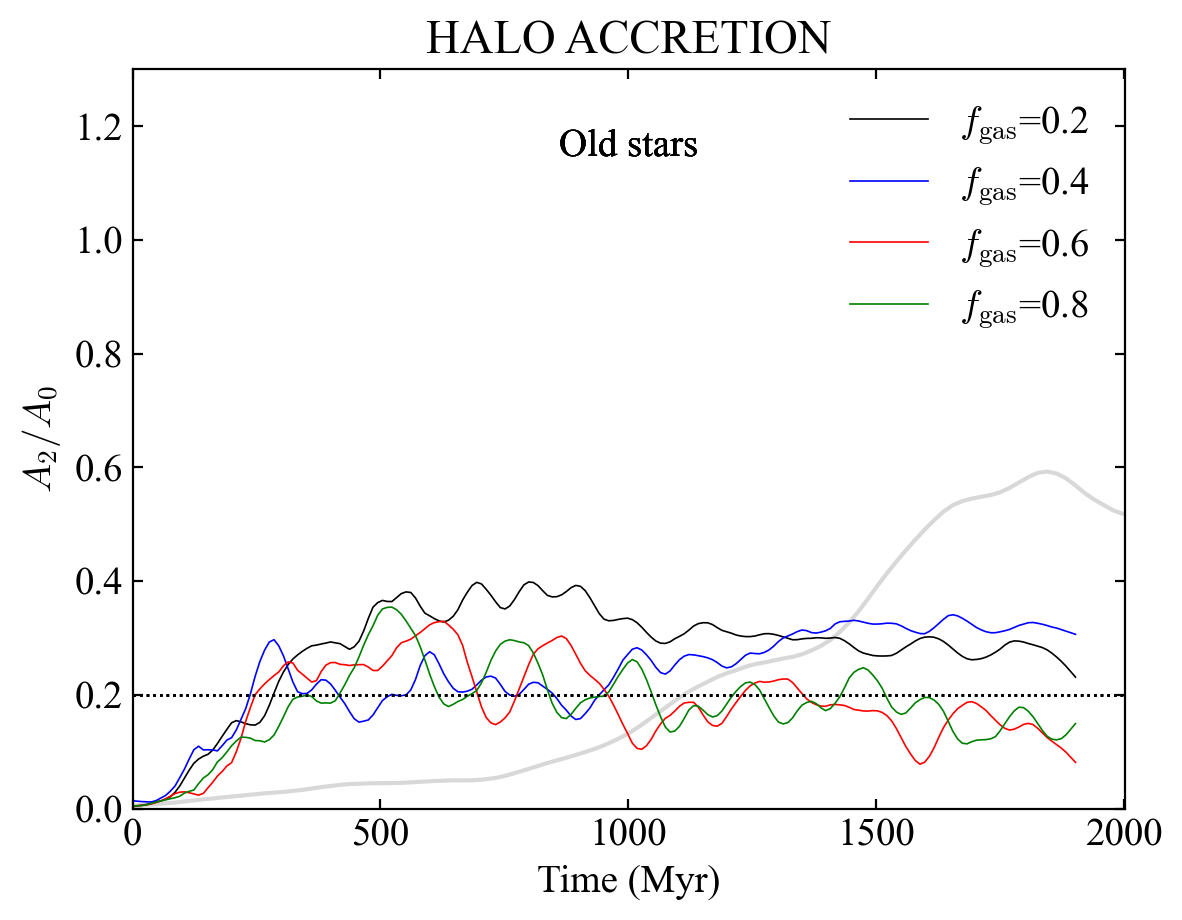}
    \caption{The evolution of $A_2/A_0$ in the surface density maps for (top) the ``no accretion" models shown in Fig.~\ref{f:lastbar}, (bottom) the ``halo accretion" models shown in Fig.~\ref{f:lastbarhalo}. These correspond to $f_{\rm disk}=0.5$ models; the coloured lines represent the different gas fractions as shown: (left) the young stellar disk, (right) the pre-existing stellar disk ($f_{\rm gas}=100\%$ not applicable). The grey line shows the bar emergence for the gas-free ($f_{\rm gas}=0$) simulation. The presence of turbulent gas speeds up the formation of the bar in all cases, but note that most bars are weaker than their gas-free counterpart, and appear to fade at the highest gas fractions. The horizontal dotted line is the widely used, minimum threshold ($A_2/A_0=0.2$) for the existence of a bar in an N-body simulation.
    }
   \label{f:a2}
\end{figure}

\begin{figure}[!htb]
    \centering
\includegraphics[width=0.6\textwidth]{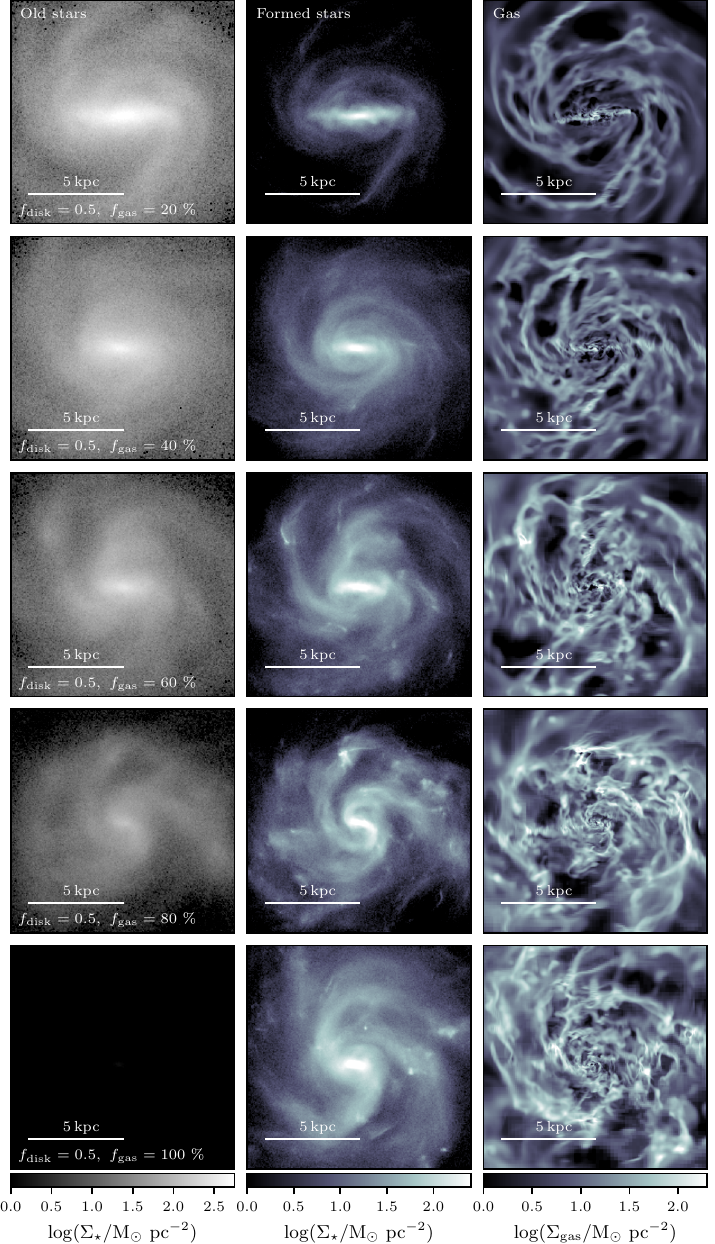}
\includegraphics[width=0.275\textwidth]{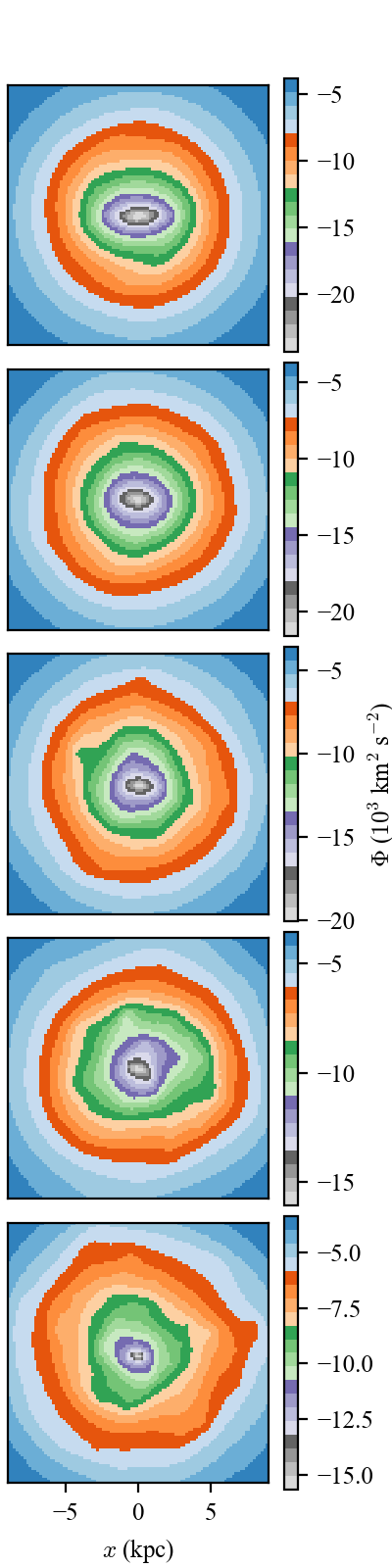}
    \caption{The five rows correspond to a single snapshot in time ($t_{\rm o}=2$ Gyr) taken from five $f_{\rm disk}=0.5$ simulations without halo accretion, in order from the top, $f_{\rm gas}=20,40,60,80,100$\%.
    The box scale is 12$\times$12 ckpc (comoving kpc). There is {\it no} halo accretion active here. The columns are: (1) surface density of pre-existing stars, (2) surface density of created stars, (3) gas surface density, (4) total gravitational potential. In the top three cases, the gas and young stellar bars survive for the full length of the simulation (2 Gyr). For the highest gas fractions, spiral arms are more diffuse, and bars are smaller and collapse to form bulges in the next timestep ($t_{\rm o}\approx 1.3$ Gyr). Radial shear flows are active in all cases, being strongest at low gas fraction (see Fig.~\ref{f:shear_bar}). These phenomena are clearly seen in the movies referenced in the main text.
    }
   \label{f:lastbar}
\end{figure}
\begin{figure}[!htb]
    \centering \includegraphics[width=0.6\textwidth]{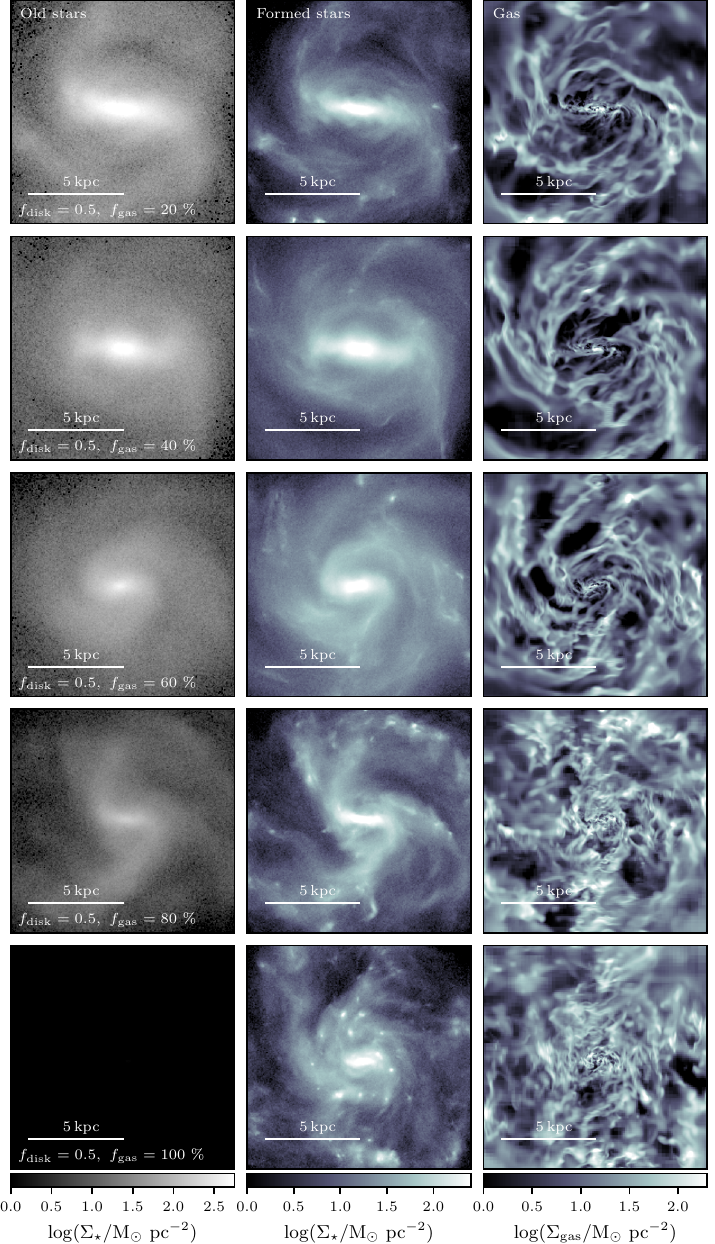}
\includegraphics[width=0.275\textwidth]{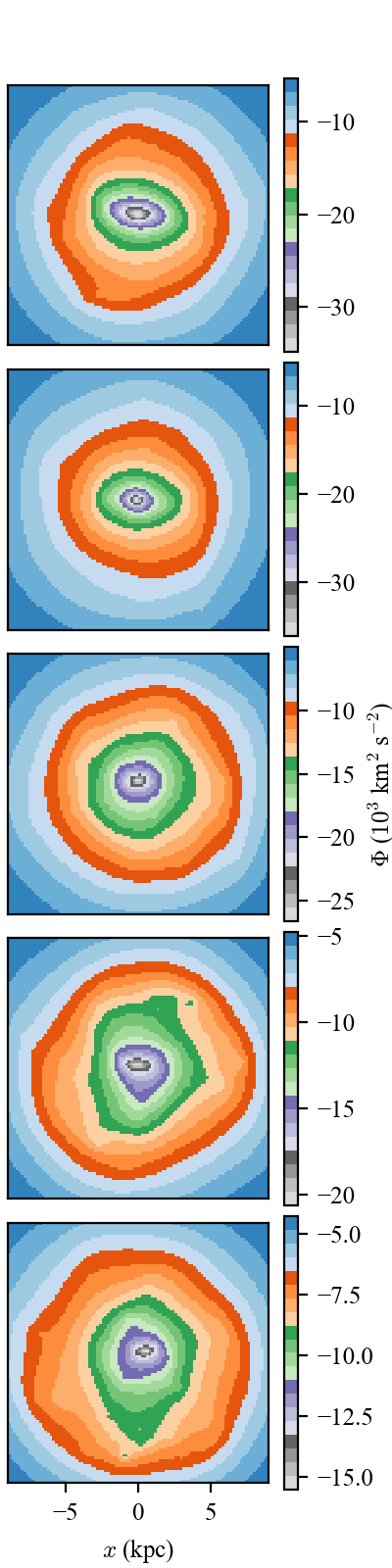}
    \caption{Same as Figure~\ref{f:lastbar}, but for simulations {\it with} halo accretion.
    }
   \label{f:lastbarhalo}
\end{figure}

\newpage
\subsection{Does gas help or hinder the formation of bar-like distortions?}
This is an age-old question that has never really been tackled in galactic dynamics, particularly in the presence of star formation and turbulent gas in the limit of high gas fraction.
Since the JWST discovery of well-developed stellar bars in turbulent, gas rich disks at $z\sim 2$ \citep{guo22,costat23}, this has become a topic of renewed interest. For earlier works that include gas, only a narrow range in gas fraction is treated and few consider prescriptions for star formation. Notable studies that do consider star-forming disks are \citet{ber07} and \citet{seo19}, but for these models, $f_{\rm gas}\lesssim 10\%$. One study suggests that the presence of gas reduces a {\it stellar} bar's lifetime \citep{vil10k} or at least weakens it \citep{ath13}. The lifetime of gaseous bars has never been considered, to our knowledge.

We believe our current framework is the ideal platform to treat bar formation in gas-rich disks. We perform simulations in the low (light) and high (heavy) disk baryon fraction limit, with a range of gas fractions. First, we show that the answer depends on whether disk baryons do or do not dominate the inner galactic potential. Gas appears to have minimal influence on bar formation when the disk is light, a conclusion that seems to hold for low or high gas fractions.

\newpage
\subsubsection{Light disks}

In \cite{bla23}, we extend the earlier work of \cite{fuj18a} to confirm the discovery of an inverse relation $-$ which we refer to as the Fujii relation $-$ between the stellar bar formation time $\tau_{\rm bar}$ and the disk baryon fraction $f_{\rm disk}$. \citet{fuj18a} found that the higher the value of $f_{\rm disk}$, the shorter the timescale for the onset of the bar instability. With sufficient disk resolution, this result is largely independent of the number of particles used to sample the density distribution \citep{fujii19,bla23}.


For $f_{\rm disk}$ evaluated at $R=R_s$, Fujii established that the dividing line occurs at $f_{\rm disk}\approx 0.3$, in the sense that smaller values lead to bar formation timescales that exceed a Hubble time. They also find that there is an asymptotic limit to the bar formation timescale such that, in the limit of high $f_{\rm disk}$, there is a finite {\it minimum} timescale for bar formation of order $1-2$ Gyr.

In the low $f_{\rm disk}$ limit ($f_{\rm disk}=0.3$), we run simulations with $f_{\rm gas}=0, 0.2, 0.4, 0.6$. These models are dubbed {\tt fd30\_fg20\_nac, fd30\_fg40\_nac} and {\tt fd30\_fg60\_nac} at our website. Since this work is primarily concerned with high-redshift discs ($z\gtrsim 3$), the expensive simulations are restricted to 2 Gyr. In all simulations shown in Fig.~\ref{f:lowfdisk}, we see that a bar does {\it not} form within 2 Gyr, thus confirming that bar formation is still suppressed in the light disc limit, even in the presence of a high gas content. {\it This is an important result.} Regardless of what makes up the disk baryons, it must dominate the local potential to form a stellar bar, or any bar-like distortion. By comparison, in heavy disks ($f_{\rm disk}\gtrsim 0.5$), internally triggered, {\it stellar} bars form in $1-2$ Gyr \citep{fuj18a,fujii19,bla23}, an issue we return to below.

\subsubsection{Heavy disks}
\label{s:heavdisks}

Our approach to exploring the role of gas is to run simulations over a range of gas fractions when the disk dominates, and to examine the bar strength and survival time, as illustrated in Fig.~\ref{f:a2}. But as the gas fraction increases, the role of any stellar bar becomes less important, so we consider the prospect of gas bars and/or young stellar bars that may form within them. 

We now examine the simulations in more depth. They are investigated using analytic methods that have become fairly routine \citep[e.g.][]{elm85,rix95}.
Bisymmetric disturbances are detected in both the surface density and kinematic maps; the latter are particularly useful in turbulent media. At each timestep, a Fourier decomposition is performed on the surface density of the simulated disk, such that
\begin{equation}
\frac{\Sigma(R,\phi)}{\Sigma_0} = \frac{1}{A_0} \sum^{\infty}_{m=0} A_m(R)e^{im[\phi_{\rm o}-\phi_m(R)]}
    \label{e:fourier}
\end{equation}
for which $A_m(R)$ and $\phi_m(R)$ are the Fourier amplitude and phase angle for the $m$th mode at a radius $R$, and $\Sigma_0$ is the central surface density. In our simulations, $m=2$ is the dominant mode when the bar emerges.

\citet{fuj18a} define $\tau_{\rm bar,0.2}$ as the timespan between the initial state of an unbarred synthetic galaxy, and the epoch at which the maximum normalised amplitude of the  quadrupole moment (Fourier $m=2$ mode), commonly referred to as $A_2/A_0$, crosses the (arbitrary) threshold value of 0.2. Once again, we adopt Fujii's $A_2/A_0$ threshold criterion.
The method used by \citet{fuj18a} to measure the Fourier amplitudes was not specified. To calculate $A_2/A_0$ and $\phi_2$, we adopt
the algorithm described in and the corresponding code provided by \cite{dehnen23}. In brief, they implement an automated identification of the bar region (subject to a number of tunable parameters) using the disk particles, and perform the discrete Fourier decomposition of the particle density within the bar region using an iterative procedure to calculate the azimuthal harmonics.

In Fig.~\ref{f:a2}, the evolution of $A_2/A_0$ is presented for a heavy disk with five different gas fractions. The models are dubbed {\tt fd50\_fg20\_nac, fd50\_fg40\_nac, fd50\_fg60\_nac, fd50\_fg80\_nac} and {\tt fd50\_fg100\_nac} at our website; in both panels, the grey line refers to the gas-free model, {\tt fd50\_fg00\_nac}. For $f_{\rm gas}=20$\%, we see from both panels that the bar onset occurs in half the time, or about 600-800 Myr depending on the stellar population considered.
While the trends are frenetic, the bar strength appears stronger and the bar onset is faster (300-600 Myr) for the young stars compared to the old stars. 
Interestingly, the bar strength appears to slowly fade for both young and old stars for the highest gas fractions, in agreement with \citet{ath13}, although the latter bars took {\it longer} to form with the inclusion of gas.

The evolutionary tracks in Fig.~\ref{f:a2} were drawn from the models presented in Figs.~\ref{f:lastbar} and ~\ref{f:lastbarhalo}. In the latter figures, which corresponds to the snapshots where the young bar is strongest in each case,
the top three rows reveal well-pronounced young stellar bars, all of which show signs of a strong radial shear flow, evident in the density map, but particularly prominent in the stellar/gas kinematics (see below). These bars all form within about 300-600 Myr and survive for the full length of the simulation ($\sim$2 Gyr). The bar length correlates with the gas fraction, such that the bar radius is about 3 ckpc (comoving kpc) for $f_{\rm gas}=20$\%, declining to about $1$ ckpc in the high $f_{\rm gas}$ limit. The bars in both of the highest gas fraction simulations collapse to form bulges shortly after the timestep shown. 

Thus we find that $f_{\rm gas}$ is found to have a secondary, but important, role in the presence of a heavy disk, and obviously a primary role when the stellar content is close to zero. The fact that turbulent gas accelerates bar formation, compared to {\em inert} gas that seems to delay the process \citep[e.g.][]{ath13,bla23}, requires an explanation. An overarching theme of the simulations, when viewed as a whole, is that higher gas fractions lead to noisier and clumpier galaxies.
Our expectation is that the increasing noise levels in the gas imparts the same or similar perturbations on the evolving stellar disk. We explore this idea in the next section.


\bigskip
\begin{figure}[!htb]
    \centering \includegraphics[width=0.49\textwidth]{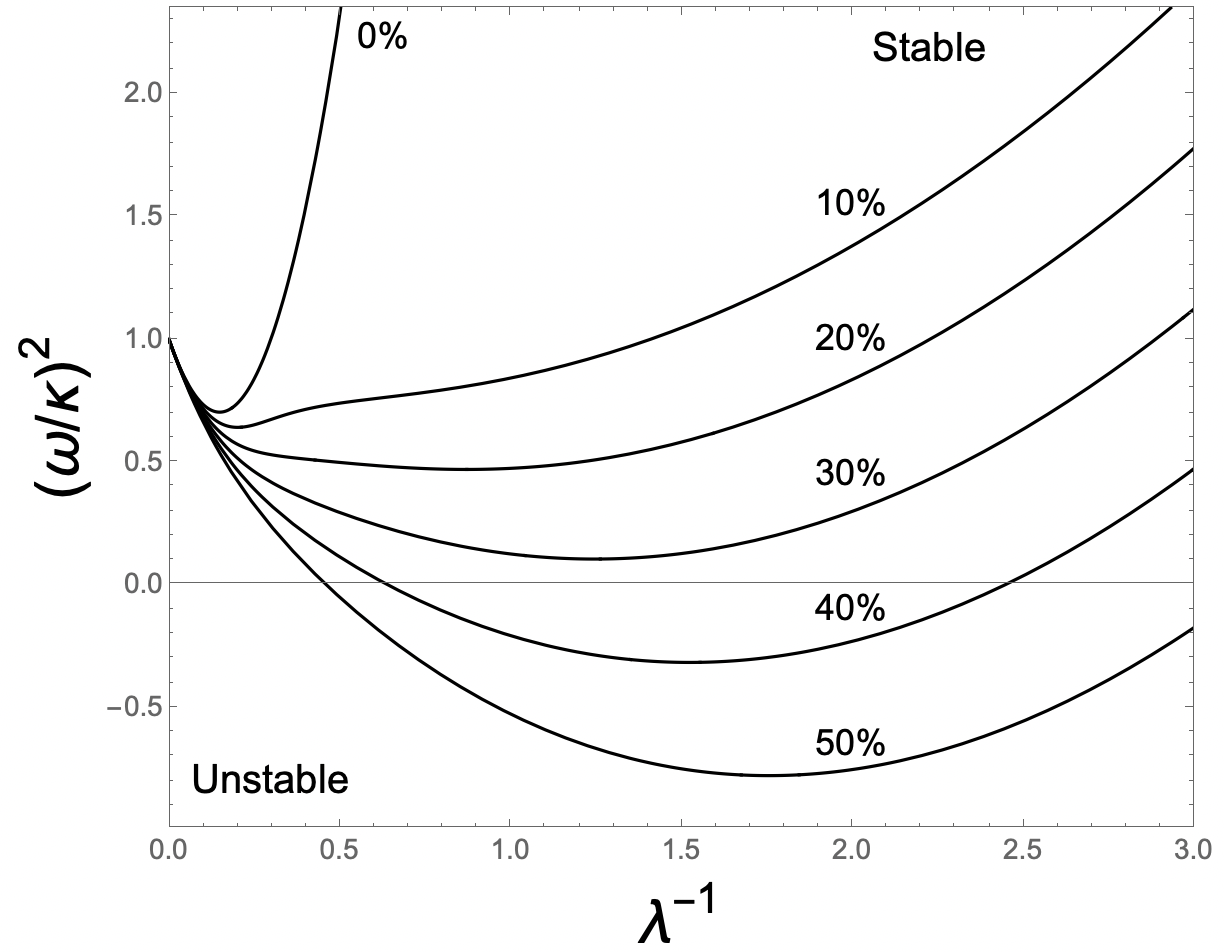}
\includegraphics[width=0.49\textwidth]{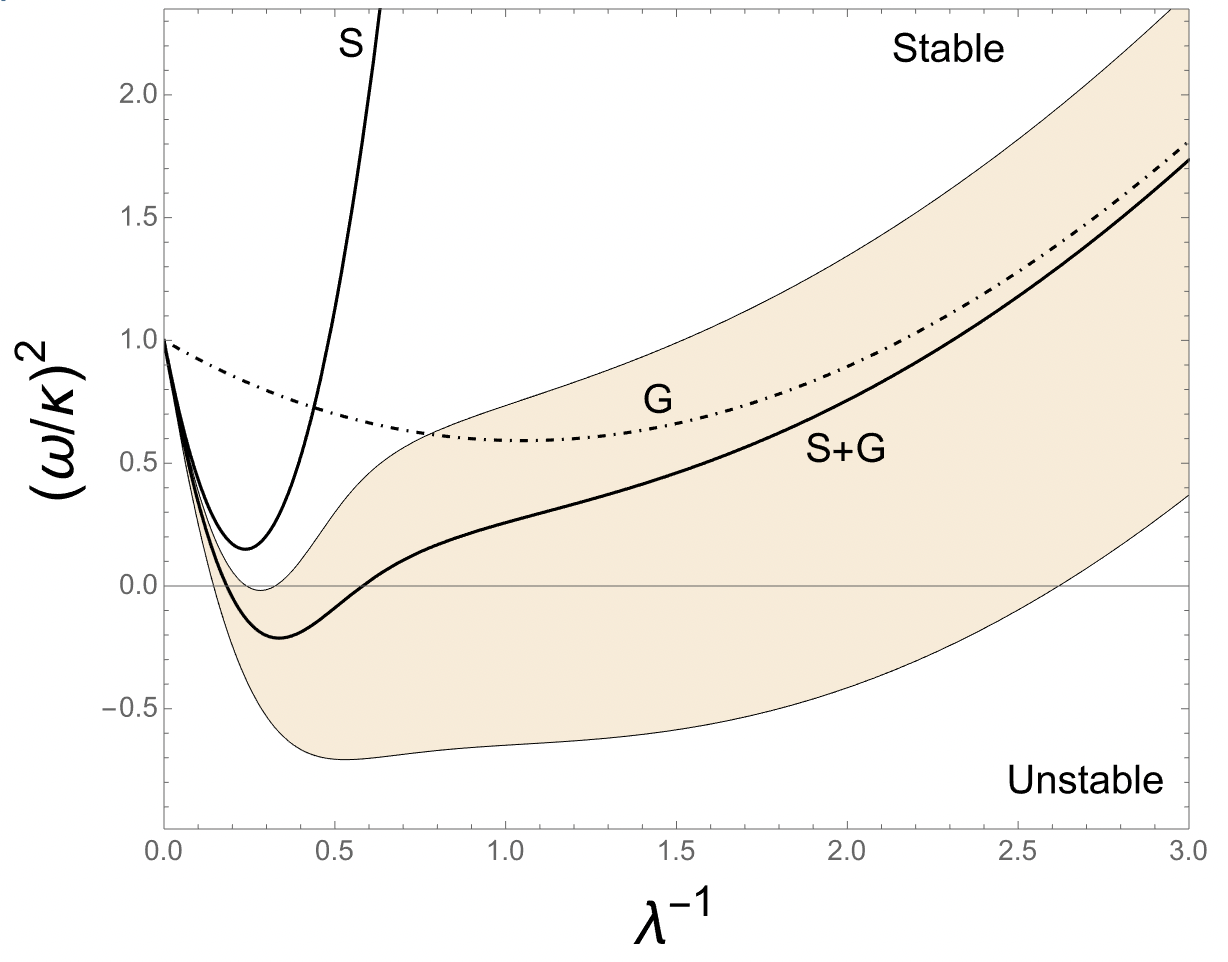}
    \caption{(Left) Square of the normalized oscillation frequency ($\omega^2/\kappa^2$) vs. inverse wavelength ($\lambda^{-1}$) of the perturbation (in units of kpc$^{-1}$). The different curves show the degree of stability for different gas fractions $f_{\rm gas}$. The upper region ($\omega^2/\kappa^2 > 0$) is largely stable; the lower region ($\omega^2/\kappa^2 < 0$) is where perturbations grow exponentially.  (Right) The different contributions from stars only (S) and gas only (G) to the overall two-fluid stability. Individually, they appear stable, but when co-existing, the two-fluid medium is more unstable (S+G). The shaded region illustrates the changing conditions in the S+G combined fluid when the local gas surface density varies by a factor of two, as in a turbulent medium.
    }
   \label{f:jog}
\end{figure}

\subsubsection{Instabilities in star-gas fluids}
\label{s:instab}

We have seen that raising the gas fraction ($f_{\rm gas}$) in the disk makes the disk increasingly unstable, and excites low order modes that lead to stellar/gas bars on shorter timescales compared to gas-free (single fluid) disks. The growth of gravitational instabilities in a fluid has played a role in galactic dynamics since the 1960s \citep[q.v.][]{bin08}. Several authors have suggested that the solar neighbourhood should be relatively stable, but even small perturbations on the assumed parameters can lead to very different growth rates \citep[e.g.][]{raf01}. This effect is amplified when one considers the gravitational interaction of two fluids, particularly when one is dynamically colder or more viscous than the other. Both fluids can be separately stable, but then rendered unstable by ``the additional gravitational self-energy in the system resulting from the gravitational interaction between the two fluids (\citealt{jog84}, see also \citealt{romeo92}).''

Perturbations propagate in space and time. In linear analysis, for any physical property under study, there is a factor $\exp[i(k r +\omega t)]$ where $\omega$ is referred to as the angular frequency and $k$ is the wavenumber of the perturbation with wavelength $\lambda$ ($k=2\pi/\lambda$). After using Poisson's equation to treat perturbations in the gravitational potential,
\citet{jog84} arrive at a quadratic dispersion relation with solutions
\begin{equation}
    \omega^2(k) = \frac{1}{2}\left((\alpha_\star+\alpha_{\rm gas})\pm \sqrt{(\alpha_\star+\alpha_{\rm gas})^2-4(\alpha_\star\alpha_{\rm gas}-\beta_\star\beta_{\rm gas})}\right).
\end{equation}
The subscripts ``$\star$" and ``gas" refer to the stellar and gas fluids; the variables are as follows:
\begin{eqnarray}
    \beta_\star &=& 2\pi G k \mu_\star {\cal R}_\star \\
    \beta_{\rm gas} &=& 2\pi G k \mu_{\rm gas} {\cal R}_{\rm gas} \\
    \alpha_\star &=& \kappa^2 + k^2 c_\star^2 -\beta_\star \\
    \alpha_{\rm gas} &=& \kappa^2 + k^2 c_{\rm gas}^2 -\beta_{\rm gas}
\end{eqnarray}
where ${\cal R}_\star$ and ${\cal R}_{\rm gas}$ are the so-called reduction factors that correct for the different vertical scaleheights of the disks (see below). The positive root describes stable oscillatory perturbations; the negative root considers the transition from stable ($\omega^2 > 0$) to unstable modes ($\omega^2 < 0$). The in-plane epicyclic frequency $\kappa$ recognizes that the differential disk rotation influences the fluid instabilities. The two fluids (with thermal
sound speeds $c_\star$ and $c_{\rm gas}$) have local surface densities $\mu_\star$ and $\mu_{\rm gas}$. The stellar ``sound speed'' is usually taken as the local radial velocity dispersion. The dispersion relation for $\omega^2$ is solved in terms of $k$, or equivalently $1/\lambda$ as we show in Fig.~\ref{f:jog}, where $\omega^2$ is made dimensionless by normalizing to $\kappa^2$.

In Fig.~\ref{f:jog} (left), the increasing trend towards instability is clear as the gas fraction $f_{\rm gas}$ increases. Here, we adopt $\mu_{\rm total} = 10^8$ M$_\odot$ kpc$^{-2}$, $\kappa=36$ km s$^{-1}$ kpc$^{-1}$, $c_\star=35$ km s$^{-1}$ and $c_{\rm gas}=5$ km s$^{-1}$, inspired by the solar neighbourhood values considered by \citet{jog84}. As presented, this figure gives the false impression that low gas fractions are everywhere stable. These curves correspond to average values and do not treat local variations.

In Fig.~\ref{f:jog} (right), we look at the two-fluid system in more detail. The gaseous (G) and stellar (S) fluids are independently stable, but when considered together (S+G), they are much less so, particularly on scales of $2-3$ kpc, due to their mutual interaction. The curves shown assume $f_{\rm gas}=20\%$ and correct for the scale-height differences between the two fluids, as presented by \citet[][Eq. (23)]{jog84}. The reduction factors are 
\begin{eqnarray}
    {\cal R}_\star &=& (1-\exp(-kh_\star))/kh_\star, \\
    {\cal R}_{\rm gas} &=& (1-\exp(-kh_{\rm gas}))/kh_{\rm gas}.
\end{eqnarray}
Here we adopt exponential scale heights of $h_\star=300$ pc and $h_{\rm gas}=75$ pc, otherwise using the same parameters above, except the total surface density is now doubled in line with \citet{jog84} and our models, i.e. $\mu_{\rm total} = 2\times 10^8$ M$_\odot$ kpc$^{-2}$. Note how the quadratic ``stars only'' curve has now moved closer to the instability region -- compared to the left panel, in line with the Toomre Q criterion in  Eq.~\ref{e:Q} and its inverse dependence on the surface density.

In Fig.~\ref{f:jog} (right), the shaded region illustrates the changing conditions arising when the local gas surface density varies by a factor of two. The shaded area grows larger as $f_{\rm gas}$ increases. In our models, the turbulent medium drives large fluctuations in the local gravitational potential. {\it These fluctuations are imposed on the stars and explain the faster onset of the bar when gas becomes important.}

We refrain from a more detailed analysis at this time, in particular, a study of 3D instabilities associated with cloud formation and their comparison with 2D instabilities affecting the disk (see e.g. \citealt{romeo10} on how turbulence scaling relations affect disc stability). Partial (Jeans-like) 3D instabilities are possible even when 2D instabilities are suppressed. In fact, this is evident when watching the development of the most massive clumps in the low $f_{\rm gas}$ simulations (cf. Fig.~\ref{f:jog}). An excellent updated review and analysis of these distinctions is given in \citet{meidt22}.

\bigskip
\begin{figure}[!htb]
    \centering
    \includegraphics[width=0.99\textwidth]{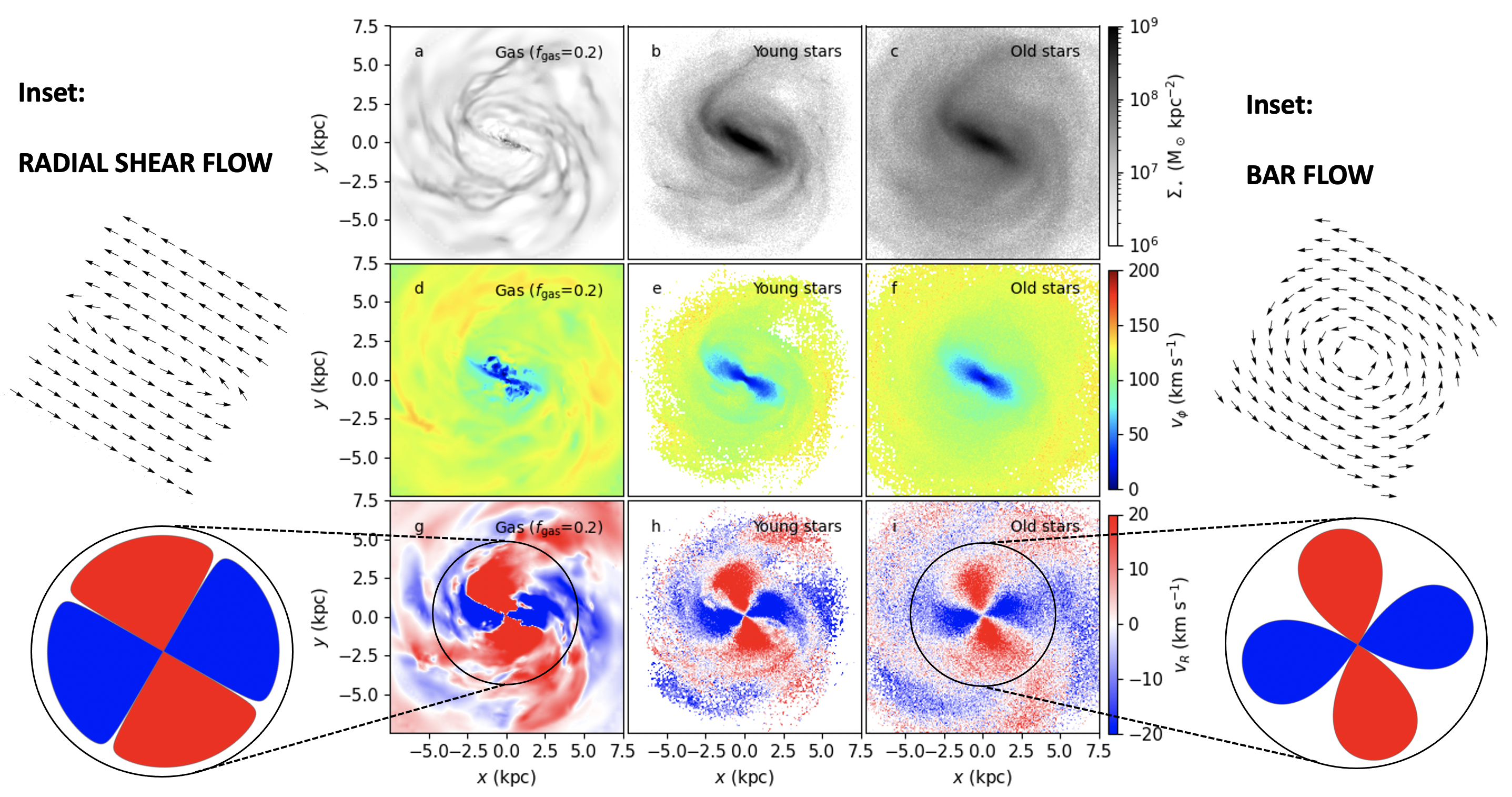}
    \caption{A collage showing the surface density and kinematic maps at ${t_{\rm o}=2}$ Gyr for our model with $f_{\rm disk}=50\%$ and $f_{\rm gas}=20\%$. In the top row, from left to right, the projected surface density is shown for gas, newly created stars and pre-existing (old) stars. In the middle row, we present the tangential velocity $v_\phi$, and in the bottom row, the radial velocity $v_R$. The developing bar is particularly prominent at late times in the stellar components. The expected quadrupolar signature (``quatrefoil'') is prominent in $v_R$ although the patterns are different for the gas and the old stars, as illustrated by the colour insets. These arise from different flow patterns as indicated: the radial shear flow on the left, and the bar flow on the right. The (saturated) encircled schematics correspond to the circular regions indicated in the bottom row.
    }
   \label{f:shear_bar}
\end{figure}

\subsubsection{Summary}

In summary, we conclude that what fundamentally drives bars to form is $f_{\rm disk}$, i.e. the {\it total} disk baryon content, with $f_{\rm gas}$ having no strong influence for light disks. Conversely, the story is very different for heavy disks, where a new mechanism appears to be at play in the presence of a high gas fraction. Gas fractions higher than about $f_{\rm gas}\gtrsim 10\%$ appear to speed up the onset of a bar, contrary to what is reported elsewhere albeit for lower gas fractions \citep{ber07,ath13}. The difference may be due to our more realistic treatment involving star formation, where the new stars contribute to the non-dissipative component in the disk. Interestingly, \citet{rob17} found that, in the presence of AGN feedback, bars form earlier for higher $f_{\rm gas}$. It is not obvious why feedback processes (AGN or star formation) are able to speed up bar formation. But, as we discuss in Sec.~\ref{s:instab}, turbulent energy tends to introduce fluctuations in the local gravitational potential, and these drive accelerated exponential growth in the density perturbations.


\bigskip
\subsection{The emergence of bar-like distortions in turbulent gas}

\subsubsection{Radial shear flow}
\label{s:radialshearflow}

The top row of Fig.~\ref{f:lastbar} presents a snapshot of the ``no accretion'' model for $f_{\rm gas}=20\%$ at $t_{\rm o}=2$ Gyr; we explore these results in more detail in Fig.~\ref{f:shear_bar}. As the movies reveal, newly formed stars emerge into what appears to be an extreme `bar-like' flow, or a radial shear flow \citep[cf.][]{Li17}.  This is best seen by watching the simulation {\tt fd50\_fg20\_nac} at our website, although a similar mechanism is clearly operating in {\tt fd50\_fg40\_nac} and {\tt fd50\_fg60\_nac}. After a few disk rotations, a differentially rotating, gas-rich turbulent disk develops a pile-up of turbulent gas in two opposing streams. The flow appears to operate with or without an underlying stellar bar, although the flow is strongest at low $f_{\rm gas}$ when there is a stellar bar present. 

In Fig.~\ref{f:shear_bar}, the radial velocity of the gas $v_R$ has a distinctive pattern. This radial shear flow is long-lived, tumbles with the disk's rotation and gives rise to a kinematic ``quatrefoil'' pattern that is more extreme than seen in normal bar-driven flows. In particular, compare the LHS and RHS insets: the vector flow patterns for both kinds of flow are also presented in Fig.~\ref{f:shear_bar} (see insets). The equivalent maps for $f_{\rm gas}=40,60,80,100\%$ are included at our website in the interests of brevity.

\begin{figure}[t]
    \centering
    \includegraphics[width=0.7\textwidth]{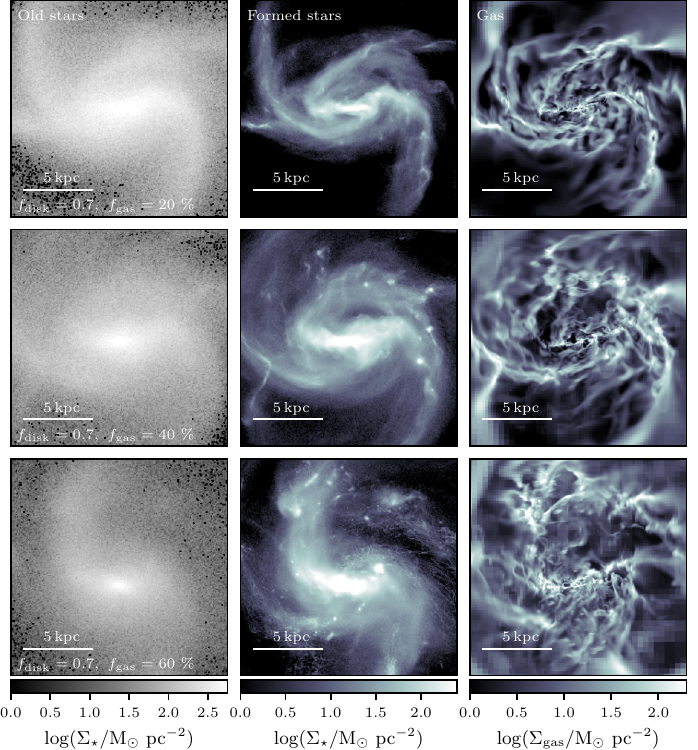}
    \caption{The three rows correspond to a single snapshot in time taken from three $f_{\rm disk}=0.7$ simulations, in order from the top, $f_{\rm gas}=20,40,60$\%. 
    The box scale is 16$\times$16 ckpc. There is {\it no} halo accretion active here. The columns are: (1) surface density of pre-existing stars, (2) surface density of created stars, (3) gas surface density. Compared to Fig.~\ref{f:lastbar}, these disks are even more dominant, and the implied gas content (by mass) is higher. Here, we observe very large bar-like distortions and strong radial shear flows that form earlier than for lighter disks. 
    }
   \label{f:lastbar2}
\end{figure}

In Fig.~\ref{f:lastbar2}, we repeat the top three models in Fig.~\ref{f:lastbar}, but this time they correspond to disks characterised by $f_{\rm disk}=0.7$, the heaviest disks we have simulated to date. These models are dubbed {\tt fd70\_fg20\_nac, fd70\_fg40\_nac} and {\tt fd70\_fg60\_nac} at our website. The bar is longer and stronger, and the radial shear flow is even more pronounced. The terminator line between the two flows is particularly evident in the top right panel in Fig.~\ref{f:lastbar}; it appears as a thin, horizontal enhancement of gas about 4 ckpc in length centred on the galaxy. The terminator is also evident in panel `d' of Fig.~\ref{f:shear_bar} as a thin blue strip with $v_\phi\approx 0$, i.e. a radial gas stream with no circular motions in the rest frame of the flow.

The roiling action of this radial shear flow is clearly seen by watching the simulation {\tt fd70\_fg20\_nac} at our website. Interestingly, \citet[][]{Li17}, in the absence of strong turbulence, may be witnessing a similar behaviour in their hydrodynamic simulations.
Heavy, turbulent disks manifest this extraordinary behaviour, leading to kinematic signatures that are somewhat different from radial bar streaming in the local Universe. These signatures may be observable in future high-resolution observations.

\begin{figure}[!htb]
    \centering
    \includegraphics[width=0.65\textwidth]{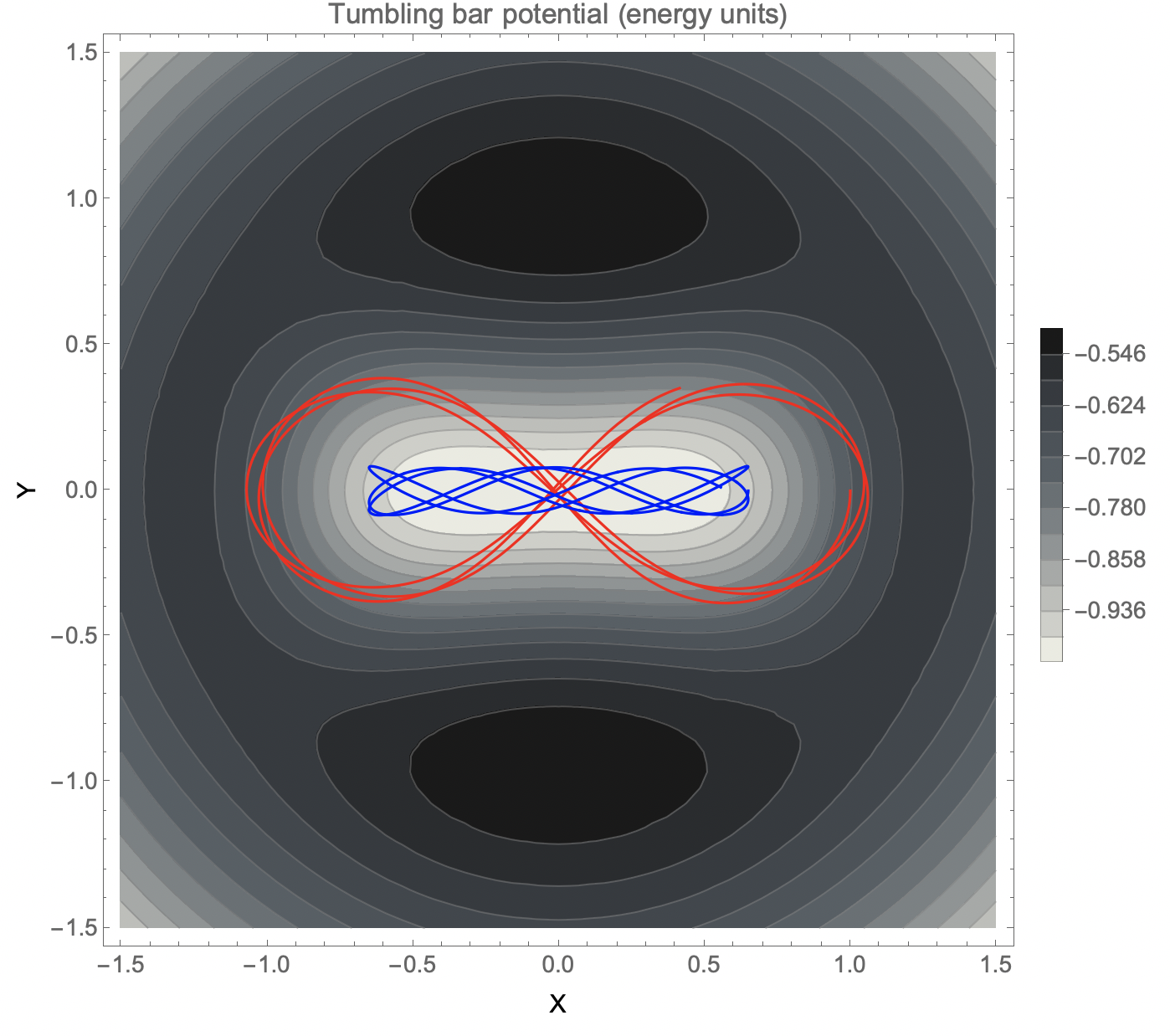}
    \caption{A toy model of a bar potential $\Phi_{\rm eff}=\Phi_{\rm o}-\Omega_b.L$ that is partly supported by gas, and partly by stars born in the gas \citep{barnes01}. The dumbbell shape of the potential aligned with the bar is typical of bars of this kind, and seen in our simulations as the bar evolves. The low energy orbit (blue) crosses the axis several times in each radial oscillation; most orbits have this form. Higher energy orbits (red) can take a similar form, or exhibit the ``bow tie'' morphology, as shown.
    }
   \label{f:cazes_bar}
\end{figure}
\subsubsection{What causes the radial shear flow?}
\label{s:shearflow}

In all simulations, the character of the stellar/gas bar is complicated, exhibiting a variety of non-axisymmetric morphologies at different evolutionary stages. Interestingly, at times the bar potential is distinctly dumbbell-shaped, as predicted by \citet{barnes01}. This may be a feature of early bars where a bulge has yet to form. In later work, we examine the stellar orbit families in more detail, but generally we observe a behaviour that is complex. In classical references \citep[e.g.][]{bin08}, we learn that many of the inner orbits are boxy, defined by three different frequencies in $R$, $\phi$ and $z$ and some orbits can move arbitrarily close to the galactic centre. Other orbits are more chaotic and can cross the bar axis several times in each orbit. The tube orbits typically lie beyond the bar region.

Fig.~\ref{f:cazes_bar} presents a toy model to illustrate this point. We examine a range of bar potentials suggested in the literature for a Hamiltonian with
the form ${\cal H}_J({\mathbf q},{\mathbf p}) = \frac{1}{2}p^2+ \Phi_{\rm eff}({\mathbf q})$ where ${\mathbf q}$ is the position vector and ${\mathbf p}$ is the momentum vector. In the absence of a strong bulge, a dumbbell-shaped potential is often seen in our simulations.
We adopt the dumbbell potential of \citet{cazes00} specifically tailored to a gaseous bar, but similar conclusions are drawn from simpler bar models. 

If ${\cal H}$ is the Hamiltonian of the system in the inertial frame, and ${\mathbf L_z}={\mathbf x}\times {\mathbf p}$ is the angular momentum about the disk's spin axis, neither of which are conserved in the tumbling potential, then we can define a conserved quantity, i.e. the Jacobi Integral ${\cal H}_J = {\cal H}-{\mathbf \Omega}_b.{\mathbf L_z}$ where  ${\mathbf \Omega}_b$ is the bar pattern speed. ${\cal H}_J$ defines the energy of the orbit within the tumbling frame for which ${\cal H}_J=\Phi_{\rm eff}$ establishes the ``zero velocity surface'' where all orbits are at their turnaround point.
In order to examine the orbits, we solve Hamilton's equations for the chosen effective potential $\Phi_{\rm eff}$ describing a Cazes bar, where
\begin{equation}
    \dot{\mathbf q} = {{\partial{\cal H}_J}\over{\partial{\mathbf p}}}, \;\;\;
    \dot{\mathbf p} = -{{\partial{\cal H}_J}\over{\partial{\mathbf q}}}
    \label{e:ham}
\end{equation}
At a fixed energy ${\cal H}_J$, we examine the main orbit families with real solutions, e.g. $(\dot{x}=0,y=0)$ at turnaround with $(x,\dot{y})$ as solutions to Eq.~\ref{e:ham}. With these initial conditions, we carry out the symplectic orbit integration as described in \citet[][Sec. 3.4]{bin08}.

This brings us to the origin of the radial shear flow. Gas wants to follow the stars which it tends to do along circularized tube orbits in the outer disk. But over the inner disk, most stellar orbits follow the bar with frequent crossings of the bar axis, as shown in Fig.~\ref{f:cazes_bar}. The turbulent, viscous gas is not able to follow the bar crossing and is deflected into the forward direction before turning back sharply, at the extremities of the bar where the bar is weaker, reversing its direction. As first noted by \citet{barnes01}, this specific potential includes the relatively unusual ``bow tie'' orbit that crosses at the galactic centre. Once again, gas attempting to cross at the centre is either accreted there or deflected forward to conserve momentum along the bar axis. Momentum is not conserved perpendicular to the bar axis.

\begin{figure}[!htb]
    \centering
    \includegraphics[width=0.99\textwidth]{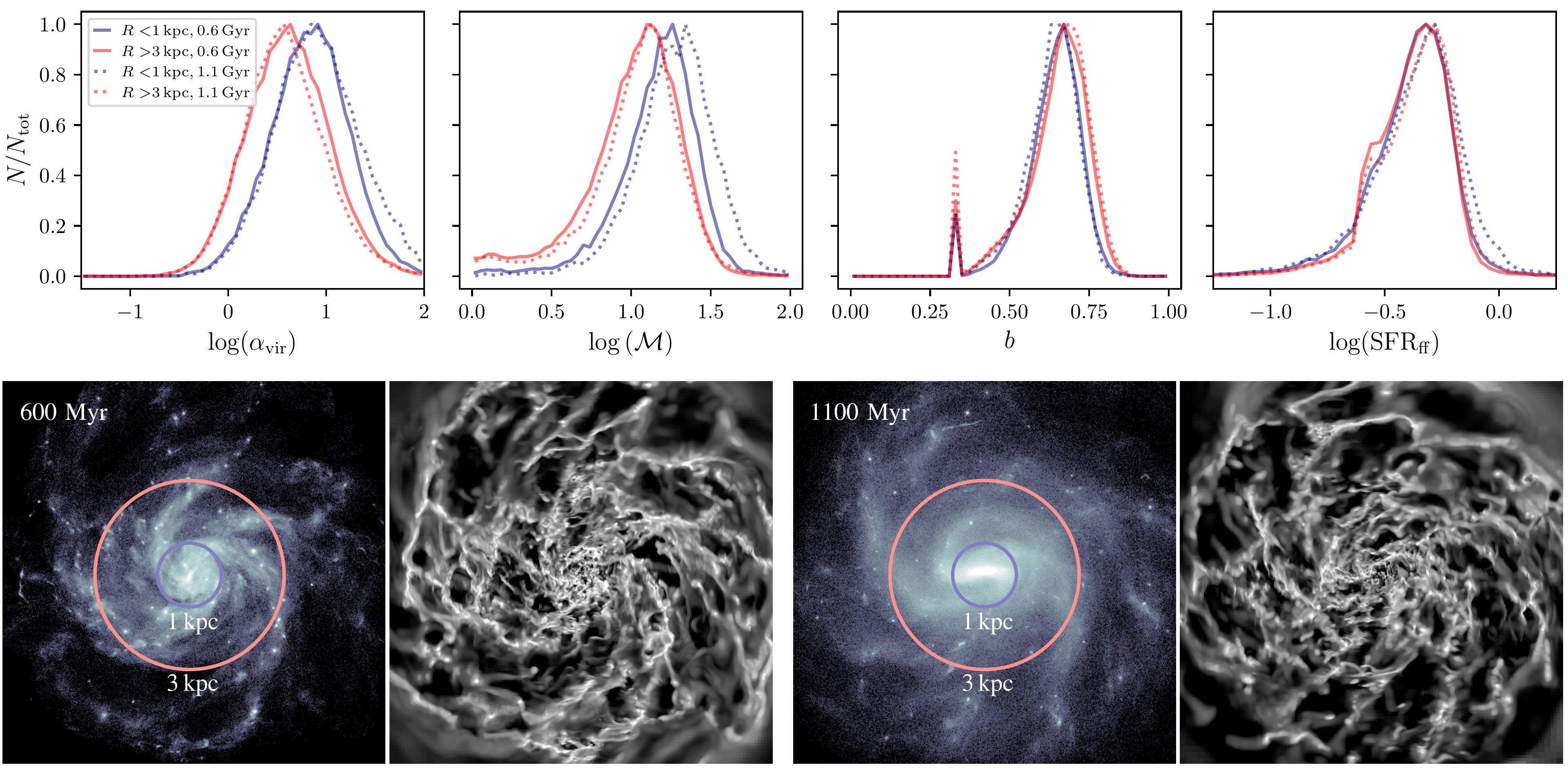}
    \caption{(Top) Key dimensionless parameters characteristic of resolved star forming regions extracted from the {\tt fd50\_fg40\_nac} simulation re-run at high resolution. From left to right, these are: $\log(\alpha_{\rm vir})$, $\log{(\cal M)}$, $b$, $\log({\rm SFR}_{\rm ff}$) $-$ see main text for discussion.
    We compare the measured parameters in two radial zones: inner ($R<1$ kpc) and outer ($1<R<3$ kpc). The solid lines show the results at 600 Myr, the dotted lines at a later time (1.1 Gyr) when a bar has formed. (Bottom) The radial zones are overlaid on the relevant time steps for young stars, next to the gas surface density maps. We find no significant change to the mode of star formation after bar formation; see also Fig.~\ref{f:OAparams2}.
    }
   \label{f:OAparams}
\end{figure}

\newpage
\subsubsection{Do bar-like gas flows help or hinder star formation?}
\label{s:hinder}

The science of how stars form is a central pillar of modern astrophysics \citep{maclow04}. Some of the most important insights have come from studies of resolved galaxies \citep{ken98}, in particular, the explicit dependence of the star formation rate (SFR) on the cold gas surface density, $\mu_{\rm gas}$ \citep{bigiel08}. Many studies invoke Toomre's famous $Q$ parameter 
\begin{equation}
Q = \frac{\kappa\sigma_{\rm gas}}{\pi G \mu_{\rm tot}}
\label{e:Q}
\end{equation}
where values larger than about unity are considered to be locally stable to gravitational collapse (self gravity). Here, once again, $\kappa$ is the local epicyclic frequency and $\sigma_{\rm gas}$ is the internal (cold) gas velocity dispersion. Note that $\mu_{\rm tot}$ integrates over the contribution from stars ($\mu_\star$), gas ($\mu_{\rm gas}$) and the underlying dark matter over the vertical scale of the gas.
But when we consider two-fluid stability (e.g. Sec.~\ref{s:instab}), a different criterion is needed \citep{jog84, romeo92}, particularly in the presence of turbulent media \citep[][]{romeo10,agertz15}. \citet{martin01} found that $Q$ {\it averaged over azimuth}, when evaluated as a function of galactic radius, roughly predicts where the star formation threshold occurs. The dependence on $Q$ likely arises from the parameter's inverse dependence on the gas surface density more than anything else.


While gas and star formation are occasionally seen along the length of stellar bars in resolved galaxies, the association is relatively rare \citep[e.g.][]{reg95,ver07}. This has led to the idea that bars effectively `quench' star formation through strong shearing motions along the bar \citep[e.g.][]{george19}. A problem with this interpretation is that the SFR surface density $\Sigma_\star$ shows a strong positive correlation with the epicyclic frequency $\kappa$ (and the Oort parameters) in local disk galaxies \citep{aouad20}. It is more likely that the gas supply has been exhausted within present-day bars through the streaming process. Today, in large relatively rare, gas-rich galaxies, streaming gas is able to form stars \citep[e.g.][]{hutte99}.

For a well-resolved sample of disk galaxies, \citet[][and references therein]{aouad20} determine the Oort parameters $A$ and $B$ (equivalent to shear and vorticity, respectively) as a function of galactic radius, such that
\begin{equation}
    A(R) = -\frac{1}{2} R \frac{d\Omega}{dR}, \;\;\;\;\;
    B(R) = -(\Omega + \frac{1}{2} R \frac{d\Omega}{dR} )
\end{equation}
where $\Omega=V_{\rm circ}/R$ is the angular frequency and $V_{\rm circ}$ is the circular velocity at a radius $R$. These are easily related to other important parameters: the epicyclic frequency is $\kappa=\sqrt{-4B\Omega}$, the magnitude of vorticity is $\omega = \vert 2B \vert = \kappa^2/2\Omega$,  and the local shear velocity is $\dot{x} = 2\delta A$ where $\delta$ is the distance between two points at different radii in the shear flow\footnote{$x$ defines an axis in the reference frame rotating at the same angular frequency $\Omega$ of the system under consideration, e.g. galaxy disk, bar or spiral arm.}. Importantly, {\it there is no compelling observational evidence for star formation suppressed by either gas shear or gas vorticity}. In fact, the opposite may be true: $\mu_\star$ has a clear positive dependence on both $B$ and $\kappa$, and only a weak dependence on $A$.


There is a substantial literature on how star formation can be enhanced through cloud collisions in regions of strong, turbulent shear and even reduced in regions of low shear \citep[e.g.][]{tan00,anath10,tasker09,dobbs15,renaud15,taka18}.
\citet{FederrathKlessen2012} reason that high Mach numbers ${\cal M} = \sigma_{\rm gas}/c_s$ in turbulent gas with sound speed $c_s$ generally {\it increase} star formation rates compared to transonic or subsonic media. (Since interstellar gas is a magnetized medium, a more general definition for ${\cal M}$ includes the magnetic pressure.) The virial parameter for a uniform spherical cloud $\alpha_{\rm vir} = 2 E_{\rm kin}/\vert E_{\rm grav}\vert$ must be kept small \citep{krum05}. High Mach numbers generally increase the SFR since they lead to smaller clumps and high densities (cf. Sec.~\ref{sec:sfrdefs}).

Thus, to offset the increased gas dispersion at high ${\cal M}$, the gravitational binding energy must be enhanced in high mass-density pockets. These are the conditions set up by shell-crossing due to eddies in a supersonically turbulent, magnetized medium. To quote \citet{elm93}, ``supersonic turbulence compresses gas at the interfaces between converging flows, and this compression lasts for a relatively long time equal to the crossing time {\it between} clumps.'' Within these eddies, turbulently-compressed clumps form and disperse continually and continuously, but the densest clumps become gravitationally unstable and collapse. The mass spectrum of post-shock clumps can be calculated \citep{elm90,elm93}; higher turbulent Mach numbers produce stronger compression and higher star-formation rates \citep{FederrathKlessen2012}.

In Fig.~\ref{f:OAparams}, we present dimensionless properties $-$ ${\cal M}$, $\alpha_{\rm vir}$, $b$, ${\rm SFR}_{\rm ff}$ $-$ measured from the {\tt fd50\_fg40\_nacc} high-resolution simulation. This is done by running the simulation with a cell based efficiency per free-fall time derived from turbulence theory (see Eq.~\ref{eq:fk12sfrff}), rather than the fixed $\epsilon_{\rm ff}$ introduced in Eq.~\ref{eq:schmidtH2}. In future work we will explore the impact of such models further. All relevant turbulence parameters presented in Fig.~\ref{f:OAparams} are  measured locally on the mesh at the time of star formation. 

These measures, which are crucial to resolve in any simulation of processes within star-forming regions, are compared in two radial zones $-$ inside and outside the bar region $-$ at two different times, as indicated. We resolve the forcing parameter $b$ for each clump describing the nature of the compression\footnote{We approximate $b$ via the the relative contribution of compressive and rotational motions to the total (local) gas flow using equation 5 in \citep{FederrathKlessenSchmidt2008}.}, typically in the range $b=1/3$ (divergence free) to $b=1$ (curl free). This parameter appears not to change either by location or in time, but more work is needed. Moreover, there is only a weak time dependence in any of the parameters. Interestingly, both $\alpha_{\rm vir}$ and ${\cal M}$ decline with increasing galactic radius, but in such a way that the SFR per freefall time, SFR$_{\rm ff}$, is essentially invariant. The normalisation of the SFR with the local freefall time is appropriate to accommodate the multi-scale nature of a turbulent medium.

\begin{figure}[!htb]
    \centering
    \includegraphics[width=0.99\textwidth]{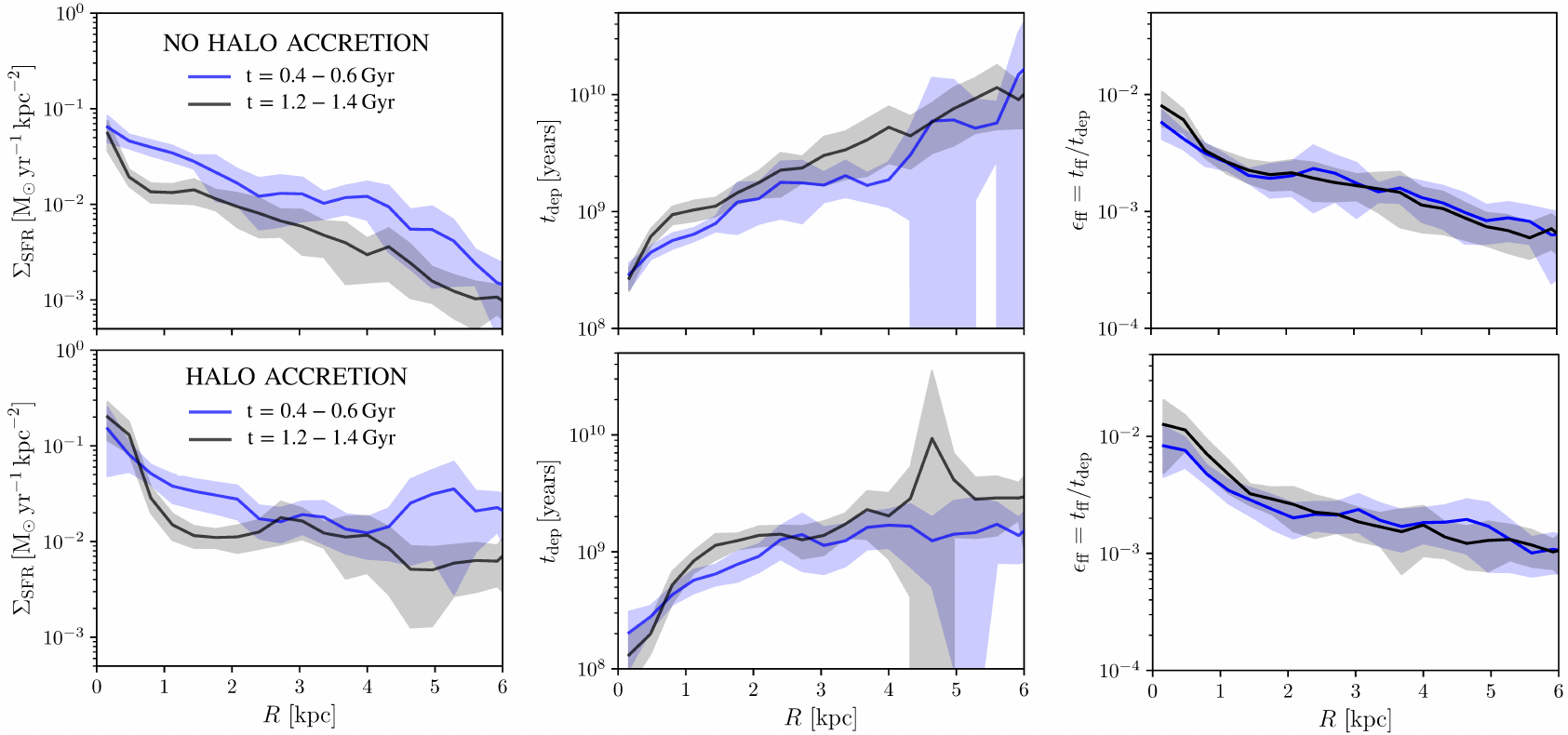}
    \caption{Azimuthally averaged radial profiles of (left) star formation rate surface density, $\Sigma_{\rm SFR}$ (middle) the gas depletion time, $t_{\rm dep}$, and (right) the characteristic star formation efficiency per free fall time, $\epsilon_{\rm ff}$ $-$ see main text for definitions.  These are the results for $f_{\rm gas}=40$\% models without halo accretion (top) and with halo accretion (bottom).  The two lines show results before ($0.4-0.6\,{\rm Gyr}$, blue) and after ($1.2-1.4\,{\rm Gyr}$, black) bar formation, with the shaded regions showing one standard deviation. Gas depletion due to star formation leads to a lowering of $\Sigma_{\rm SFR}$ over time, with a slight increase in $t_{\rm dep}$. We find no significant change to the mode of star formation after bar formation, as indicated by the dimensionless $\epsilon_{\rm ff}$ profiles; see also Fig.~\ref{f:OAparams}.
    }
   \label{f:OAparams2}
\end{figure}

In Fig.~\ref{f:OAparams2}, we show the radial profiles of the star formation rate surface density $\Sigma_{\rm SFR}$, the gas depletion time and the average star formation efficiency per free fall time, $\epsilon_{\rm ff}=t_{\rm ff}/t_{\rm dep}$, where $t_{\rm ff}$ is the azimuthally averaged gas free-fall time. Overall, as seen in the right hand panel, there is little evolution in the star formation efficiency by location or in time. We conclude that the radial shear flow or bar flow neither promotes nor suppresses star formation. We can overload a bar-like potential with gas and still form stars relatively efficiently. The streaming motions simply concentrate the star forming regions. This is broadly consistent with the findings of a recent survey of resolved molecular gas in 12 barred galaxies
\citep{diaz21}, although a related study claims evidence for a modest decline in the star forming efficiency \citep{mae23}.

Thus, it is plausible that the shear flows arising in turbulent gas disks at early times provided the necessary conditions for forming stars in a central bar-like configuration. This bar-like structure is predicted to be associated with dense gas and young, luminous stars, or a post-starburst population if enough time has elapsed, i.e. of order a few disk rotations. This is what is seen in all of our Milky Way progenitor simulations.

\begin{figure}[!htb]
    \centering
\includegraphics[width=0.9\textwidth]{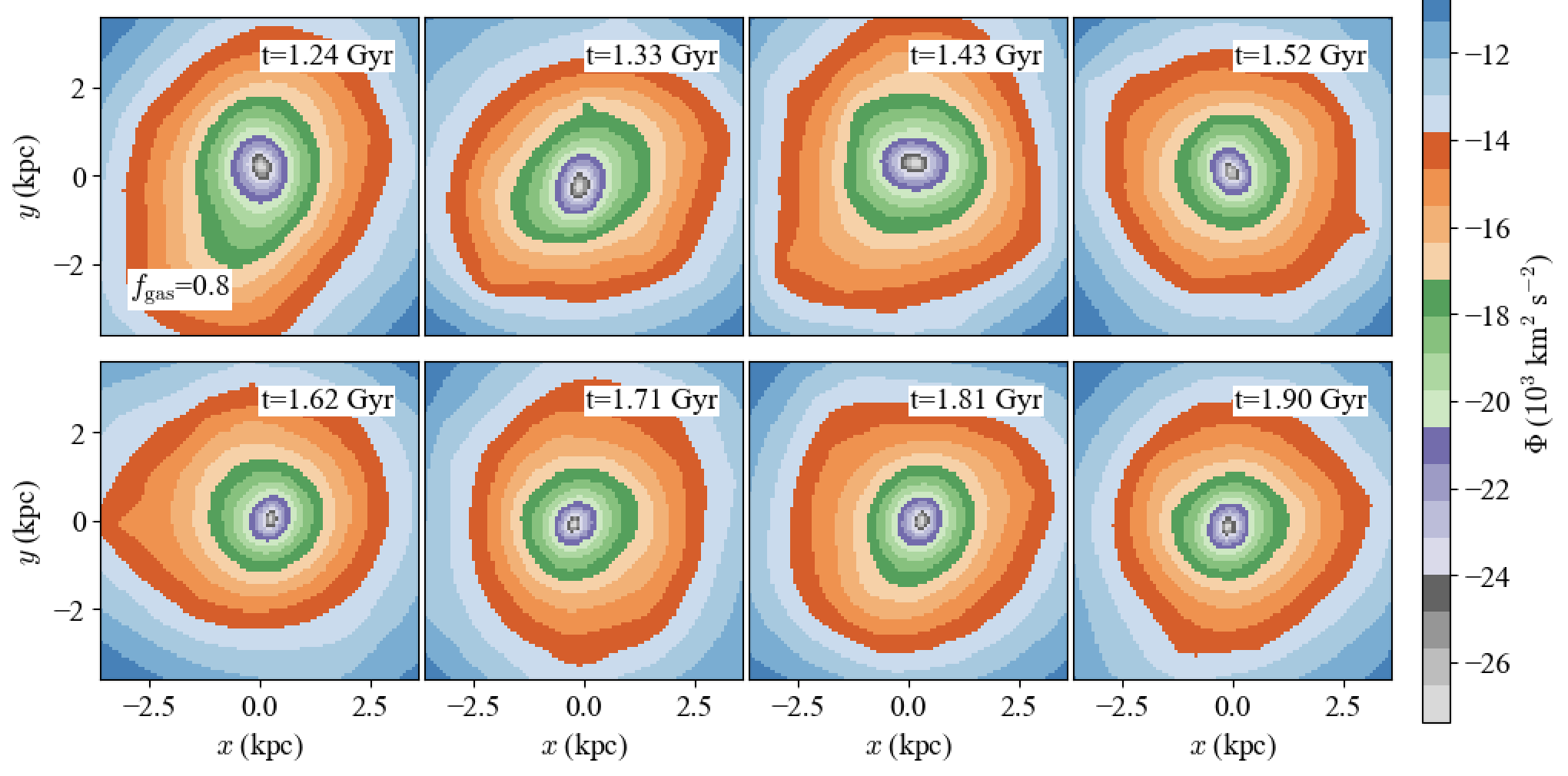}
    \caption{The evolution of the disk gravitational potential $\Phi$ from the $f_{\rm gas}=80\%$ model (with halo accretion), {\tt fd50\_fg80\_acc}. Here, we show its later stages (top left to bottom right) after the central bar has begun to dissolve and on its way to forming a compact central bulge with a half mass radius of $r_b \approx 0.5$ ckpc.
    }
   \label{f:bar_bulge}
\end{figure}

\subsubsection{Gaseous bars and their evolution}

While we have a basic understanding of how stellar bars form, it is much less obvious why gaseous bars emerge at high gas fractions? Cold, rotationally-supported gas disks with self-gravity are low-entropy systems that are highly susceptible to radial instabilities, much as with cold stellar disks.
To quote from \cite{barnes01}:
``Just as cold, axisymmetric stellar dynamical configurations are known to be dynamically unstable toward a bisymmetric instability if they are sufficiently self-gravitating, the same is true for fluid configurations.''
Part of the process appears to be young stars being launched into the disk\footnote{It is unknown whether a turbulent gas-rich disk would form a bar without star formation, although this outcome seems feasible \citep{cazes00}. But, in our view, this is a philosophical question given that the high surface density of gas demands star formation to occur, and so we do not consider this case.}. Indeed, this is what is actually being observed in the simulations, e.g. Figs.~\ref{f:lastbar}, \ref{f:lastbarhalo} and \ref{f:lastbar2}. The build-up of the young stellar bar is well defined, whereas the turbulence disperses the build-up of gas along the same axis. But the shear flow is the region where most stars are born. In Figs.~\ref{f:lastbar} and \ref{f:lastbarhalo}, the disk's central gravitational potential is bar-like even when the circumnuclear regions appear noisy.

What we see is that the turbulent disk forms discrete clumps and these launch dissolving star clusters into the disk. The young stars return energy to the cluster gas, which becomes overpressured and is forced out by a combination of radiation and wind action. Few of our clusters survive for more than 100~Myr or so, in line with \citet{bla10}. For example, in the simulation {\tt fd50\_fg100\_nac}, star clusters emerge and disappear once the disk has settled down from the initial burst phase. These stars diffuse into the general disk for which the total baryon content dominates the local galactic potential. While this remains true, the stellar disk is susceptible to bar instabilities. Thus, it appears to be the young stars that lead the way to bar formation.

Gas-rich bars are occasionally observed in nearby galaxies, as discussed in Sec.~\ref{s:hinder}, but always in association with dominant stellar bars \citep{aalto99,hutte99,kohno08}. Nuclear gas bars are relatively common but appear to be associated with $x_2$ orbits embedded within the bar-supporting $x_1$ orbit families \citep{ath92}. 
To date, we are unaware of any claims of a gas-dominated bar residing in a disk potential in the local Universe. The situation may be very different at high redshift, as discussed in the Introduction, with bar-like structures claimed in gas-dominated disks \citep[e.g.][]{tsukui24,huang23}. In the early universe, gas disks are likely to have formed before star formation commenced in the gas, as judged from the relative thinness of stellar disks at all epochs. 

Are there fossil signatures in old stellar bars today that can separate the two different evolutionary paths? We have seen that gaseous bars emerge in a few rotation periods in turbulent gas disks, particularly at high fractions. It is interesting to contemplate what these become and how these differ from stellar bars that emerge from stellar disks. At the highest gas fractions, the gas-dominated bars collapse to form compact central bulges after 1.2$-$1.5 Gyr. This is a novel mechanism for bulge formation and deserves further study. Interestingly, bar formation dominated by radial shear flows do not appear to form box/peanut bulges, as can be seen from any of the $x-z$ or $y-z$ side elevations at our movie website. 

In the prescient \citet{barnes01} study, they find that a high fraction of stars born in a gas bar are injected into `bowtie orbits' and these build up a highly flattened dumbbell-shaped gravitational potential. \citet{barnes01} refer to this as the ``Cazes bar'' \citep{cazes00} that was the focus of our study in Sec.~\ref{s:shearflow}. We provide an example of a bowtie (centre crossing) orbit in Fig.~\ref{f:cazes_bar}. The dumbbell shape is seen along the full length of the bar, unlike what is seen in the Milky Way for example where a {\it vertical} central box/peanut bulge has taken hold. The latter form through vertical instabilities and have been widely discussed since their discovery \citep{comb90}. In the Barnes study, there are surprisingly few prograde $x_1$ orbits in the gas bar compared to those that normally dominate N-body bars. Whether this is an artefact of their analytic model or symptomatic of gaseous bars remains to be seen.

\bigskip
\begin{figure}[!htb]
    \centering
    \includegraphics[width=0.6\textwidth]{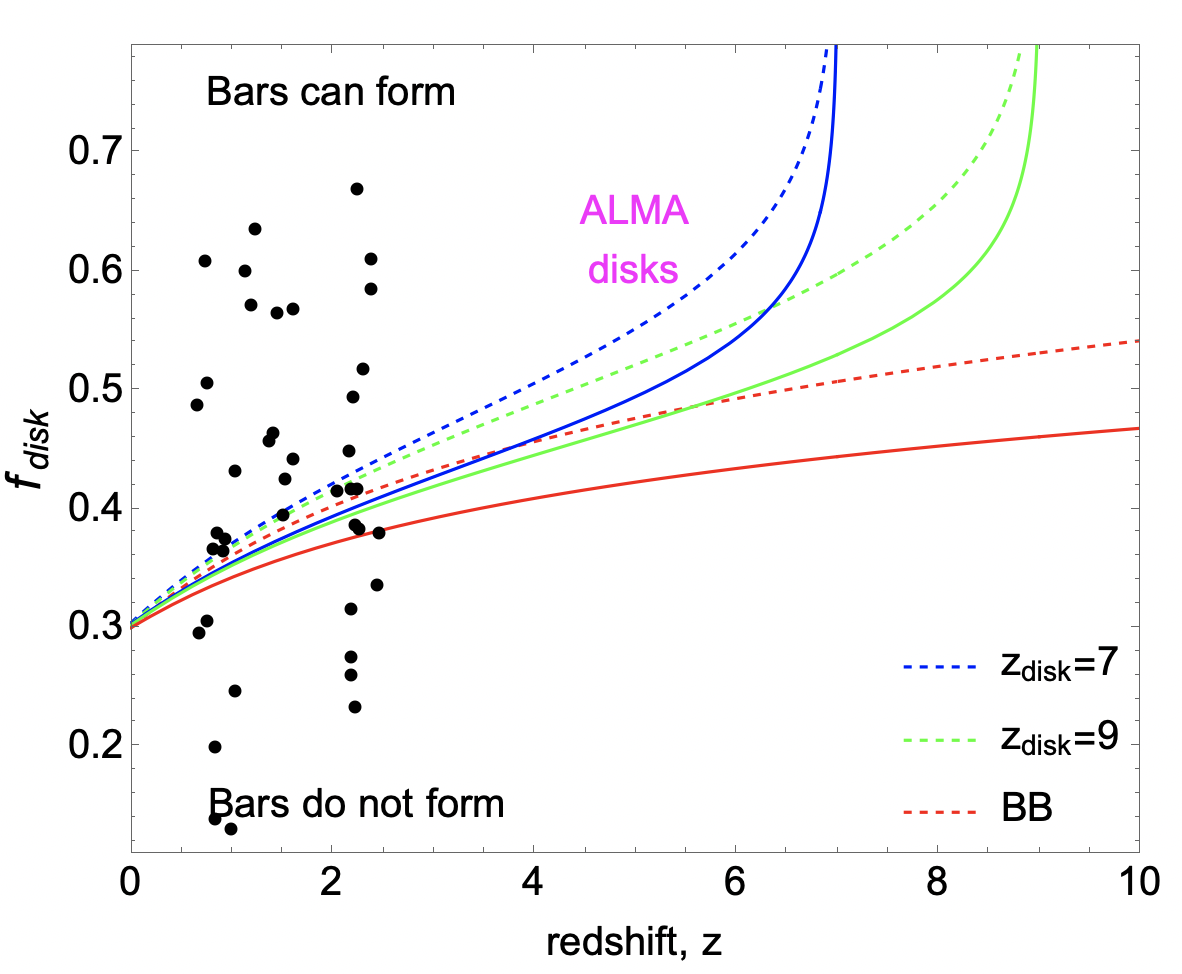}
    \caption{The inferred values of disk mass fraction $f_{\rm disk}$ plotted against the measured redshifts for each galaxy in \citet{Price2021}; this is currently the only catalogue with computed $f_{\rm disk}$ data. The expected location of the highest-redshift ALMA galaxies is also indicated. The six overlaid curves are the dividing lines between bar formation and insufficient time for bars to form since the epoch of disk formation $z_{\rm disk}$ for three different disk onset times: (blue) $z_{\rm disk}=7$, (green) $z_{\rm disk}=9$, (red) Big Bang. Disk galaxies that fall above the curves are able to form a bar in the time span available. The dashed lines are for the gas-free models; the solid lines are for the turbulent gas disk models ($f_{\rm gas} \gtrsim 40\%$) supported by star formation. The formulae for all curves are given in the text.
    }
   \label{f:fdisktime}
\end{figure}

\bigskip
\section{Discussion: broader implications} \label{s:discuss}

\subsection{Implications for high-redshift bars}

In Fig.~\ref{f:fdisktime}, we present an updated version of the diagram first presented in \citet{bla23} in light of the results from Sec.~\ref{s:heavdisks}. Given that cool ALMA disks are now being observed to $z\approx 6.5$ \citep{neel23}, we show only ``bar $-$ no bar'' dividing lines for $z=7$ and higher. The dashed lines refer to our earlier gas-free models; the solid lines refer to our new gas-rich models. Objects that fall above the lines have sufficient time to form a bar-like deformation in stars or in gas; objects that fall below the dividing lines do not.

The data points are extracted from \citet{Price2021}; this is presently the only published list of disk mass fractions beyond $z=1$ because it requires high-quality, spatially-resolved (sub-kpc) kinematic data. Such information is now becoming possible with ALMA data and so we anticipate many more $f_{\rm disk}$ data points in the near future, with estimates extending to higher redshifts. The ``ALMA disks'' label refers to the region populated in the new unpublished study by F. Roman de Oliveira (2023, personal communication). In the upper regions, we anticipate that stellar bars will be more common at lower redshift, and gaseous bars will dominate at higher redshift given their accelerated formation times.

In Fig.~\ref{f:fdisktime}, the three dashed curves for the gas-free model are as follows:
\begin{eqnarray}
f_{\rm disk}(z) &=& 0.30\, -0.0719424 \ln  (-0.0559584 + g(z))\;\;\;\;\;\;\;\;\;\;\;\; z_{\rm disk} = 7 \\
f_{\rm disk}(z) &=& 0.30\, -0.0719424 \ln (-0.0400557 + g(z))\;\;\;\;\;\;\;\;\;\;\;\; z_{\rm disk} = 9 \\
f_{\rm disk}(z) &=& 0.311\, -0.0719424 \ln (1.220\; g(z))\;\;\;\;\;\;\;\;\;\;\;\;\;\;\;\;\;\;\;\;\;\;\;\; z_{\rm disk} = \infty
\end{eqnarray}
where
\begin{equation}
g(z) = 0.819698 \sinh^{-1}\frac{1.54591} {(z+1)^{3/2}}.
\end{equation}
The small adjustment at $z=0$ reflects the slightly higher halo mass compared to the low halo mass model in \citet{bla23}. The three solid curves for the gas-rich models ($f_{\rm gas} \gtrsim 40$\%) are as follows:
\begin{eqnarray}
f_{\rm disk}(z) &=& 0.30\, -0.0\dot{5} \ln  (-0.0559584 + g(z))\;\;\;\;\;\;\;\;\;\;\;\;\;\;\;\;\;\;\;\;\; z_{\rm disk} = 7 \\
f_{\rm disk}(z) &=& 0.30\, -0.0\dot{5} \ln (-0.0400557 + g(z))\;\;\;\;\;\;\;\;\;\;\;\;\;\;\;\;\;\;\;\;\; z_{\rm disk} = 9 \\
f_{\rm disk}(z) &=& 0.314\, -0.0\dot{5} \ln (1.220\; g(z))\;\;\;\;\;\;\;\;\;\;\;\;\;\;\;\;\;\;\;\;\;\;\;\;\;\;\;\;\;\;\;\;\; z_{\rm disk} = \infty .
\end{eqnarray}
Here we have taken an average across all gas-rich models. The updated curves from \citet{bla23} are constrained by the new $f_{\rm disk}=0.3,0.5$ and 0.7 models (Table~\ref{t:mod}) and so are approximately correct. In future, more sampling in $f_{\rm disk}$ would be useful to refine the curves further.

\begin{figure}[!htb]
    \centering
    \includegraphics[width=0.55\textwidth]{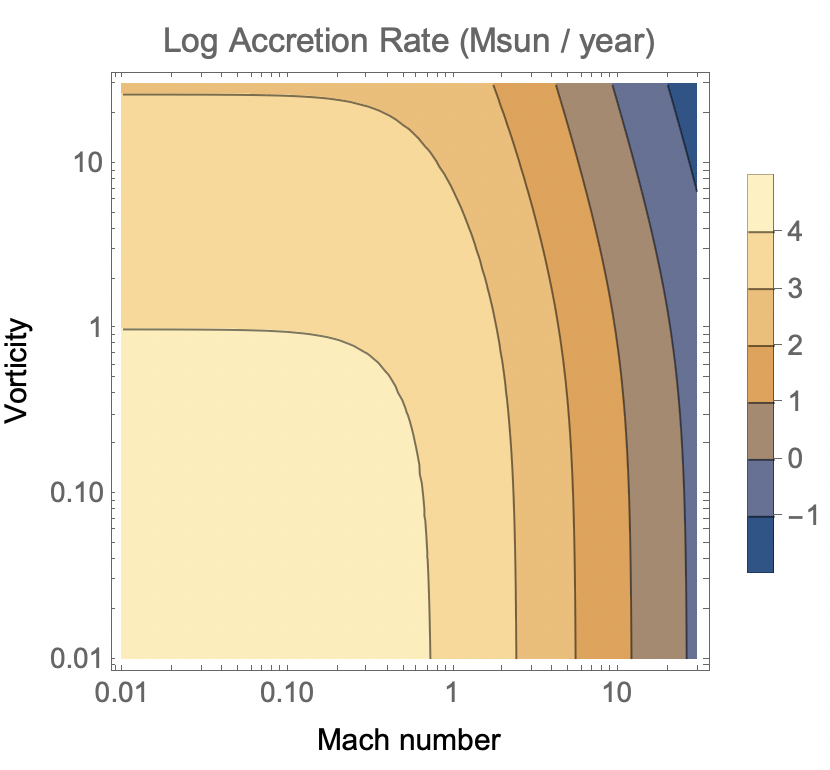}
    \caption{Bondi-Hoyle accretion rate for a supermassive black hole ($M_\bullet = 10^9$ M$_\odot$) as a function of the Mach number ${\cal M}$ and normalised vorticity $\omega_\star$ of the infalling gas (see text for formulae). The contours show the accretion rate in log units of M$_\odot$ yr$^{-1}$. The region within the lowest contour (bottom left) is essentially the Bondi maximum accretion rate (${\cal M}=0$).
    }
   \label{f:fmachvort}
\end{figure}

\bigskip
\subsection{Implications for feeding nuclear activity}

A remarkable aspect of high-redshift, turbulent ALMA disks is that some of them host powerful quasars \citep{farina22,walter22,tsukui21,tsukui23}. Our work has identified interesting themes that provide new avenues on the broader question of how supermassive black holes interact with their environment. 
The radial shear flow and its subsequent collapse to a central bulge (high $f_{\rm gas}$) is a new phenomenon that should be investigated at higher spatial resolution using magnetodydrodynamics \citep[e.g.][]{abram13}. Stellar and associated gas bars have long been associated with nuclear activity \citep[q.v.][]{shlos89,shlos90}. Indeed, these authors even suggest that, when $f_{\rm gas} \gtrsim 20$\% over the inner disk, a gas bar can form at a different orientation to the main bar, and this can increase the flow of gas onto the nucleus.

Classically, quasar power is considered in the context of the Bondi-Hoyle accretion rate \citep{edgar04} given by
\begin{eqnarray}
    \dot{M}_\bullet &=& {{4\pi\rho_{\infty}(GM_\bullet)^2}\over{(c^2_\infty+v^2_\infty)^{3/2}}}\\
    &=& {{4\pi\rho_{\infty}(GM_\bullet)^2}\over{{c_\infty^3(1+{\cal M}^2)^{3/2}}}}
    \label{e:bh}
\end{eqnarray}
where $(\rho_\infty, c_\infty, v_\infty)$ are the density, sound and wind speed of the upwind flow far from the source, $M_\bullet$ is the accretor mass, and ${\cal M}$ is the sonic Mach number as before. (We have ignored additional factors, e.g. ratio of specific heats, that are of order unity - see \citet{ruff94}.) Numerous studies find that the Bondi-Hoyle formula substantially overestimates the accretion rate. But these estimates drop rapidly when one considers gas angular momentum and vorticity, instabilities and magnetic fields, and more realistic environments like turbulent media. Such considerations reveal that there are different regimes, e.g. a circumnuclear disk or torus that forms beyond the Schwarzschild radius, which modifies the ongoing accretion both through its geometry and subsequent feedback \citep[e.g.][]{abram81}.

There is already substantial work on how turbulent media supply gas to a supermassive black hole \citep[q.v.][]{krum05a}. A useful approach is to examine the Bondi-Hoyle accretion rate in the presence of vorticity ($\nu = \vert \nabla \times v \vert$). This simple framework overlooks the fact that both momentum and mass are transferred to the accretor with the passage of time. The flow is also unstable, particularly in the region of the converged flow behind the accretor. Moreover, the accretion rate cannot be increased indefinitely because, at some point, the radiation pressure supplied by the accretion disk will cut off the gas supply (Eddington limit). 

A helpful analysis is supplied by \citet{krum05a} who consider the dependence of the accretion rate on the nature of the approaching orbit. Low vorticity gas falls on essentially a radial orbit, gas accretes at the Bondi rate ($\approx r_B/c_s$) and the Bondi-Hoyle formula is unchanged. Gas approaching the accretor with high vorticity (equivalently, specific angular momentum $\ell$) accretes at a much slower rate. In brief, the dimensionless vorticity $\nu_\star \approx 1$ describes the transition between efficient accretion and suppressed accretion. The transition vorticity $\nu_\star = 1$ corresponds to gas arriving at an impact parameter equal to the Bondi radius ($r_B=GM_\bullet/c_\infty^2$) travelling at the Keplerian velocity ($=\sqrt{GM_\bullet/r_B}$), with specific angular momentum $\ell_\infty \approx \nu_\star c_s r_B$ \citep[see also][]{abram81}. Here $c_s=c_\infty$ for an isothermal gas; \citet{krum05} argues that the choice of the equation of state does not change the flow pattern or the accretion rate very much. The simplified approximation reduces to 
\begin{equation}
    \dot{M}_\bullet = 4\pi r_B^2 \rho_\infty c_s {\cal F}(\nu_\star)
    \label{e:mrk}
\end{equation}
where the piecewise function ${\cal F}$ is supplied elsewhere \citep{krum05a}. 

Thus, for a supermassive black hole with mass $M_\bullet = 10^9$ M$_\odot$, typical of high-redshift quasars, the accretion depends on both ${\cal M}$ (see Eq.~\ref{e:bh}) and the gas vorticity  (Fig.~\ref{f:fmachvort}). In producing Fig.~\ref{f:fmachvort}, we divide Eq.~\ref{e:mrk} by the Bondi accretion rate (${\cal M}=0$) \citep[see][]{krum06}. From our initial analysis, we find that the inner regions have substantial levels of turbulence, with $\nu_\star \gg 1$ and Mach numbers ${\cal M}=5-30$ (Fig.~\ref{f:OAparams}). The expected accretion rates are only $\dot{M}_\bullet\sim 0.01-1$ M$_\odot$ yr$^{-1}$, orders of magnitude below what is needed to explain the observations ($\dot{M}_\bullet \sim 10-100$ M$_\odot$ yr$^{-1}$) \citep{farina22}. At present, our simulations only reach to parsec scales and a different approach, in addition to MHD, is needed to go further \citep[cf.][]{fed21,lee14,cun12}.  What appears to be a radial shear flow was identified by \citet{Li17} in their simulations, but they found an inner gas ring developed in the presence of a massive central object.

\begin{figure}[!htb]
    \centering
\includegraphics[width=0.8\textwidth]{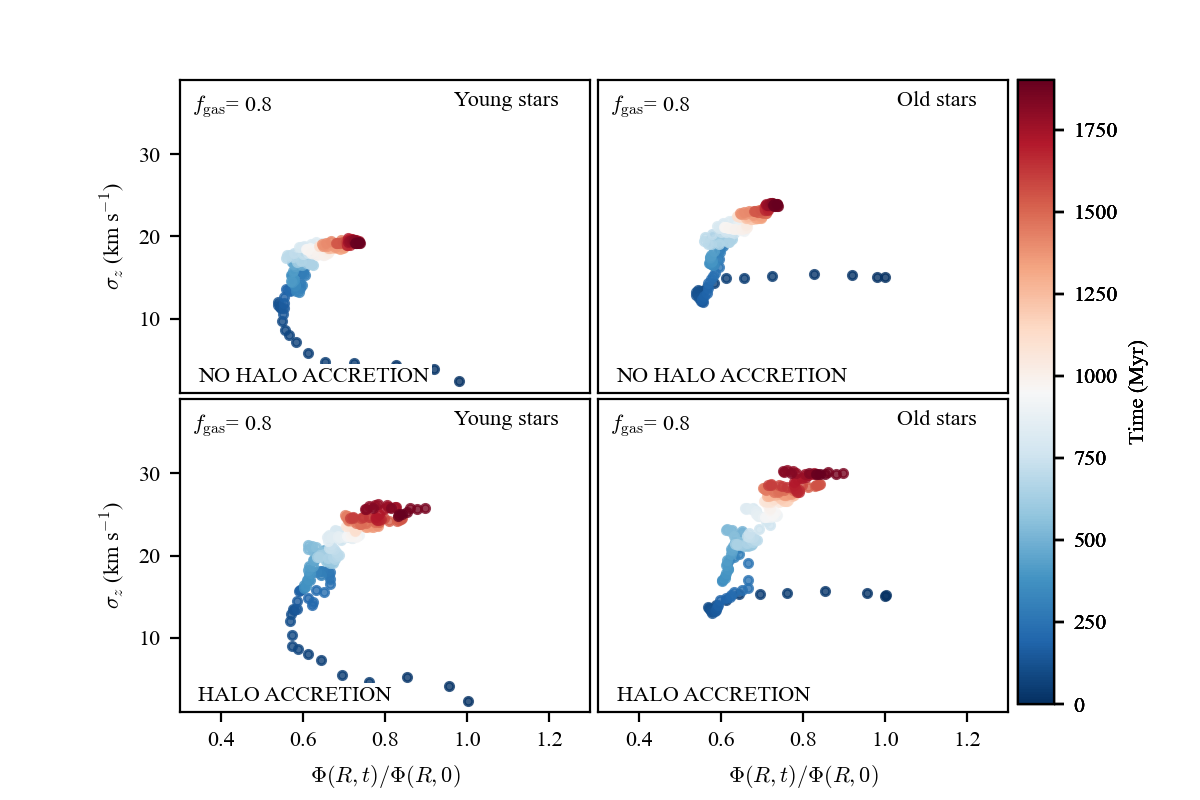}
    \caption{The evolution of the disk in the high gas fraction limit ($f_{\rm gas}=80\%$) for the (top) ``no accretion'' and (bottom) ``halo accretion'' simulations. (These are four magnified plots taken from Fig.~\ref{f:disp_evol}.) The evolving disk potential is shown against the vertical dispersion for (left) young stars, (right) pre-existing stars. The colour coding indicates time passing, with blue to red indicating early to late times. As shown in Fig.~\ref{f:pots0}, the disk gravitational potential loses a lot of baryon mass due to circulating winds, and then recovers this gas at later times. In the main text, we show how the evolving kinematic dispersion can be understood dynamically.
    }
   \label{f:disk_evol}
\end{figure}

\medskip
\subsection{Implications for disk evolution imposed by baryon mass loss}
\label{s:diskmassloss}
In Fig.~\ref{f:pots0}, we presented the evolution of the total
disk potential $\Phi(R,t)$ normalised to the starting potential $\Phi(R,0)$.
For $f_{\rm gas} < 0.5$, the disk potential is fairly constant, but above this limit, the loss of gas mass in circulating winds leads to a substantial weakening of the disk potential by up to 50\%. This is an unsettling aspect of trying to match the properties of high-redshift, gas-rich disks. The measured levels of SFR surface density are sufficient to disrupt the disk substantially or even entirely, assuming the strong mechanical coupling is correct. We can reduce the rate of mass loss from the disk by lowering the efficiency of feedback coupling. But the coupling efficiency used ($\sim 10\%$) is consistent with most contemporary cosmological simulations \citep[e.g.][]{Hopkins2018,agertz2021}.

How does the weakening of the disk potential influence the disk's dynamical evolution, in particular, the 3D stellar kinematics? Interestingly, this issue does not appear to have been addressed to date. In Fig.~\ref{f:disk_evol}, we show how the vertical dispersion evolves with time and with the change in disk potential $\Phi$. To do this, a radial profile is determined for $\sigma_z$ and $\Phi$ in the range $0 < R < 5 R_{\rm disk}$ for each time step, and a median value is computed for that profile.
Our analysis shows that, for the old stars, the kinematic dispersions $\sigma_R$ and $\sigma_z$ both decline with time during the most active mass loss phase but then regather when the disk mass increases subsequently. The young stellar dispersions remain relatively constant as the disk weakens, before increasing as the disk potential grows again through reaccretion. All the while, the ratio $\sigma_z/\sigma_R$ remains roughly constant. How are we to understand this?

To understand what happens to the {\it stellar} disk when a major part of the baryons are lost, we consider the galactic potential of an axisymmetric exponential disk. 
The mass loss is sufficiently slow (i.e. a few percent per rotation period) that we can treat the vertical action of an ensemble of stars as a conserved quantity, i.e. an adiabatic invariant. The vertical action $J_z$ can be related to a disk's vertical kinematic dispersion $\sigma_z$ through the vertical disk frequency, $\Omega_z$.
For all actions ($i=R,\phi,z$), we may write
\begin{equation}
    2\pi J_i=\oint \dot x_i\,{\rm d} x_i={1\over\Omega_i}\int_0^{2\pi}(\dot x)^2\,{\rm d}\theta_i={2\pi\over\Omega_i}\ex{v^2_i}.
\end{equation}
Thus the time average of a star's squared velocity component is related to the action $J_i$ through the associated frequency $\Omega_i$ such that $\ex{v_i^2}=\Omega_i J_i$ \citep{bla19}. Passing from this result for time averages for individual stars to population averages over the stars that reach a given
place is non-trivial, but it generally follows that
\begin{equation}
    \sigma_i^2/\sigma_j^2=\ex{\Omega_i J_i}/\ex{\Omega_j J_j}
\end{equation}
where $\ex{.}$ is an appropriate average. More broadly, we may write
\begin{equation}
    \sigma_z/\sigma_R \approx \sqrt{\ex{\Omega_z/\Omega_R}}\sqrt{\ex{J_z/J_R}} .
\end{equation}


First, in order to understand why the dispersions decline during the mass loss phase, we consider a simple {\it axisymmetric} disk model \citep{bin08}
\begin{equation}
\Phi_{\rm o}(R,z) = {{v_c^2}\over{2}} \ln (R^2+\frac{z^2}{q^2})
\end{equation}
for which the circular rotation curve ($z=0$) has a constant circular velocity $v_\phi(R)=v_c$. This is the equation for an isothermal sphere that has been flattened by a factor $q<1$ 
along the $z$-axis. More complex treatments arrive at the same general conclusion \citep{sharma13}. For a rotating system, the effective gravitational potential is given by 
\begin{equation}
\Phi_{\rm eff}(R,0) = \Phi_{\rm o}(R,0) + {{L_z^2}\over{2R^2}}
\end{equation}
for which $L_z$ is the conserved disk angular momentum about the $z$-axis.


The vertical and radial frequencies are defined by
\smallskip
\begin{eqnarray}
\Omega_z^2\; &=&\; \vert{{\partial^2 \Phi_{\rm eff}}\over{\partial z^2}}\vert_{z=0} \;\;\approx \frac{v_c^2}{q^2 R^2}\\
\Omega_R^2\; &=&\; \vert{{\partial^2 \Phi_{\rm eff}}\over{\partial R^2}}\vert_{z=0} \;\;\approx \frac{v_c^2}{R^2} + O(\frac{1}{R^3})
\end{eqnarray}
where both are evaluated in the plane at an arbitrary radius. The explicit appearance of $q$ arises from flattening
an isothermal spheroid, causing $\sigma_z$ to increase due to adiabatic compression; 
there is negligible contribution to $\sigma_R$.
If we assume that the disk keeps its shape as mass is
lost slowly from the system, and the actions are broadly conserved, at a fixed radius, both $\Omega_z$ and $\Omega_R$ must be reduced given the slower rotation speed. Thus, as observed in the simulations, both stellar dispersions decline as disk mass is lost, and the ratio $\sigma_z/\sigma_R$ is approximately conserved. Stars move outwards and upwards, reaching progressively higher disk heights and disk radii. In practice, this ratio can be influenced by other internal processes \citep{ida93}. The dispersions $\sigma_\phi$ and $\sigma_R$ are largely coupled by the epicyclic motion. 

Interestingly, for all gas fractions, about half of the young stars are born in a very cold, thin disk, which is where the bar streaming and radial shear flow operate most effectively. A comparable fraction of stars appear to be born far from the mid-plane in the extended turbulent disk \citep[see also][]{vandonkelaar22}. Much of the disk thickening appears to arise from these two processes, at least at early times although other processes are probably operating in a cosmological setting \citep[e.g.][]{bird21,mcclus24}. We consider these other processes in later papers.

\subsection{Implications for Galactic archaeology}
\label{s:galarch}

The archaeological record contains a great deal of information about how the Milky Way's stellar dispersions evolve as a function of location, stellar abundance and in time \citep{hay15n,aumer16,haywood18}. Since the launch of ESA Gaia, stellar dispersions can be determined for all axes, i.e. $R$, $\phi$ and $z$ \citep{sharma21}. Cosmological simulators have come to appreciate the extraordinary richness contained within multidimensional kinematic data \citep[e.g.][]{mcclus24}.

In Sec.~\ref{s:sigmaz}, we showed how the vertical dispersions $\sigma_z$ in stars and gas evolve within the turbulent disk simulations, but similar data can be extracted for all axes, as we show elsewhere. Our full suite of models have important implications for galactic archaeology \citep[e.g.][]{hay15n}, which we explore in an upcoming paper for the GALAH DR4 data release (Bland-Hawthorn et al 2024, in prep.). Our earlier work demonstrates that there is no discernible contribution from insufficient numerical resolution in our approach \citep{bla21e} and so we believe the changing kinematic dispersions reflect dynamical processes \citep[e.g.][]{ida93,aumer16,sharma21}. 

In the last section, we looked at how the stellar disk is expected to evolve in the presence of mass loss through circulating wind flows, particularly at high gas fraction. Generally, the stellar dispersions are expected to decline at all radii where the disk baryons are reduced. {\it But the converse is also true} $-$ a slow increase in disk mass has the opposite effect in the sense that the kinematic dispersions increase with time. Inter alia, this has implications for the old $\alpha$-rich disk in the Milky Way, which has a vertical dispersion of about $\sigma_z^{\rm T} \approx 50$ km s$^{-1}$ today \citep{bla16a}, a factor of two larger than the more massive $\alpha$-poor disk. Such a large value for a rotating disk does {\it not} need to reflect its intrinsic dispersion at birth. 

The $\alpha$-rich disk preceded the more massive $\alpha$-poor disk, which has built up mostly through quiescent accretion over past $9-10$ Gyr, and shares the same gravitational potential. (The last significant merger event before the disrupting Sgr dwarf observed at the current epoch appears to date back to about $z\sim 2$.) Thus, the build-up of the
$\alpha$-poor disk over billions of years may have increased the $\alpha$-rich {\it in situ} dispersion significantly from an initially low value, say, $\sigma_z^{{\rm T}_i} \approx 25-30$ km s$^{-1}$ as a consequence of conserved action. This possibility appears to have been overlooked to date.


In Fig.~\ref{f:disp_evol}, the message is that halo accretion significantly increases the gas and stellar dispersions $-$ at all radii and for all time $-$ by almost a factor two in some cases, compared to the no-accretion models. Within an aperture of $R=2.2 R_{\rm disk}$, for the ``no halo accretion'' models, we have an average value of $\langle\sigma_z\rangle \approx 18$ km s$^{-1}$ at the highest $f_{\rm gas}$, declining to $\langle\sigma_z\rangle \approx 6$ km s$^{-1}$ for the most gas-poor model. For the halo accretion models, taken at the same time ($t_{\rm o} \approx 1$ Gyr), the vertical dispersions are $\langle\sigma_z\rangle \approx 30$ km s$^{-1}$ and $\langle\sigma_z\rangle \approx 13$ km s$^{-1}$ respectively. The ionized gas dispersions over the same apertures are factors of $2-3$ times higher. These average values compare favourably with the warm (e.g. H$\alpha$) and cold (e.g. CO) gas diagnostics for the range of values seen in galaxies matched in total baryonic mass \citep[e.g.][]{ejdet22}. In upcoming papers, we carry out detailed comparisons of our simulations with the latest high-redshift results \citep[e.g.][]{tsukui23,neel23}.

\begin{figure}[!htb]
    \centering
    \includegraphics[width=0.4\textwidth]{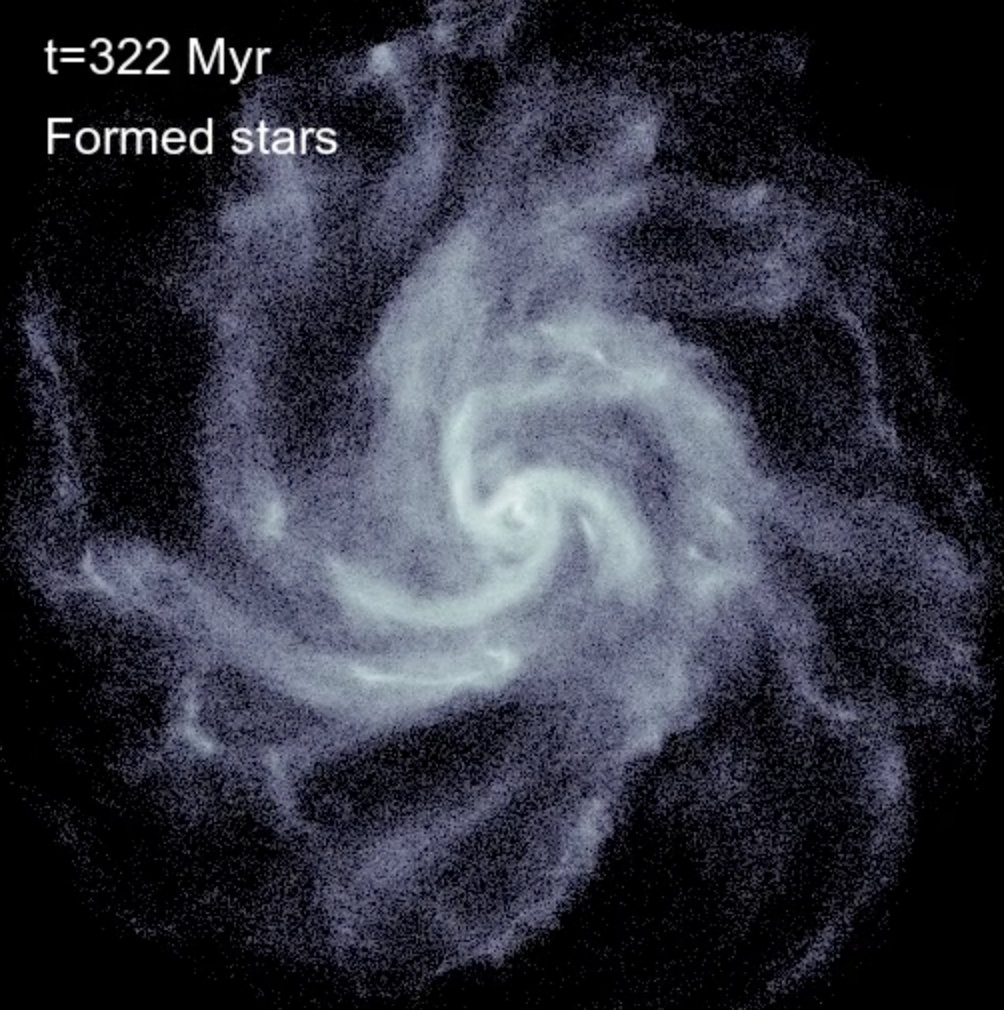}
    \includegraphics[width=0.4\textwidth]{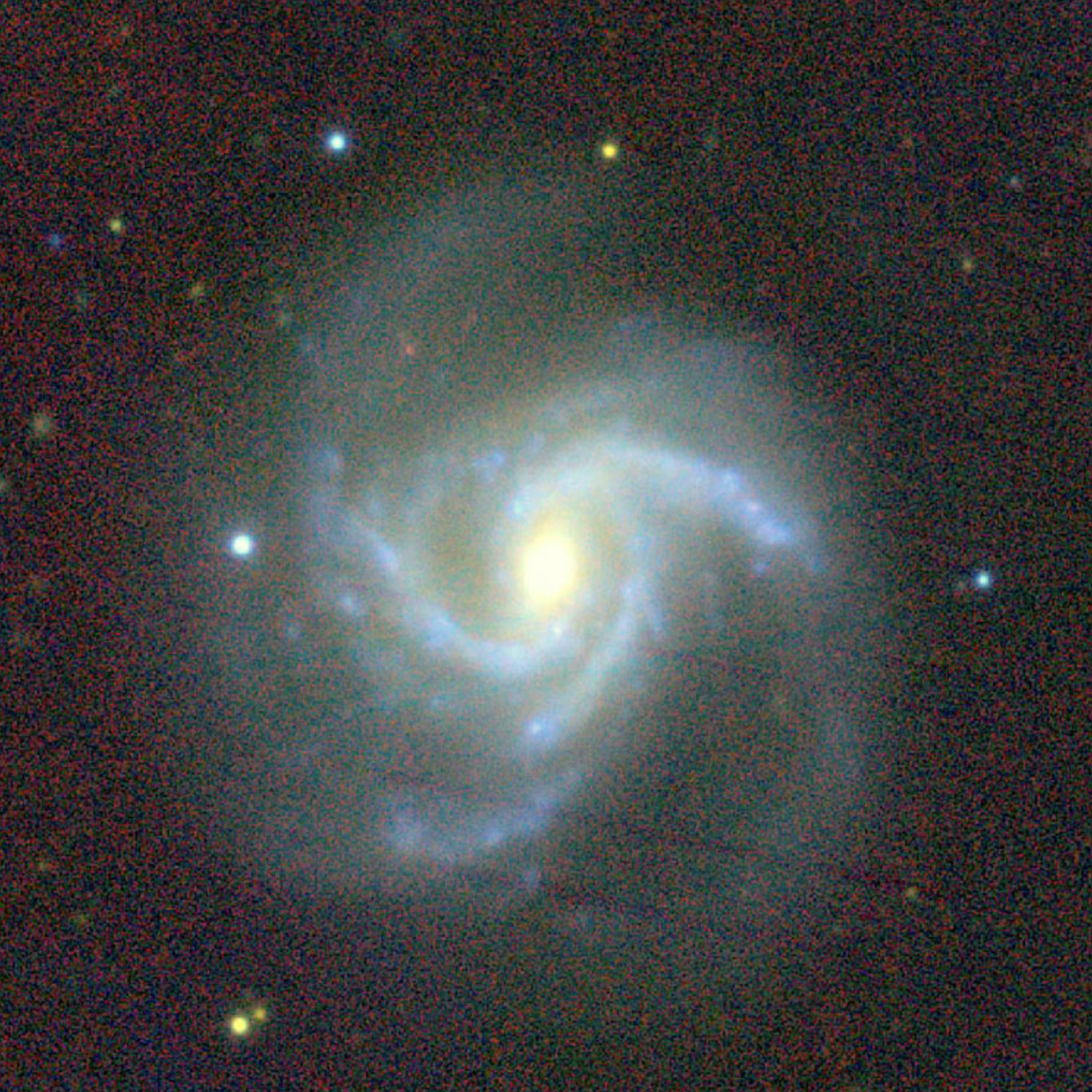}
    \caption{(Left) Three-arm spirals are a common occurrence in the early stages of gas-rich simulations for which disks dominate the local gravitational potential. The model uses $f_{\rm disk}=0.5$ with a gas fraction of 40\%. We ran the simulation with three different random seeds, and 3-arm spirals are observed to be transitory (of order a rotation period) in each case. (Right) They are occasionally seen in nearby galaxies and at least one is claimed at high redshift (see text). The example shown here is NGC 7309 taken from C. Seligman's atlas at \url{https://cseligman.com}.
    }
   \label{f:3arm}
\end{figure}

\subsection{Possible manifestations of turbulent gas disks}


The most obvious manifestations of turbulent galaxies at high redshift are the observed high gas fractions, elevated star formation rates and broadened gas kinematics seen in ALMA and JWST surveys. These have been the main themes of our paper. 
To date, all of the enhanced star-forming disks at high redshift exhibit larger velocity dispersions than for local disks, regardless of the emission diagnostic used \citep{ejdet22}, and presumably these reflect enhanced levels of turbulent energy.
In future observations, these same galaxies are expected to have very active circumgalactic media that may be observable in ionized emission lines or in warm dust emission, and almost certainly in absorption-line diagnostics along quasar sightlines, for example \citep{tum17}. It is unlikely that we can probe the stellar kinematics in the same way, except for fossil signatures that live on in local stellar populations, discussed briefly in the last section. But our simulations show a few manifestations that are worth contemplating in future observations. 

An interesting feature of our gas-rich models is the occasional appearance of 3-arm ($m=3$) spirals before the two-arm spiral and/or bar-like shear flow sets in. For example, 3-arm spirals emerge, albeit temporarily, in all three versions of {\tt fd50\_fg40\_nac}; we show one such timestep in Fig.~\ref{f:3arm} (left). In the near field, there are spectacular 3-arm spirals known \citep{elme92,hanc19}, but this manifestation is relatively rare across disk galaxies; we show one example in Fig.~\ref{f:3arm} (right). The few studies that do exist agree that the 3-arm spiral is likely to be transitory and reflects a global instability, particularly if it is confined within the $m=3$ resonances, i.e. $\Omega_s \pm \kappa/3$, where $\Omega_s$ is the spiral arm pattern speed, which appears to be the case. It can arise from an $m=1$ mode (i.e. lopsided disturbance) interacting with an $m=2$ mode (e.g. tidal perturbation). To our knowledge, such behaviour has not been modelled before, although $m=3$ kinematic patterns have been considered \citep{canz93}. We raise the issue here because a 3-arm spiral has been suggested in a turbulent, gas-rich disk at $z=2.2$ \citep{Law12}. In brief, our models appear to confirm that this manifestation is relatively shortlived and indicative of large-scale instabilities early in the lifecycle of the disk.
\begin{figure}[!htb]
    \centering
\includegraphics[width=0.3\textwidth]{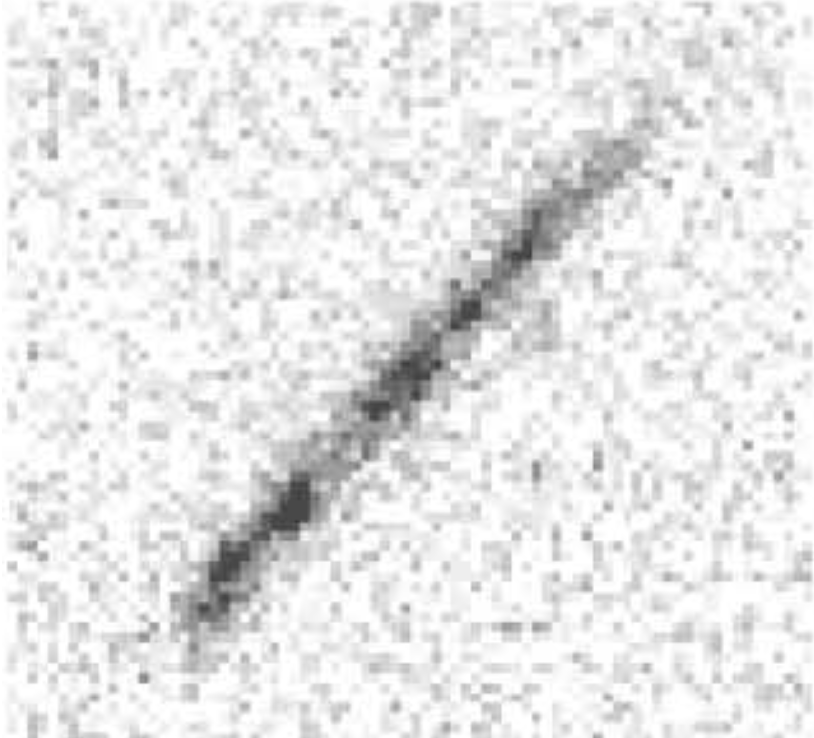}
\includegraphics[width=0.32\textwidth]{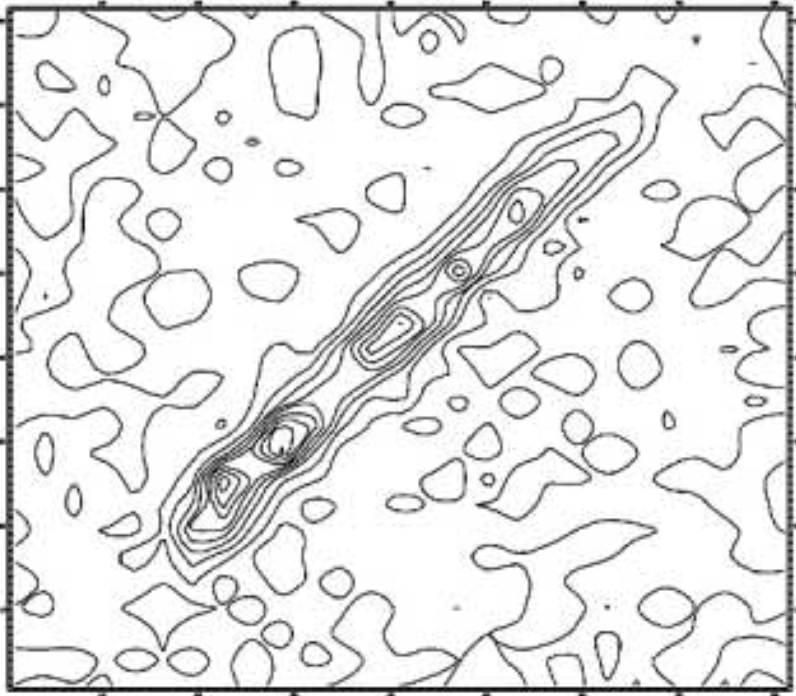}
\includegraphics[width=0.29\textwidth]{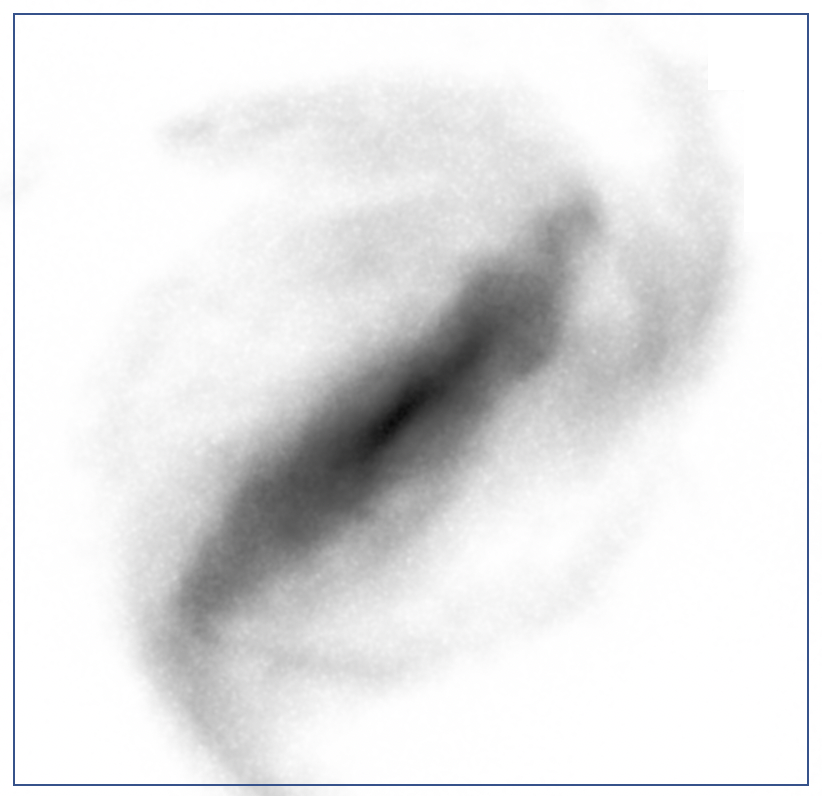}
    \caption{Example kiloparsec-scale `chain galaxy' object from \citet{elme04} where the $V$ band intensity is shown (left) in grayscale and (middle) in contours;
    a more up-to-date collage is presented in \citet[][their figure 17]{pand24}. The contact layer in the radial shear flow produces stars along a narrow corridor (right); only the young stars are shown here; radiation transfer in dense gas (and matched instrument response) is needed to properly compare to the left and middle figures. This is a single timestep ($t_{\rm o} = 1.677$ Gyr) taken from {\tt fd50\_fg20\_nac} for illustration purposes. The few chain galaxies that have kinematic signatures show either weak or no evidence for velocity gradients along their axis, as observed in the simulated shear flow. Are some chain galaxies powered by radial shear flows in turbulent disks?}
   \label{f:chain}
\end{figure}

Since their discovery with the Hubble Space Telescope \citep{cow95}, `chain galaxies' have defied an easy explanation \citep{elme04}. These enigmatic objects were recently revisited in a new JWST study \citep{pand24}. There are of order a hundred linear objects at $z=1-3$ identified to date and these may comprise a heterogeneous class of objects, e.g. interacting and/or edge on clumpy galaxies \citep[e.g.][]{dekel09b,agertz09}, or lensed systems, but even so, many are difficult to explain, e.g. the example shown in Fig.~\ref{f:chain}. They are typically at the detection limit of the HST in most bands, are sufficiently bright in the restframe blue bands to suggest active star formation, and are of order a few kiloparsecs in linear dimension. Intriguingly, few show signs of a velocity gradient consistent with an edge-on disk undergoing circular rotation \citep[e.g.][]{bunk00}. It is plausible that at least some of these sources may arise from star formation confined to a radial shear flow. This could explain the lack of a velocity gradient along the chain axis in some sources. But at higher spatial resolution, the shearing effect could be visible perpendicular to the flow (see Fig.~\ref{f:shear_bar}). Further work is needed to establish if these reside within an extended cool gas disk since the radial shear flow cannot exist in isolation.

\bigskip
\section{Main results} \label{s:main}
The surprising and compelling new evidence for well-developed disks (stars$+$gas) at early times (up to at least $z\approx 7$) is one of the most pressing problems in galaxy formation studies. The first hydro/N-body cosmological simulations tended to overload the central regions of galaxies with baryons. This was known as the ``overcooling problem'' and led to the introduction of feedback mechanisms to disperse the baryons over larger radial scales \citep[e.g.][]{ben10a}. Just how the models are to be fixed to ensure dominant central baryons with net rotation at early times is not at all clear \citep{kre21,gur22}.


In \cite{bla23}, we set out to investigate and extend important work \citep{fuj18a,fujii19} that deserves wider attention in light of new results for high-redshift disks in recent years \citep[e.g.][]{rizzo20,Price2021}.
Even without any external interaction, given enough time, all substantial disks succumb to bar instabilities eventually.
We refer to the exponential dependence of the bar formation time ($\tau_{\rm bar}$) as a function of $f_{\rm disk}$ as the `Fujii relation.' 
Above some limiting value of $f_{\rm disk}$ ($\approx 0.30\pm 0.05$),
the bar formation time scales exponentially fast, with $1 < \tau_{\rm bar} < 2$ Gyr for most models. We find that the presence of a bar in a high-redshift disk puts a lower limit on $f_{\rm disk}$ for a given redshift. Fig.~\ref{f:fdisktime} is particularly useful because $f_{\rm disk}$ can be estimated independently from the disk kinematics \citep[e.g.][]{gen20,forster2020,Price2021}.

In our new work, we explore the evolution of gas-rich disks in the early Universe in the presence of strong turbulence driven by the energy output from star formation. We examine in detail the distinctive signatures of both no halo accretion and halo accretion models. In the former case, the gas supply and star formation rate decline fairly rapidly over an exponential timescale of about a Gyr; in the latter case, both decline once again but over a longer timescale.

\medskip
Here are the summary points of our work to date where we include the earlier results in the first two items:
\begin{itemize}
\item For internally triggered stellar bars, there is a clear dependence between the disk mass fraction ($R\lesssim 2.2 R_{\rm disk}$) and the onset time of the stellar bar. After that, the bar is long-lived for gas-free simulations.
\item If the disk baryons dominate the inner galaxy, stellar bars are likely to be common at redshifts currently surveyed by JWST and ALMA imaging ($z\lesssim 7$).
\item If the disk baryons are sub-dominant, the bar onset time is very long ($t>5$ Gyr) such that the bar is unlikely to form at high redshift; the fraction of baryons in gas does not alter this conclusion.
\item If the disk baryons dominate the inner galaxy, the bar onset time does depend on the gas fraction. At $f_{\rm gas}=20$\%, the turbulent gas appears to speed up bar formation by 50\% or more. For higher gas fractions, the bar onset time is at least a factor of two shorter than the gas-free case.
\item In all cases, the stellar bars (in the presence of gas) are weaker than their gas-free counterpart. The bar strength and bar length appear to be inversely related to the gas fraction $f_{\rm gas}$.
\item As in observations, ``gas bars'' exist but are mostly stochastic and intermittent features. The observed bar-like objects seen with ALMA (see Sec. 1) may arise from young stars on bar-like orbits heating up the associated dust and molecules.
\item In the high $f_{\rm gas}$  ($>70\%$) limit, a (mostly gas) bar emerges but eventually collapses to form a bulge.
\item For all gas-rich models, a radial shear flow is active. These produce the distinctive quadrupolar (quatrefoil) pattern in the gas kinematics, particularly at low $f_{\rm gas}$, which may be visible in future ALMA observations.
\item The predicted gas dispersions in the cool ($\lesssim 10^3$~K) gas are typically $\sigma_z \lesssim 30$ km s$^{-1}$ for both the halo accretion and no accretion models. The warm gas dispersions are consistently $2-3$ times higher than their cool counterparts.
\end{itemize}

\bigskip
\section{Next steps} \label{s:next}
Very little is known about the character, formation or evolution of early galactic disks ($z > 3$). These objects are currently at the observational limits of our most powerful telescopes and appear to be different in character from their low-redshift counterparts. We have attempted to provide a framework for advancing the discussion of these enigmatic objects. 

A major omission at the present time is that there are only a handful of high-redshift galaxies with combined JWST and ALMA imaging \citep[e.g.][]{Wu23,amvro24}, or combined JWST imaging and integral field spectroscopy \citep{forster2020}.
This issue will need to be addressed in the coming years. We anticipate that there will be large samples of galaxies with estimated $f_{\rm disk}$ values in the near future, out to the highest ALMA redshifts. JWST imaging will be needed to determine the baryon mass fraction in stars compared to gas, although significant dust corrections may be required. We fully anticipate that early disks will show evidence of blue, star-forming bars, nuclear disks and spiral arms, even while dominated by gas turbulence. Central stellar bulges are expected to be relatively small until later times ($z<2$).

We have made the case for tracing departures from axisymmetry as evidence for the importance of baryons over dark matter in the inner regions of disk galaxies. This goes to the heart of how galaxies form and evolve at early times. Stellar bars and spiral arms are now seen well beyond $z\sim 2$; this was quite unexpected. Adding to the mystery, these manifestations were thought to be unlikely to occur in the presence of strong gas turbulence. 
All of the simulations show long-term evolution in their baryonic properties and distributions, as was demonstrated in Sec.~\ref{s:galarch}. There are aspects of the existing simulations that we explore in more detail in later work, e.g. metal production, secular evolution (e.g. migration), 3D stellar dispersion (e.g. disk thickening), and the long-term evolution of the models within a CDM hierarchy.

A key step is to understand better how our controlled experiments fit within the context of cosmological evolution. We have begun to consider the response of turbulent disks to external forces, in particular, the strong dynamical impulse from a merging system. Can the radial shear flow survive a merger interaction (A. Wetzel, personal communication)? An interesting prospect is that the radial shear flow is able to feed an active galactic nucleus, particularly at late times and high gas fractions when the gas bars collapse to a central bulge. In turn, how does the turbulent disk respond to strong AGN feedback?

Our early results point the way to future refinements and developments, especially in forward modelling to improve comparison with observations.  We can use our simulations to interpret the limited multi-wavelength data for high-redshift disks. Specifically, we now post-process the multiphase simulations to address specific emission lines targetted by ALMA. These are particularly useful for establishing whether the inferred star formation rates, gas and dust fractions and total masses, etc. are mutually consistent. This line of analysis will benefit greatly from recent developments in non-equilibrium chemistry networks coupled to on-the-fly multifrequency radiation transport in \ramses\ \citep[PRISM:][]{katz22}. Moreover, the primary source of ionization or heating can be inferred from the relative strength of C$^+$, O and O$^+$ lines, inter alia. These lines have been used to infer gas masses but, in most instances, the line strengths are a reflection of cooling rates and how much energy was dumped into these lines initially \citep{apple17}.

The predicted line strengths can guide our understanding of whether the simulated feedback processes coupling to the gas reflect the actual physical conditions. For example, in our attempt to model ALMA galaxy disks, we are forced to high gas fractions in 
dominant disks. But as we have seen, this can lead to a catastrophic loss of disk baryons in extrema, at least in our models, although the baryons are re-accreted in some models. Is it really true that some ALMA disks lay the seeds of their own destruction, or does this evolutionary path solve the longstanding mystery of superthin disc galaxies \citep{goad81,ban13,ossa23}?

Arguably, the most urgent improvement is to incorporate magnetohydrodynamics, accompanied by cosmic ray heating \citep{farcy22}, with good spatial resolution while operating on galactic scales. There are solid theoretical reasons for believing that magnetic fields influence the star formation prescriptions (Sec.~\ref{s:sfr}). Moreover, it is feasible that, at even higher resolution, denser filaments emerge leading to a lower covering fraction and filaments that are harder to accelerate \citep[e.g.][]{coop08,coop09}, thus a lower overall coupling efficiency. This fundamental limitation has yet to be addressed by any of the cosmological simulations targetting galaxies in the early Universe. Rather than ``feedback-free'' galaxy evolution \citep[e.g.][]{dekel23}, there is a case for weakening the strong coupling by a substantial factor. This is a problem that extends across to all cosmological simulations, and needs to be investigated.

\section{Acknowledgments}
JBH, TTG and OA wish to thank Professors Paola Di Matteo and Misha Haywood for hosting them at the Paris and Meudon Observatories, for their hospitality and excellent research culture. We received valuable feedback from Ken Freeman, Mark Krumholz, Fran\c{c}ois Boulanger, Andrew Wetzel, Takafumi Tsukui, Emily Wisnioski, Francesca Rizzo, Filippo Fraternali and Fernanda Roman de Oliveira. We are grateful to Eugene Vasiliev for continuous assistance with \agama\ and to Walter Dehnen for help with his code. JBH is indebted to James Binney for long walks, pub lunches and inspired conversations, and to Merton College and the Beecroft building for providing a stimulating research environment. 

TTG acknowledges financial support from the Australian Research Council (ARC) through an Australian Laureate Fellowship awarded to JBH. OA acknowledges support from the Knut and Alice Wallenberg Foundation, the Swedish Research Council (grant 2019-04659) and the Swedish National Space Agency (SNSA Dnr 2023-00164). CF acknowledges funding provided by the Australian Research Council (Discovery Project DP230102280), and the Australia-Germany Joint Research Cooperation Scheme (UA-DAAD).

The computations and data storage were enabled by two facilities: (i) the National Computing Infrastructure (NCI) Adapter Scheme, provided by NCI Australia, an NCRIS capability supported by the Australian Government; and (ii) LUNARC, the Centre for Scientific and Technical Computing at Lund University (resource allocations LU 2023/2-39 and LU 2023/12-6).

Finally, we are indebted to an insightful referee who encouraged us to think harder about the implications of this work, in addition to improving our overall presentation.

\bibliographystyle{aasjournal}


\end{document}